\documentclass[journal]{IEEEtran}

\usepackage{cite}
\ifCLASSINFOpdf
  \usepackage[pdftex]{graphicx}
\else
  \usepackage[dvips]{graphicx}
\fi

\usepackage[cmex10]{amsmath}
\usepackage{amsmath,amsfonts,amssymb}
\usepackage{mdwmath}
\usepackage[ruled]{algorithm2e}
\usepackage{array}
\usepackage{fixltx2e}
\usepackage{bm}
\usepackage{stfloats}
\usepackage{graphicx}
\usepackage{caption}
\usepackage{mdwtab}
\usepackage{subfigure}
\usepackage{booktabs}
\usepackage{multirow}
\usepackage{etoolbox}
\usepackage{makecell}
\usepackage{xcolor}
\usepackage[bottom]{footmisc}
\usepackage{mdframed}

\newtheorem{lemma}{Lemma}

\newtheorem{remark}{Remark}
\newcommand{\e}{\begin{equation}}
\newcommand{\ee}{\end{equation}}
\newcommand{\eqn}{\begin{eqnarray}}
\newcommand{\eeqn}{\end{eqnarray}}

\begin{document}
% paper title
\title{Terahertz Ultra-Massive MIMO-Based\\
Aeronautical Communications in\\
Space-Air-Ground Integrated Networks}
\author{
Anwen Liao, Zhen Gao,~\IEEEmembership{Member,~IEEE}, Dongming Wang,~\IEEEmembership{Member,~IEEE}, Hua Wang, Hao Yin, \\
Derrick Wing Kwan Ng,~\IEEEmembership{Fellow,~IEEE}, and Mohamed-Slim Alouini,~\IEEEmembership{Fellow,~IEEE} %
%\vspace*{-5.0mm}
\thanks{The codes and some other associated materials of this work may be available at https://gaozhen16.github.io.}

\thanks{A. Liao, Z. Gao, and H. Wang are with School of Information and Electronics,
 Beijing Institute of Technology, Beijing 100081, China (E-mails: \{liaoanwen, gaozhen16,
 wanghua\}@bit.edu.cn).}
\thanks{D. Wang is with the National Mobile Communications Research Laboratory,
 Southeast University, Nanjing 210096, China (e-mail: wangdm@seu.edu.cn). }
\thanks{H. Yin is with Institute of China Electronic System Engineering Corporation,
 Beijing 100141, China (E-mail: \mbox{yinhao@cashq.ac.cn}).}
\thanks{D. W. K. Ng is with the School of Electrical Engineering and Telecommunications,
 University of New South Wales, Sydney, NSW 2052, Australia (e-mail: w.k.ng@unsw.edu.au). }
\thanks{M.-S. Alouini is with the Electrical Engineering Program, Division of Physical
 Sciences and Engineering, King Abdullah University of Science and Technology, Thuwal,
 Makkah Province, Saudi Arabia (E-mail: slim.alouini@kaust.edu.sa).}
}

%\markboth{Journal of \LaTeX\ Class Files,~Vol.~x, No.~x, September~2019}%
%{Liao \MakeLowercase{\textit{et al.}}: Manuscript of IEEEtran.cls for IEEE Journals}

\maketitle

\begin{abstract}
 The emerging space-air-ground integrated network has attracted intensive research and necessitates reliable and efficient aeronautical communications.
 This paper investigates terahertz Ultra-Massive (UM)-MIMO-based aeronautical communications and proposes an effective channel estimation and tracking scheme, which can solve the performance degradation problem caused by the unique {\emph{triple delay-beam-Doppler squint effects}} of aeronautical terahertz UM-MIMO channels.
 Specifically, based on the rough angle estimates acquired from navigation information, an initial aeronautical link is established, where the delay-beam squint at transceiver can be significantly mitigated by employing a Grouping True-Time Delay Unit (GTTDU) module (e.g., the designed {\emph{Rotman lens}}-based GTTDU module).
 According to the proposed prior-aided iterative angle estimation algorithm, azimuth/elevation angles can be estimated, and these angles are adopted to achieve precise beam-alignment and refine GTTDU module for further eliminating delay-beam squint.
 Doppler shifts can be subsequently estimated using the proposed prior-aided iterative Doppler shift estimation algorithm.
 On this basis, path delays and channel gains can be estimated accurately, where the Doppler squint can be effectively attenuated via compensation process.
 For data transmission, a data-aided decision-directed based channel tracking algorithm is developed to track the beam-aligned effective channels.
 When the data-aided channel tracking is invalid, angles will be re-estimated at the pilot-aided channel tracking stage with an equivalent sparse digital array, where angle ambiguity can be resolved based on the previously estimated angles.
 The simulation results and the derived Cram\'{e}r-Rao lower bounds verify the effectiveness of our solution.
\end{abstract}


\IEEEpeerreviewmaketitle

\section{Introduction}\label{S1}

 Terahertz (THz) communication is expected to play a pivotal role in the future Sixth Generation (6G) wireless systems, which promise to provide ubiquitous connectivity with broader and deeper coverage~\cite{Yang_Netw19}.
 THz-band (spectrum ranges from 0.1 to 10 THz) is envisioned to offer significantly larger
 bandwidths than millimeter-Wave (mmWave) for supporting up to tens of Gigahertz (GHz) ultra-broadband
 and Terabit per second (Tbps) ultra-high peak data rate \cite{Akyildiz_PC14,HanC_TWC15,Akyildiz_CM18}.
 Meanwhile, THz communications can be conducive to realize the Ultra-Massive Multiple-Input Multiple-Output
 (UM-MIMO)-based transceivers equipped with tens of thousands of antennas (even the Uniform Planar Array (UPA) with size of $1024\! \times\! 1024$ \cite{Slim_JSAC19}),
 which can effectively combat the severe path loss of THz signals and further extend the communication range
 using beamforming techniques \cite{Akyildiz_NCN16,Akyildiz_ADN19,Akyildiz_Access20}. Therefore, THz UM-MIMO technique has been
 emerging as a promising candidate for the 6G mobile communication systems \cite{Yang_Netw19}.
 However, due to the severe atmospheric molecular absorption (such as water vapor) and rain attenuation \cite{Akyildiz_Access20,HanC_CM18},
 the applications of THz communications are restricted to short-link distance \cite{Akyildiz_WC14,HanC_TSP16,HanC_TTHz16}.
 Fortunately, those atmospheric molecule absorption and rain attenuation mainly occur in the troposphere,
 and these negative factors can be largely mitigated due to the negligible absorption in the stratosphere and above \cite{Xu_TAP19,Slim_OJCS20,Saeed_PC20}.

\begin{figure*}[!tp]
%\vspace{-5mm}
\begin{center}
 \includegraphics[width=1.2\columnwidth,keepaspectratio]{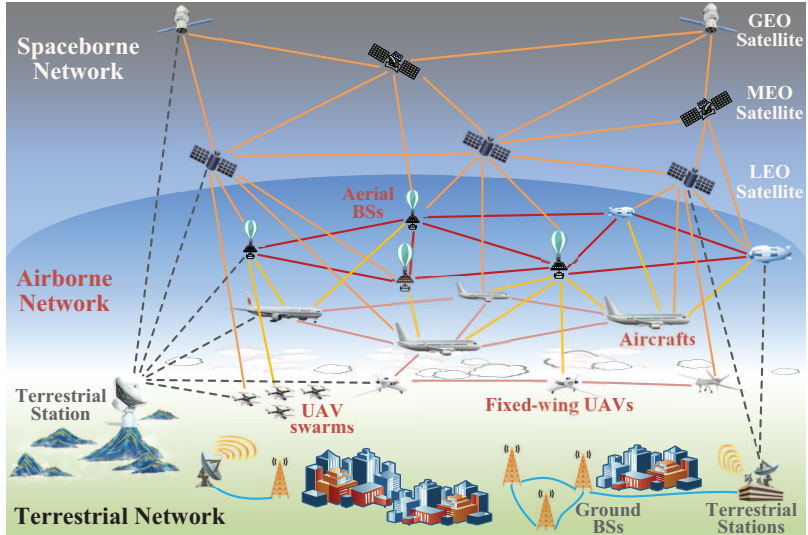}
\end{center}
 \captionsetup{font = {footnotesize}, singlelinecheck = off, justification = raggedright, name = {Fig.}, labelsep = period}%
\caption{Typical SAGIN includes spaceborne, airborne, and terrestrial networks, where numerous LEO satellites,
 aerial BSs, aircrafts, and UAVs together constitute the aeronautical \emph{ad hoc} network \cite{Hanzo_VTM19,Zhang_ProcIEEE19}.}
 \label{FIG1}
%\vspace{-6mm}
\end{figure*}

 On the other hand, the ambitious 6G is poised to seamlessly integrate space-air networks with terrestrial mobile cellular networks.
 Against this background, the concept of Space-Air-Ground Integrated Network (SAGIN)
 is conceived and has attracted intensive research \cite{Liu_Tutor18,Hanzo_VTM19}.
 As shown in Fig.~\ref{FIG1}, a typical SAGIN consists of three layers including spaceborne, airborne,
 and terrestrial networks \cite{Hanzo_VTM19}.
 The Geostationary Earth Orbit (GEO), Medium Earth Orbit (MEO), and Low Earth Orbit (LEO) satellites
 that operate at different altitudes constitute the spaceborne network.
 In the airborne network, aerial Base Stations (BSs) such as balloons and airships can jointly serve various aircrafts
 and Unmanned Aerial Vehicles (UAVs).
 In particular, numerous LEO satellites, aerial BSs, aircrafts, and UAVs can constitute the
 aeronautical \emph{ad hoc} network to achieve the goal of ``Internet above the clouds''~\cite{Zhang_JSAC18,Zhang_ProcIEEE19},
 which necessitates THz UM-MIMO technique to support the reliable and efficient aeronautical communications\footnote{In general, civil aircrafts spend most of their flight time at the bottom of the stratosphere, where the relatively stable flight state is convenient for the establishment of THz communication links. Therefore, the aeronautical communications studied in this paper can be mainly aimed at the aircrafts flighted at the stratospheric.}.

 To guarantee the Quality-of-Service (QoS) for THz UM-MIMO-based aeronautical communications,
 reliable Channel State Information (CSI) acquisition at the transceiver is indispensable \cite{Ogbe_TAES19}.
 However, due to the high-speed mobility of flying aircrafts/UAVs and the wobbles of aerial BSs,
 these aerial communication links exhibit the dramatically fast time-varying fading characteristics,
 which make accurate channel estimation and tracking rather challenging.
 To acquire the accurate estimate of fast time-varying channel, some channel estimation
 and  tracking schemes \cite{Gong_JSAC19,ZhangS_Tcom19M,ZhangS_Tcom19N} were proposed to
 reduce the training overhead caused by frequent channel estimation. In \cite{Gong_JSAC19},
 a data-aided channel tracking scheme is proposed to estimate and track the partial channel
 coefficients of angle domain channels using lens antenna array. By exploiting the sparsity
 of the virtual channel vector in angle domain, the virtual channel parameters based on
 first order auto regressive model were estimated and tracked using the expectation
 maximization-based sparse Bayesian learning framework in \cite{ZhangS_Tcom19M,ZhangS_Tcom19N}.
 Moreover, by acquiring the dominant channel parameters including the Angle of
 Arrivals/Departures (AoAs/AoDs), Doppler shifts, and channel gains, rather than the
 complete MIMO channel matrix, some multi-stage channel estimation solutions were proposed
 in \cite{Qin_TVT18,Cheng_WCL19} enabling fast channel tracking for narrow-band mmWave MIMO
 systems. Note that these schemes above just consider the channel estimation and tracking for
 common mmWave systems. In \cite{GaoXY_TVT17}, a priori-aided THz channel tracking scheme
 with low pilot overhead was proposed to predict and track the physical direction of
 Line-of-Sight (LoS) component of the time-varying massive MIMO channels in THz beamspace
 domain. For the dynamic indoor short-range THz communications, the authors
 in \cite{Peng_TVT17} proposed an AoA estimation method based on Markov process and
 Bayesian inference, where the forward-backward algorithm is implemented to carry out
 the Bayesian inference.

 However, the aforementioned channel estimation solutions are difficult to be applied to the aeronautical THz UM-MIMO systems
 due to the unprecedentedly ultra-large array aperture, ultra-broad band, and ultra-high velocity.
 {\emph{Compared with the sub-6 GHz or mmWave massive MIMO systems with limited aperture and bandwidth,
 the aeronautical THz UM-MIMO channels present the unique \textbf{triple delay-beam-Doppler squint effects}}}.
 To be specific, adopting the UPA form, the UM-MIMO arrays mounted on the transceiver of aerial BSs or aircraft can be equipped with
 up to hundreds of antennas in the single horizontal or vertical dimension,
 resulting in the ultra-large array aperture even in a small physical size.
 If the direction of arrival is not perpendicular to the array, we can observe different propagation delays at different antennas for the same received signal filling this array aperture.
 Moreover, this delay gap can be as large as multiple symbol periods due to the usage of ultra-broadband THz communications.
 This indicates that the inter-symbol-interference can be non-negligible even for the LoS link,
 and this phenomenon is termed as the {\emph{delay squint effect}} of THz UM-MIMO (also named as
 spatial-frequency wideband effects in \cite{GaoFF_CM18,GaoFF_TSP18} and
 aperture fill time effect in radar systems \cite{Krozer_TMTT10}), which is an inevitable challenge for THz UM-MIMO systems.
 Meanwhile, this delay squint effect can further introduce the {\emph{beam squint effect}},
 where the beam direction is a function of the operating frequency.
 This is primarily because radio waves at different frequencies would accumulate different phases
 given the same transmission distance, while the adjacent antenna spacing is designed according to
 the central carrier frequency. Hence, beam squint effect would pose undesired beam directions
 for the signals at marginal carrier frequencies. Furthermore, the high-speed mobility of aeronautical communications
 causes large Doppler shift and the Doppler shift is also frequency-dependent
 for aeronautical THz UM-MIMO with very large bandwidth. This phenomenon is called {\emph{Doppler squint effect}}.
 Therefore, the aeronautical THz UM-MIMO systems present {\emph{triple delay-beam-Doppler squint effects}}.
 However, recent researches mainly focus on the impact of beam squint effect on mmWave or THz systems \cite{Heath_TWC19,Hanzo_TWC20,GaoFF_TWC19,GaoFF_TSP19,GaoFF_TSP20}.
 To be specific, the impact of beam squint on compressive subspace estimation and the optimality of frequency-flat beamforming was studied in \cite{Heath_TWC19}.
 By projecting all frequencies to the central frequency and constructing the common analog Transmit Precoding (TPC) matrix for all subcarriers, several hybrid TPC schemes were proposed in \cite{Hanzo_TWC20} to design the analog and digital TPC matrices and mitigate the beam squint effect.
 The channel estimation schemes were proposed to exploit the characteristics of mmWave channels affected by beam squint for estimating the wideband mmWave massive MIMO channels \cite{GaoFF_TWC19,GaoFF_TSP19,GaoFF_TSP20}, where the beam squint effect is not mitigated.
 To sum up, the triple squint effects are seldom considered in state-of-the-art channel estimation and hybrid beamforming solutions
 \cite{Gong_JSAC19,ZhangS_Tcom19M,ZhangS_Tcom19N,Qin_TVT18,Cheng_WCL19,GaoXY_TVT17,
 GaoFF_CM18,GaoFF_TSP18,Peng_TVT17,Heath_TWC19,Hanzo_TWC20,GaoFF_TWC19,GaoFF_TSP19,GaoFF_TSP20}
 and can dramatically degrade the data transmission performance of THz UM-MIMO-based aeronautical communications.
 Consequently, an efficient signal processing paradigm for channel estimation and data transmission is invoked
 for enabling aeronautical THz UM-MIMO technique.
 
\begin{figure}[!tp]
%\vspace{-5mm}
\begin{center}
 \includegraphics[width=0.9\columnwidth,keepaspectratio]{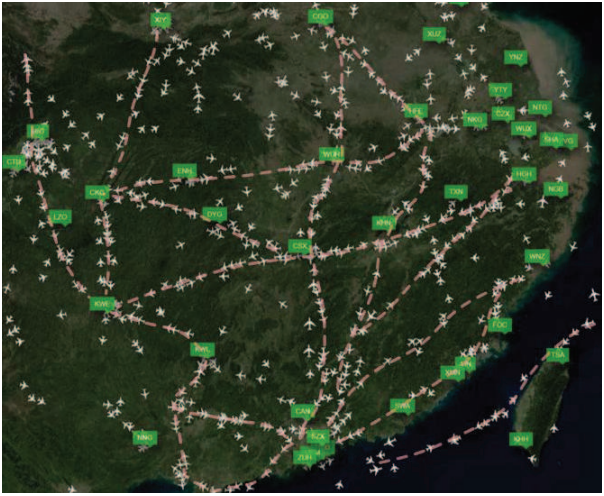}
\end{center}
 \captionsetup{font = {footnotesize}, singlelinecheck = off, justification = raggedright, name = {Fig.}, labelsep = period}%
\caption{A real-time flight tracking snapshot of civil aircrafts in south China,
 where the aircrafts generally fly along their fixed routes\protect\footnotemark.}
 \label{FIG2}
%\vspace{-6mm}
\end{figure}
\footnotetext{This real-time snapshot can be found on the website URL link:
 https://flightadsb.variflight.com/tracker/112.761836,29.084716/6.}

 In this paper, we mainly investigate the THz UM-MIMO-based aeronautical communication links connecting aircraft and aerial BSs in SAGIN\footnote{The proposed signal processing solution can also be applied to the space-space/space-air links between the UAVs and multiple aerial BSs, or between aircrafts/UAVs and multiple LEO satellites, etc, and the transmission links between the terrestrial stations built on high-altitude mountains and space-air networks. Furthermore, the research on space-ground or air-ground communications in SAGIN is beyond the scope of this paper, and it may be an important research direction of future work.}, where the practical triple squint effects of aeronautical THz UM-MIMO channel with LoS link will be considered.
 Specifically, for the airborne network in Fig.~\ref{FIG1}, the trajectories
 of aircrafts are usually regular along their fixed routes, as shown in Fig.~\ref{FIG2}.
 Based on this fact, the aerial BSs can be deployed near these trajectories to ensure
 that multiple aircrafts or UAVs can communicate with multiple aerial BSs for constituting
 the aeronautical \emph{ad hoc} network.
 Since there are few other scatterers in the stratosphere except high altitude platforms for THz aeronautical communications, we mainly focus on the THz UM-MIMO channel with only LoS component between the aerial BS and the aircraft in this paper.
 More specifically, we consider that
 multiple aerial BSs can jointly serve a high-speed mobile aircraft through respective THz LoS
 links, and different aerial BSs can be cooperated via THz backbone links connecting
 different aerial BSs or the air-to-ground backbone links.
 To combat the multipath effect at the receiver of aircraft caused by multiple THz LoS links, the Orthogonal Frequency-Division Multiplexing (OFDM) technique will be applied to this aeronautical communication system\footnote{To meet the high quality-of-service requirement for hundreds of people in the aircraft simultaneously, the relatively complicated high-order modulation methods, i.e., OFDM and Quadrature Amplitude Modulation (QAM), can be utilized to enhance the data transmission rate and throughput in this paper. Moreover, due to the high Peak-to-Average Power Ratio (PAPR) in OFDM systems, Discrete Fourier Transform-Spread-OFDM (DFT-S-OFDM) technique is also the potential alternative for THz UM-MIMO-based aeronautical communication systems.}.
 Among the THz links aforementioned, the THz UM-MIMO-based aeronautical communication links
 connecting the aircrafts and aerial BSs are the most challenging to be established due to
 their fast time-varying fading characteristics.
 On the one hand, by exploiting the prior information (e.g., positioning, flight speed and direction,
 and posture information) at aerial BSs and aircrafts, some rough channel parameter estimates
 (e.g., angle and Doppler shift) can be acquired for facilitating the link establishment.
 On the other hand, these rough channel parameter estimates are not accurate enough for data transmission.
 Particularly, due to the exceedingly long link distance and extremely narrow beamwidth of aeronautical THz UM-MIMO,
 a slight deviation of angle parameter resulted from the positioning accuracy error and the posture rotation
 of antenna arrays mounted on transceiver would lead to the undesired beam pointing.
 Therefore, how to effectively leverage the prior information above to establish and track
 the fast time-varying links is vital for THz UM-MIMO-based aeronautical communications.

\begin{figure*}[!tp]
%\vspace{-5mm}
\begin{center}
 \includegraphics[width=1.2\columnwidth,keepaspectratio]{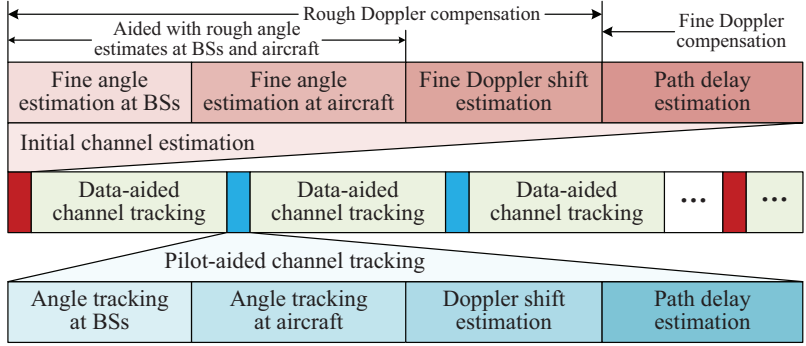}
\end{center}
 \captionsetup{font = {footnotesize}, singlelinecheck = off, justification = raggedright, name = {Fig.}, labelsep = period}%
 \caption{Frame structure of the proposed channel estimation and tracking solution.}
 \label{FIG3}
%\vspace{-6mm}
\end{figure*}

 The proposed channel estimation and tracking solution can be divided into three stages,
 including the initial channel estimation for link establishment, data-aided channel tracking,
 and pilot-aided channel tracking. The frame structure is shown in Fig.~\ref{FIG3},
 and the details are presented as follows:

$\blacktriangleright$ At the initial channel estimation stage, by utilizing the rough angle estimates acquired
 according to the positioning and flight posture information, the rough transmit beamforming
 and receive combining can be achieved to establish the THz UM-MIMO link, where the impact
 of delay-beam squint effects on both the transmitter and receiver can be significantly mitigated
 by employing a Grouping True-Time Delay Unit (GTTDU) module with low hardware cost.

$\blacktriangleright$ After the link establishment, the fine estimates of azimuth/elevation angles at both the
 transmitter and receiver, Doppler shifts, and path delays at the receiver are then obtained,
 where the rough Doppler shift estimates are utilized to compensate the received signals
 for improved parameter estimation. For the fine azimuth/elevation angle estimation,
 the UM hybrid array can be equivalently considered as a low-dimensional fully-digital array by
 employing a reconfigurable Radio Frequency (RF) selection network with dedicated connection pattern.
 In this way, the accurate estimates of azimuth/elevation angles at BSs and aircraft
 can be separately acquired using the proposed prior-aided iterative angle estimation algorithm.
 These fine angle estimates can be used not only
 to achieve the more precise beam alignment, but also to refine the GTTDU module
 at the transceiver for further eliminating the delay-beam squint effects.
 Meanwhile, thanks to the large beam alignment gain and the sufficient receive
 Signal-to-Noise Ratio (SNR), the Doppler shifts can be accurately estimated based on the
 proposed prior-aided iterative Doppler shift estimation algorithm, where the Doppler
 squint effect can be attenuated vastly by compensating the received signals with the
 rough Doppler shift estimates. On this basis, path delays and channel gains can be estimated
 subsequently, where Doppler squint effect can be also attenuated vastly via fine compensation process.

$\blacktriangleright$ At the data transmission stage, a Data-Aided Decision-Directed (DADD)-based channel tracking algorithm
 is developed to track the beam-aligned effective channels, where the correctly decoded data
 will be regarded as the known signals to estimate channel coefficients.

$\blacktriangleright$ The pilot-aided channel tracking is proposed when the data-aided channel tracking is ineffective.
 At this stage, an equivalent fully-digital sparse array will be formed by reconfiguring the connection
 pattern of the RF selection network, where the angle ambiguity issue derived from sparse array can be
 addressed with the aid of the previously estimated angles at the receiver. Once the precise beam
 alignment is achieved again, the Doppler shift and path delay estimation can be executed similar
 to the initial channel estimation stage, and then the transceiver will enter the data transmission
 stage again.

 The main contributions of our proposed scheme are summarized as the following aspects:
\begin{itemize}
\item
 THz UM-MIMO-based aeronautical communication channels exhibit the huge spatial dimension and very fast time-variability.
 To reduce the training overhead, we propose a parametric channel estimation and tracking solution.
 At the stages of initial channel estimation and pilot-aided channel tracking, by exploiting the
 proposed prior-aided iterative angle and Doppler shift estimation algorithms, the proposed solution can acquire
 the fine estimates of channel angles, Doppler shifts, and path delays, whereby some rough channel parameter estimates
 are leveraged to improve the estimated accuracy and reduce the pilot overhead. At the data transmission stage,
 to further save the pilot overhead, the proposed DADD-based channel tracking algorithm can reliably track the fast
 time-varying channel gains of the effective beam-aligned link.
\item
 The proposed scheme can effectively overcome the unique triple delay-beam-Doppler squint effects of aeronautical THz
 UM-MIMO communications. Note that this triple squint effects are rarely observed and investigated in the sub-6 GHz
 or mmWave massive MIMO systems due to the limited aperture and bandwidth. To cope with the delay-beam squint effects,
 we propose the low-cost GTTDU module at the transceiver, which can compensate the signal transmission delays at different
 antenna group with the aid of navigation information.
 In this way, the delay-beam squint effects can be significantly
 mitigated and the sufficient receive SNR can be guaranteed to establish the THz link.
 Also, the designed Rotman lens-based GTTDU module in Section~VIII provides a feasible implementation architecture of the tunable TTD module based Phase Shift Network (PSN), which would be a potential direction for the future research work.
 Furthermore, by utilizing the proposed prior-aided iterative angle and Doppler shift estimation algorithms to further mitigate the impact of beam and Doppler squint effects, the fine angle and Doppler shift estimates can be acquired for the following data transmission.
\item
 We introduce a reconfigurable RF selection network to obtain the equivalent low-dimensional fully-digital
 array by designing the dedicated connection pattern. On this basis, the robust array signal processing techniques
 such as Two-Dimensional Unitary ESPRIT (TDU-ESPRIT) \cite{Haardt_SSD_TSP98,Liao_Tcom19}
 can be utilized to accurately estimate and track the azimuth/elevation angles at the transceiver.
 Particularly, by reconfiguring the connection pattern of the RF selection network, the equivalent fully-digital
 sparse array can be obtained for improved angle estimation accuracy at the pilot-aided channel tracking stage,
 where angle ambiguity issue can be addressed well based on the previously estimated angles.
\item
 The Cram\'{e}r-Rao Lower Bounds (CRLBs) of dominant channel parameters are derived based on the effective received signal models. Particularly, at the pilot-aided channel tracking stage, the CRLBs of angles are derived
 to theoretically verify the improved estimation accuracy by employing the sparse array. Simulations results have
 the good tightness with the analytical CRLBs, which testifies the good performance of the proposed scheme.
\end{itemize}

 The remainder of this paper is organized as follows.
 Section~\ref{S2} introduces the system model, including the signal transmission and channel models with triple squint effects.
 The initial channel parameter estimation stage, including the estimations of azimuth/elevation angles at BSs and aircraft, Doppler shifts, path delays, and channel gains, is illustrated in Section~\ref{S3}.
 The DADD-based channel tracking and the pilot-aided channel tracking methods are proposed in Sections~\ref{S4} and~\ref{S5}, respectively.
 Section~\ref{S6} presents the performance analysis on CRLB and computational complexity.
 The numerical evaluations is given in Section~\ref{S7}.
 Finally, Section~\ref{S8} concludes this paper.

 Throughout this paper, boldface lower and upper-case symbols denote column vectors and matrices, respectively. % italic
 $(\cdot)^*$, $(\cdot)^{\rm T}$, $(\cdot)^{\rm H}$, $(\cdot)^{-1}$, and $|\cdot|$ denote the conjugate, transpose, Hermitian transpose,
 matrix inversion, and modulus operators, respectively.
 ${\| {\bm{a}} \|_2}$ and ${\| {\bm{A}} \|_F}$ are the ${\ell_2}$-norm of ${\bm{a}}$ and the Frobenius norm of ${\bm{A}}$, respectively.
 The Kronecker and Hadamard product operations are denoted by $\otimes$ and $\circ$, respectively.
 $\left\langle \bm{a},\bm{b} \right\rangle$ expresses the inner product of vectors $\bm{a}$ and $\bm{b}$.
 $\bm{0}_n$ and $\bm{I}_n$ denote the vector of size $n$ with all the elements being $0$ and the $n\times n$ identity matrix, respectively.
% and $\bm{O}_{m\times n}$ is the null matrix of size $m\times n$,
% while $\bm{1}_n$ denotes the vector of size $n$ with all the elements being $1$.
 $|{\cal Q}|_c$ is the cardinality of the set ${\cal Q}$, and $\{{\cal Q}\}_n$ denotes the $n$th element of the ordered set ${\cal Q}$.
 $[\bm{a}]_{{\cal Q}}$ denotes the sub-vector containing the elements of $\bm{a}$ indexed in the ordered set ${\cal Q}$.
 $[\bm{a}]_m$ and $[\bm{A}]_{m,n}$ denotes the $m$th element of $\bm{a}$
 and the $m$th-row and the $n$th-column element of $\bm{A}$, respectively.
 $\text{diag}(\bm{a})$ is the diagonal matrix with the elements of $\bm{a}$ at its diagonal entries.
 $\partial(\cdot)$ and ${\partial^2}(\cdot)$ are the first- and second-order partial derivative operations, respectively.
 Finally, $\mathbb{E}(\cdot)$ and $\Re \{\cdot\}$ denote the expectation and real part of the argument, respectively.

\section{System Model}\label{S2}

\begin{figure*}[!tp]
%\vspace{-5mm}
\begin{center}
 \includegraphics[width=1.7\columnwidth, keepaspectratio]{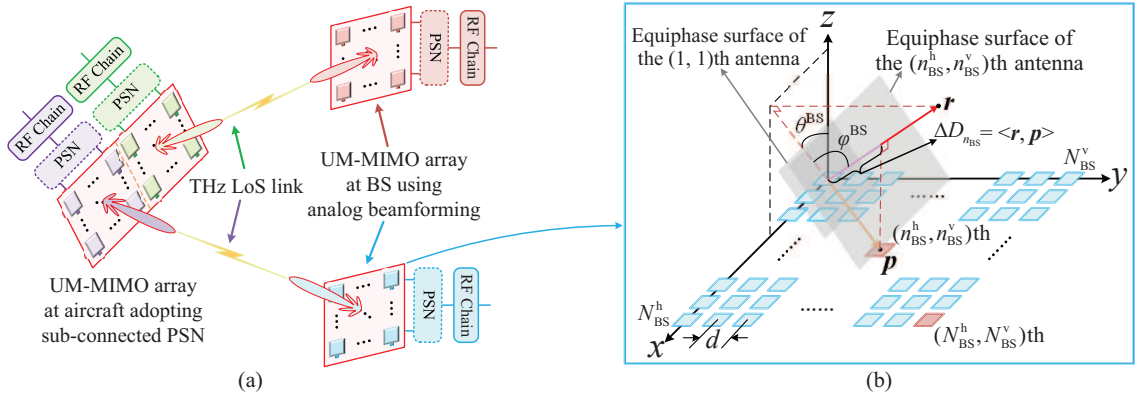}
\end{center}
\captionsetup{font = {footnotesize}, name = {Fig.}, labelsep = period} %, singlelinecheck = on, justification = raggedright
\caption{The structure diagram of the antenna arrays at transceiver:
 (a) $L\! =\! 2$ BSs that use analog beamforming communicate with aircraft adopting sub-connected PSN
 through respective LoS links, and (b) takes the UPA at BS with size of $N_{\rm BS}^{\rm h}\! \times\! N_{\rm BS}^{\rm v}$
 as an example to illustrate the delay squint effect of THz UM-MIMO array.}
%\vspace{-6mm}
\label{FIG4}
\end{figure*}

 In this section, we will formulate the signal transmission and channel models with LoS link for THz UM-MIMO-based aeronautical communications, where the full-dimensional UM-MIMO channel model using UPAs involves azimuth and elevation angles \cite{Liao_Tcom19,HanC_TVT17}.
 Fig.~\ref{FIG4}(a) depicts the specific scenario that $L$ aerial BSs jointly serve an aircraft through respective THz LoS links.
 The aerial BSs and aircraft adopt the hybrid beamforming structure with a sub-connected PSN \cite{Akyildiz_CM18,HanC_CM18},
 where the sub-connected PSNs at BSs can be simplified as analog beamforming to serve the assigned aircraft.
 The specific configurations of these antenna arrays are as follows.
 The total number of antennas at BS arrays is $N_{\rm BS}\! =\! N_{\rm BS}^{\rm h}N_{\rm BS}^{\rm v}$,
 where $N_{\rm BS}^{\rm h}$ and $N_{\rm BS}^{\rm v}$ are the numbers of antennas in horizontal and vertical directions, respectively.
 Due to the sub-connected PSN adopted at aircraft, we define ${\widetilde I}_{\rm AC}^{\rm h}$
 ($M_{\rm AC}^{\rm h}$) and ${\widetilde I}_{\rm AC}^{\rm v}$ ($M_{\rm AC}^{\rm v}$)
 as the numbers of subarrays (antennas within each subarray) in horizontal and vertical directions, respectively;
 while $N_{\rm AC}^{\rm h}\! =\! {\widetilde I}_{\rm AC}^{\rm h}M_{\rm AC}^{\rm h}$ and
 $N_{\rm AC}^{\rm v}\! =\! {\widetilde I}_{\rm AC}^{\rm v}M_{\rm AC}^{\rm v}$
 are the numbers of antennas in horizontal and vertical directions of array, respectively.
 Then, the total numbers of antennas in each subarray and the whole antenna array are
 $M_{\rm AC}\! =\! M_{\rm AC}^{\rm h}M_{\rm AC}^{\rm v}$ and $N_{\rm AC}\! =\! N_{\rm AC}^{\rm h}N_{\rm AC}^{\rm v}$, respectively.
 Clearly, the aircraft and BS are equipped with $L\! =\! {\widetilde I}_{\rm AC}^{\rm h}{\widetilde I}_{\rm AC}^{\rm v}$
 RF chains and only one RF chain, respectively,
 and each subarray and the corresponding RF chain mounted on aircraft are assigned to one BS.

 According to the frame structure in Fig.~\ref{FIG3}, the azimuth/elevation angles at BSs and aircraft are estimated in the Uplink (UL) and Downlink (DL), respectively, and OFDM with $K$ subcarriers is adopted.
 The UL baseband signal $y_{{\rm UL},l}^{[m]}[k]$ received by the $l$th BS at the $k$th
 subcarrier of the $m$th OFDM symbol can be expressed as
\begin{align}\label{eq_y_ul} % eq 1
 y_{{\rm UL},l}^{[m]}[k] =& \sqrt{P_l} \bm{q}_{{\rm RF},l}^{\rm H} \bm{H}_{{\rm UL},l}^{[m]}[k] \bm{P}_{\rm RF}
 \bm{P}_{\rm BB}^{[m]}[k] \bm{s}_{\rm UL}^{[m]}[k] \nonumber\\
 &+ \bm{q}_{{\rm RF},l}^{\rm H} \bm{n}_{{\rm UL},l}^{[m]}[k] ,
\end{align}
 where $1\! \le\! l\! \le\! L$, $1\! \le\! k\! \le\! K$, and $P_l$ is the transmit power.
 In (\ref{eq_y_ul}), $\bm{q}_{{\rm RF},l}\! \in\! \mathbb{C}^{N_{\rm BS}}$
 is the analog combining vector of the $l$th BS,
 $\bm{P}_{\rm RF}\! \in\! \mathbb{C}^{N_{\rm AC}\!\times\! L}$ and
 $\bm{P}_{\rm BB}^{[m]}[k]\! \in\! \mathbb{C}^{L\!\times\! L}$
 are the analog and digital precoding matrices at aircraft, respectively,
 while $\bm{H}_{{\rm UL},l}^{[m]}[k]\! \in\! \mathbb{C}^{N_{\rm BS}\!\times\! N_{\rm AC}}$
 is the UL effective baseband channel matrix, $\bm{s}_{\rm UL}^{[m]}[k]\! \in\! \mathbb{C}^{L}$
 is the transmitted signal vector, and $\bm{n}_{{\rm UL},l}^{[m]}[k]\! \in\! \mathbb{C}^{N_{\rm BS}}$
 is the complex Additive White Gaussian Noise (AWGN) vector with the covariance $\sigma_n^2$,
 i.e., $\bm{n}_{{\rm UL},l}^{[m]}[k]\! \sim\! {\cal CN}\! \left(\bm{0}_{N_{\rm BS}},\sigma_n^2\bm{I}_{N_{\rm BS}}\right)$.
 Similarly, the DL baseband signal vector $\bm{y}_{\rm DL}^{[n]}[k]\! \in\! \mathbb{C}^{L}$
 received by aircraft at the $k$th subcarrier of the $n$th OFDM symbol is given by
\begin{align}\label{eq_y_dl} % eq 2
 \bm{y}_{\rm DL}^{[n]}[k] =&\!\
 (\bm{W}_{\rm BB}^{[n]}[k])^{\rm H} \bm{W}_{\rm RF}^{\rm H}
 \Big(\sum\limits_{l=1}^{L}{\sqrt{P_l} \bm{H}_{{\rm DL},l}^{[n]}[k] \bm{f}_{{\rm RF},l}
 s_{{\rm DL},l}^{[n]}[k]} \nonumber\\
 &+ \bm{n}_{\rm DL}^{[n]}[k] \Big) ,
\end{align}
 where $\bm{W}_{\rm RF}\! \in\! \mathbb{C}^{N_{\rm AC}\!\times\! L}$ and
 $\bm{W}_{\rm BB}^{[n]}[k]\! \in\! \mathbb{C}^{L\!\times\! L}$
 are the analog and digital combining matrices at aircraft, respectively,
 $\bm{f}_{{\rm RF},l}\! \in\! \mathbb{C}^{N_{\rm BS}}$ is the analog precoding vector of the $l$th BS,
 while $\bm{H}_{{\rm DL},l}^{[n]}[k]\! \in\! \mathbb{C}^{N_{\rm AC}\!\times\! N_{\rm BS}}$
 is the DL effective baseband channel matrix, and $s_{{\rm DL},l}^{[n]}[k]$ and
 $\bm{n}_{{\rm DL},l}^{[n]}[k]\! \in\! \mathbb{C}^{N_{\rm AC}}$ are the transmitted
 pilot signal (or the modulated/coded data) and the AWGN vector (similar to
 $\bm{n}_{{\rm UL},l}^{[m]}[k]$), respectively.

 To illustrate the delay squint effect of THz UM-MIMO channels, we take the antenna array at BS as an example as shown in Fig.~\ref{FIG4}(b).
 Specifically, the first $(1,1)$th antenna element
 can be regarded as the reference point, and define ${\bm{r}}\! =\! \left( \sin(\theta_l^{\rm BS})
 \cos(\varphi_l^{\rm BS}),\sin(\varphi_l^{\rm BS}),\cos(\theta_l^{\rm BS})\cos(\varphi_l^{\rm BS}) \right)$
 as the unit direction vector, where $\theta_l^{\rm BS}$ and $\varphi_l^{\rm BS}$
 are the azimuth and elevation angles associated with the $l$th BS, respectively.
 Defining the $(n_{\rm BS}^{\rm h},n_{\rm BS}^{\rm v})$th antenna as the $n_{\rm BS}$th antenna with
 $n_{\rm BS}\! =\! (n_{\rm BS}^{\rm v}\! -\! 1)N_{\rm BS}^{\rm h}\! +\! n_{\rm BS}^{\rm h}$,
 its direction vector relative to the reference antenna is ${\bm{p}}\! =\!
 \left( (n_{\rm BS}^{\rm h}\!-\!1)d,(n_{\rm BS}^{\rm v}\!-\!1)d,0 \right)$,
 where $d$ denotes the adjacent antenna spacing with half-wavelength.
 The wave path-difference between the $n_{\rm BS}$th antenna and the first antenna, denoted by $\Delta\!{D_{n_{\rm BS}}}$,
 is equal to the distance between the equiphase surfaces of these two antennas, i.e.,
 $\Delta\!{D_{n_{\rm BS}}}\! =\! \left\langle \bm{r},\bm{p} \right\rangle\! =\!
 (n_{\rm BS}^{\rm h}\!-\!1)d\sin(\theta_l^{\rm BS})\cos(\varphi_l^{\rm BS})\!+\!(n_{\rm BS}^{\rm v}\!-\!1)d\sin(\varphi_l^{\rm BS})$.
 Denoting $\tau_l^{[n_{\rm BS}]}$ as the transmission delay from the $n_{\rm BS}$th antenna to the first antenna for the $l$th BS,
 we can obtain $\tau_l^{[n_{\rm BS}]}\! =\! \Delta\!{D_{n_{\rm BS}}}/c$ with $c$ being the speed of light.
 Note that $\tau_l^{[n_{\rm BS}]}$ is related to the antenna index and the azimuth/elevation angles.
 When the signal direction is not perpendicular to the array and $n_{\rm BS}$ is large,
 $\tau_l^{[n_{\rm BS}]}$ can be even larger than the symbol period $T_s$\footnote{We consider an extreme scenario that
 the impinging signal comes from the diagonal direction of UPA of size $(n_{\rm BS}\! +\! 1)\! \times\! (n_{\rm BS}\! +\! 1)$,
 and those $(n_{\rm BS}\! +\! 1)$ diagonal antennas consist of the Uniform Linear Array (ULA) of size $(n_{\rm BS}\! +\! 1)$ with $\sqrt{2}d$ antenna spacing. When angle $\theta_l^{\rm BS}\! = \! 60^\circ$, carrier frequency $f_c\! =\! 0.1\, {\rm THz}$, and bandwidth $f_s\! = \! 1\, {\rm GHz}$ for the typical THz UM-MIMO aeronautical communication scenario, $n_{\rm BS}\! =\! 200$ antennas will make its filling time satisfy $\tau _l^{[n_{\rm BS}]}\! =\! \frac{\sqrt{2}n_{\rm BS}\sin(\theta_l^{\rm BS})}{2f_c}\! \approx \! 1.225\,T_s$.}, which compels higher demands on
 the signal processing at the receiver, especially for the analog or hybrid beamforming architecture.
 Therefore, the delay squint effect needs to be taken into account for aeronautical THz UM-MIMO systems.

 Considering the channel reciprocity in time division duplex systems,
 we focus on the formulation of DL channel matrix next.
 According to the channel model in \cite{Akyildiz_TWC11,GaoFF_TSP19},
 define the DL passband channel matrix in the spatial-delay domain as % continuous
 $\bm{\bar H}_{{\rm DL},l}^{(t)}(\tau)\! \in\! \mathbb{C}^{N_{\rm AC}\!\times\! N_{\rm BS}}$ at time $t$ corresponding to the $l$th BS,
 whose the ($n_{\rm AC},n_{\rm BS}$)th element, i.e., $[\bm{\bar H}_{{\rm DL},l}^{(t)}(\tau)]_{n_{\rm AC},n_{\rm BS}}$,
 can be expressed as
\begin{align}\label{eq_H_tau_dl_mn} % eq 3
 &[\bm{\bar H}_{{\rm DL},l}^{(t)}(\tau)]_{n_{\rm AC},n_{\rm BS}} \nonumber\\
 &= \sqrt{G_l} \alpha_l e^{\textsf{j} {2\pi\psi_l t}}
 \delta\big( \tau - \tau_l - \underbrace{(\tau_l^{[n_{\rm AC}]} + \tau_l^{[n_{\rm BS}]})}_{\bm{Delay\,\,squint}} \big) ,
\end{align}
 where $1\! \le\! n_{\rm AC}\! \le\! N_{\rm AC}$, $1\! \le\! n_{\rm BS}\! \le\! N_{\rm BS}$,
 $G_l$ and $\alpha_l\! \sim\! {\cal CN}(0,\sigma_\alpha^2)$ are the large-scale fading gain of communication link and
 the channel gain\footnote{Due to the negligible frequency-dependent attenuation of THz communication links
 (e.g., atmospheric molecular absorption) in the stratosphere and above \cite{Slim_OJCS20,Xu_TAP19},
 the channel gain $\alpha_l$ can be modeled as a frequency flat coefficient,
 which is different from the frequency-dependent channel coefficient in \cite{Akyildiz_TWC11}.}, respectively,
 $\psi_l\! =\! \underline{v}_l/\lambda_c$ denotes the Doppler shift with $\underline{v}_l$ and $\lambda_c$ being the relative radial velocity
 and carrier wavelength, respectively, $f_c$ is the corresponding carrier frequency,
 $\tau_l^{[n_{\rm AC}]}$ denotes the transmission delay between the $n_{\rm AC}$th antenna
 ($n_{\rm AC}\! =\! (n_{\rm AC}^{\rm v}\!-\!1)N_{\rm AC}^{\rm h}\!+\!n_{\rm AC}^{\rm h}$, and
 it also the $(n_{\rm AC}^{\rm h},n_{\rm AC}^{\rm v})$th antenna of UPA at aircraft) and its reference point,
 and $\delta(\cdot)$ and $\tau_l$ are the Dirac impulse function and the path delay, respectively.
 After some algebraic transformations, the DL spatial-frequency channel matrix $\bm{H}_{{\rm DL},l}^{[n]}[k]$ in (\ref{eq_y_dl}) at the $k$th subcarrier of the $n$th OFDM symbol can be expressed as
\begin{align}\label{eq_H_k_dl_1} % eq 4
 \bm{H}_{{\rm DL},l}^{[n]}[k] =& \sqrt{G_l} \alpha_l
 e^{\textsf{j} {2\pi \psi_{l,k} (n-1)T_{\rm sym}}} e^{-\textsf{j} {2\pi \left({\textstyle{k-1 \over K}}-{\textstyle{1 \over 2}}\right)f_s \tau_l}} \nonumber\\
 &\times \bm{A}_{{\rm DL},l}[k] ,
\end{align}
 where $T_{\rm sym}$ and $f_s$ denote the duration time of an OFDM symbol and system bandwidth, respectively,
 $\psi_{l,k}\! =\! \psi_{z,l}\! +\! {\textstyle{\underline{v}_l \over c}}({\textstyle{{k\!-\!1} \over K}}\!-\!{\textstyle{1 \over 2}})f_s$ is the frequency-dependent Doppler shift at the $k$th subcarrier with $\psi_{z,l}$ being the Doppler shift of the central carrier frequency $f_z$ (wavelength $\lambda_z$) and ${\textstyle{\underline{v}_l \over c}}({\textstyle{{k\!-\!1} \over K}}\!-\!{\textstyle{1 \over 2}})f_s$ being the {\emph{\textbf{Doppler squint part}}} due to the large bandwidth in THz communications, and $\bm{A}_{{\rm DL},l}[k]\! \in\! \mathbb{C}^{N_{\rm AC}\!\times\! N_{\rm BS}}$ is the DL array response matrix associated
 with the array response vectors at aircraft and the $l$th BS, given by
\begin{align}\label{eq_A_dl_k1} % eq 5
 \bm{A}_{{\rm DL},l}[k] =& \underbrace{\left( \bm{a}_{\rm AC}(\mu_l^{\rm AC},\nu_l^{\rm AC})
 \bm{a}^{\rm H}_{\rm BS}(\mu_l^{\rm BS},\nu_l^{\rm BS}) \right)}_{\bm{A}_{{\rm DL},l}} \nonumber\\
 &\circ \underbrace{\left( \bm{\bar a}_{\rm AC}(\mu_l^{\rm AC},\nu_l^{\rm AC},k)
 \bm{\bar a}^{\rm H}_{\rm BS}(\mu_l^{\rm BS},\nu_l^{\rm BS},k)
 \right)}_{\bm{\bar{A}}_{{\rm DL},l}[k]\ {(\bm{Beam\,\,squint\,\,component})}},
\end{align}
 where $\mu_l^{\rm AC}\! =\! \pi\sin(\theta_l^{\rm AC}) \cos(\varphi_l^{\rm AC})$
 ($\mu_l^{\rm BS}\! =\! \pi\sin(\theta_l^{\rm BS})\cos(\varphi_l^{\rm BS})$) and
 $\nu_l^{\rm AC}\! =\! \pi\sin(\varphi_l^{\rm AC})$ ($\nu_l^{\rm BS} \! =\! \pi\sin(\varphi_l^{\rm BS})$)
 are the horizontally and vertically virtual angles at aircraft (the $l$th BS), respectively,
 $\bm{A}_{{\rm DL},l}$ is the conventional DL array response matrix without beam squint effect at aircraft and BS,
 and $\bm{\bar{A}}_{{\rm DL},l}[k]$ is the corresponding array response squint matrix considering beam squint effect.
 In (\ref{eq_A_dl_k1}), $\bm{a}_{\rm AC}(\mu_l^{\rm AC},\!\nu_l^{\rm AC})\! =\!
 \bm{a}_{\rm v}(\nu_l^{\rm AC},\!N_{\rm AC}^{\rm v})\!\otimes\!
 \bm{a}_{\rm h}(\mu_l^{\rm AC},\!N_{\rm AC}^{\rm h})$
 and $\bm{a}_{\rm BS}(\mu_l^{\rm BS},\!\nu_l^{\rm BS})\! =\!
 \bm{a}_{\rm v}(\nu_l^{\rm BS},\!N_{\rm BS}^{\rm v})\!\otimes\!
 \bm{a}_{\rm h}(\mu_l^{\rm BS},\!N_{\rm BS}^{\rm h})$
 are the general array response vectors at aircraft and the $l$th BS \cite{Liao_Tcom19}, respectively,
 and $\bm{\bar a}_{\rm AC}(\mu_l^{\rm AC},\nu_l^{\rm AC},k)\! =\!
 \bm{\bar a}_{\rm v}(\nu_l^{\rm AC},N_{\rm AC}^{\rm v},k)\!\otimes\!
 \bm{\bar a}_{\rm h}(\mu_l^{\rm AC},N_{\rm AC}^{\rm h},k)$
 and $\bm{\bar a}_{\rm BS}(\mu_l^{\rm BS},\nu_l^{\rm BS},k)\! =\!
 \bm{\bar a}_{\rm v}(\nu_l^{\rm BS},N_{\rm BS}^{\rm v},k)\!\otimes\!
 \bm{\bar a}_{\rm h}(\mu_l^{\rm BS},N_{\rm BS}^{\rm h},k)$
 are the frequency-dependent array response squint vectors, respectively.
 Moreover, the vectors at aircraft, i.e., the horizontal/vertical steering vectors
 $\bm{a}_{\rm h}(\mu_l^{\rm AC},N_{\rm AC}^{\rm h})$ and
 $\bm{a}_{\rm v}(\nu_l^{\rm AC},N_{\rm AC}^{\rm v})$, and the horizontal/vertical
 steering squint vectors $\bm{\bar a}_{\rm h}(\mu_l^{\rm AC},N_{\rm AC}^{\rm h},k)$ and
 $\bm{\bar a}_{\rm v}(\nu_l^{\rm AC},N_{\rm AC}^{\rm v},k)$ can be further written as
\begin{align} % eqs 6,7,8,9
 &\bm{a}_{\rm h}(\mu_l^{\rm AC},N_{\rm AC}^{\rm h}) =
 \left[ 1 ~ e^{\textsf{j}\mu_l^{\rm AC}} ~ \cdots ~
 e^{\textsf{j}(N_{\rm AC}^{\rm h}-1)\mu_l^{\rm AC}} \right]^{\rm T}, \label{eq_a_AC_h}\\
 &\bm{a}_{\rm v}(\nu_l^{\rm AC},N_{\rm AC}^{\rm v}) =
 \left[ 1 ~ e^{\textsf{j}\nu_l^{\rm AC}} ~ \cdots ~
 e^{\textsf{j}(N_{\rm AC}^{\rm v}-1)\nu_l^{\rm AC}} \right]^{\rm T}, \label{eq_a_AC_v}\\
 &\bm{\bar a}_{\rm h}(\mu_l^{\rm AC},N_{\rm AC}^{\rm h},k) \nonumber\\
 &= \Big[ 1 ~ e^{\textsf{j}\left({\textstyle{{k-1} \over K}}-{\textstyle{1 \over 2}}\right){\textstyle{f_s \over f_z}}\mu_l^{\rm AC}} ~
 \cdots ~ e^{\textsf{j}\left({\textstyle{{k-1} \over K}}-{\textstyle{1 \over 2}}\right)
 {\textstyle{f_s \over f_z}}(N_{\rm AC}^{\rm h}-1)\mu_l^{\rm AC}} \Big]^{\rm T}, \label{eq_a_AC_h_k}\\
 &\bm{\bar a}_{\rm v}(\nu_l^{\rm AC},N_{\rm AC}^{\rm v},k) \nonumber\\
 &= \Big[ 1 ~ e^{\textsf{j}\left({\textstyle{{k-1} \over K}}-{\textstyle{1 \over 2}}\right)
 {\textstyle{f_s \over f_z}}\nu_l^{\rm AC}} ~ \cdots ~ e^{\textsf{j}
 \left({\textstyle{{k-1} \over K}}-{\textstyle{1 \over 2}}\right)
 {\textstyle{f_s \over f_z}}(N_{\rm AC}^{\rm v}-1)\nu_l^{\rm AC}} \Big]^{\rm T}. \label{eq_a_AC_v_k}
\end{align}
 Note that the vectors at BSs, i.e., $\bm{a}_{\rm h}(\mu_l^{\rm BS},N_{\rm BS}^{\rm h})$,
 $\bm{a}_{\rm v}(\nu_l^{\rm BS},N_{\rm BS}^{\rm v})$,
 $\bm{\bar a}_{\rm h}(\mu_l^{\rm BS},N_{\rm BS}^{\rm h},k)$,
 and $\bm{\bar a}_{\rm v}(\nu_l^{\rm BS},N_{\rm BS}^{\rm v},k)$,
 have the similar definitions and expressions to (\ref{eq_a_AC_h})-(\ref{eq_a_AC_v_k}),
 and their details are omitted for simplicity.
 The detailed derivation of DL channel matrix $\bm{H}_{{\rm DL},l}^{[n]}[k]$ can be found in Appendix~A.

 Similar to (\ref{eq_H_k_dl_1}), the UL spatial-frequency baseband channel matrix
 $\bm{H}_{{\rm UL},l}^{[m]}[k]$ in (\ref{eq_y_ul}) at the $k$th subcarrier of the $m$th OFDM symbol
 corresponding to the $l$th BS can be formulated as
\begin{equation}\label{eq_H_k_ul} % eq 10
 \bm{H}_{{\rm UL},l}^{[m]}[k] = \sqrt{G_l} \alpha_l e^{\textsf{j} {2\pi \psi_{l,k} (m-1)T_{\rm sym}}} \bm{A}_{{\rm UL},l}[k] ,
\end{equation}
 where the UL array response matrix $\bm{A}_{{\rm UL},l}[k]\! \in\! \mathbb{C}^{N_{\rm BS}\!\times\! N_{\rm AC}}$ is
\begin{align}\label{eq_A_ul_k} % eq 11
 \bm{A}_{{\rm UL},l}[k] =& \underbrace{\left( \bm{a}_{\rm BS}(\mu_l^{\rm BS},\nu_l^{\rm BS})
 \bm{a}^{\rm H}_{\rm AC}(\mu_l^{\rm AC},\nu_l^{\rm AC}) \right)}_{\bm{A}_{{\rm UL},l}} \nonumber\\
 &\circ \underbrace{\left( \bm{\bar a}_{\rm BS}(\mu_l^{\rm BS},\nu_l^{\rm BS},k)
 \bm{\bar a}^{\rm H}_{\rm AC}(\mu_l^{\rm AC},\nu_l^{\rm AC},k)
 \right)}_{\bm{\bar{A}}_{{\rm UL},l}[k]\ {(\bm{Beam\,\,squint\,\,component})}} .
\end{align}

\section{Initial Channel Estimation}\label{S3}

 As shown in Fig.~\ref{FIG3}, at the initial channel estimation stage, the fine azimuth/elevation angles
 at BSs and aircraft, Doppler shifts, and path delays are estimated successively.
 At this stage, according to the positioning and flight posture information acquired in aeronautical systems,
 some rough channel parameter estimates (e.g., angle and Doppler shift) can be utilized to establish the initial THz UM-MIMO link.
 Due to the positioning accuracy error and the posture rotations of antenna arrays mounted on aerial BSs and aircraft,
 these rough channel parameter estimates are not accurate enough for data transmission.
 Therefore, the accurate acquisition of dominant channel parameters is still indispensable.

\begin{figure}[!tp]
%%\vspace{-5mm}
\begin{center}
 \includegraphics[width=0.98\columnwidth, keepaspectratio]{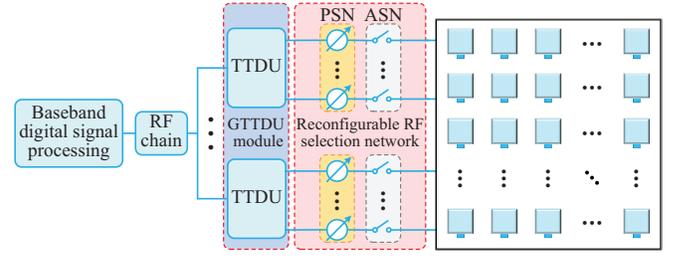}
\end{center}
\setlength{\abovecaptionskip}{-1mm}
\captionsetup{font = {footnotesize}, singlelinecheck = off, name = {Fig.}, labelsep = period} %, justification = raggedright
\caption{The transceiver structure corresponding to one RF, where this RF chain connects with the antenna array via the GTTDU module and the reconfigurable RF selection network consisting of a sub-connected PSN and an ASN.}
%\vspace{-6mm}
\label{FIG5}
\end{figure}

\begin{figure}[!tp]
%\vspace{-5mm}
\begin{center}
 \includegraphics[width=1.0\columnwidth, keepaspectratio]{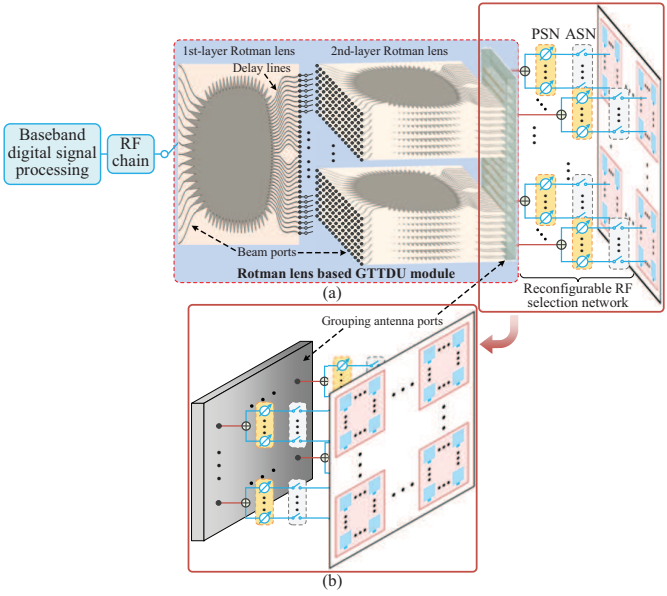}
\end{center}
\captionsetup{font = {footnotesize}, singlelinecheck = off, name = {Fig.}, labelsep = period} %, justification = raggedright
\caption{(a) A feasible transceiver structure corresponding to one RF, where the Rotman lens-based GTTDU module can be utilized to implement the practical tunable TTDU module \cite{Kodheli_CST20}; and (b) the other side elevation drawing of a part of RF front-end that includes the grouping antenna ports, reconfigurable RF selection network, and THz UM-MIMO array.
The beam ports of the first-layer and second-layer Rotman lenses steer the horizontal and vertical directions, respectively. The total number of grouping antenna ports is consistent with that of antenna groups in the previous GTTDU module.
This elaborated cascading two-layer Rotman lenses are equivalent to the wideband phase shifters of the tunable TTD module, which can be utilized to eliminate the beam squint effect.}
%\vspace{-6mm}
\label{FIG6}
\end{figure}

 To overcome the delay-beam squint effects of THz UM-MIMO array, the fully-digital array architecture
 with each antenna equipping a dedicated RF chain is preferred, but the involved prohibitive hardware cost and power consumption make it impracticable.
 Moreover, the aforementioned hybrid beamforming and channel estimation schemes \cite{Heath_TWC19,Hanzo_TWC20,GaoFF_TWC19,GaoFF_TSP19,GaoFF_TSP20} utilize some signal processing methods to attenuate the impact of delay-beam squint effects on the results, rather than eliminate these effects during signal transmission.
 Therefore, those processing methods are only suitable for the terrestrial mmWave or THz cellular networks with abundant scatterers, where the receiver in short-distance transmission (at most hundreds of meters) can receive the signals affected by delay-beam squint effects.
 However, for THz UM-MIMO-based aeronautical communication systems that rely on the long-distance transmission of LoS link (up to hundreds of kilometers) without supernumerary scatterers, the receiver will most likely fail to receive the signals at marginal carrier frequencies due to the very narrow pencil beam and (even slight) delay-beam squint effects.
 Except for the indispensable signal processing, the transceivers of aeronautical communication systems should be elaborately designed to eliminate the delay-beam squint effects and ensure that all carrier frequencies within effective bandwidth can establish a reliable THz communication link.
 A common treatment of delay-beam squint effects is to design the transceiver based on the TTDU module \cite{CM08,LiYe_JSAC17}.
 The optimal TTDU module is made up of numerous true-time delay units, and each unit is assigned to its dedicated antenna \cite{Akyildiz_VTC18}, where the detailed designs of these tunable TTDUs can be found in \cite{JLT18,LinF_TMTT19}.
 Nevertheless, the excessively high hardware complexity and cost of this optimal module prompt us to design a sub-optimal implementation of TTDU module, i.e., GTTDU module based transceiver structure\footnote{Since the TTDU/GTTDU module is difficult to tackle multiple path signals in the analog domain simultaneously, the proposed transceiver structure and the subsequent solution for THz aeronautical communications cannot be directly applied in terrestrial vehicular communication scenarios, where the non-LoS components caused by various scatterers are ubiquitous.} as shown in Fig.~\ref{FIG5}.
  From Fig.~\ref{FIG5}, we observe that except for the antenna array, this transceiver structure contains a GTTDU module and a reconfigurable RF selection network involving a sub-connected PSN and an Antenna Switching Network (ASN) \cite{Han_JSAC20}, where this ASN can control the active or inactive state of the antenna elements to form different connection patterns of the RF selection network at the angle estimation stage.
  In this GTTDU module, a TTDU can be shared by a group of antennas and this imperfect hardware limitation can be handled by the subsequent signal processing algorithms well.
 Observe that although the delay squint effect for the whole UM array can be non-negligible, this effect
 for antennas within a group is mild. Hence, the GTTDU module can mitigate the delay squint effect among
 the antennas in different groups, and the residual phase deviations of these antennas within each group
 can be further eliminated using their respective phase shifters.
 Furthermore, to illustrate the feasibility of the transceiver designed in Fig.~\ref{FIG5}, we propose a potential implementation of transceiver structure involving the Rotman lens-based GTTDU module in Fig.~\ref{FIG6}, where the cascading two-layer Rotman lenses can be utilized to implement the full-dimensional beamforming \cite{Kodheli_CST20}. The Rotman lens based GTTDU module is a practical photonic implementation \cite{Rotman_Proc16}, and this design employs the optical properties of electromagnetic waves to achieve the tunable TTD module \cite{GaoY_TVT17,WangX_TAP18}, which provides a prospective direction for our future research work.

 When the acquired angle information is
 accurate enough, the impact of delay squint effect would be significantly mitigated using this GTTDU module.
 To be specific, based on the prior information acquired from navigation information, the rough estimates of
 azimuth and elevation angles at BSs (aircraft) can be defined as $\{ {\widetilde \theta}_l^{\rm BS} \}_{l=1}^{L}$
 ($\{ {\widetilde \theta}_l^{\rm AC} \}_{l=1}^{L}$) and $\{ {\widetilde \varphi}_l^{\rm BS} \}_{l=1}^{L}$
 ($\{ {\widetilde \varphi}_l^{\rm AC} \}_{l=1}^{L}$), respectively, and the corresponding
 horizontally and vertically virtual angles are
 $\{ {\widetilde \mu}_l^{\rm BS} \}_{l=1}^{L}$ ($\{ {\widetilde \mu}_l^{\rm AC} \}_{l=1}^{L}$)
 and $\{ {\widetilde \nu}_l^{\rm BS} \}_{l=1}^{L}$ ($\{ {\widetilde \nu}_l^{\rm AC} \}_{l=1}^{L}$), respectively.
 According to $\bm{H}_{{\rm DL},l}^{[n]}[k]$ in (\ref{eq_H_k_dl_1}), we present the expression of
 the DL channel matrix after ideal TTDU module processing in the following lemma,
 denoted by $\bm{\widetilde H}_{{\rm DL},l}^{[n]}[k]$, which is proved in Appendix~B.
\begin{lemma}\label{lemma1} %
 According to the rough angle estimates above, the antenna transmission delays of THz UM-MIMO arrays at BSs and aircraft
 can be compensated using the ideal TTDU module, and the compensated DL spatial-frequency
 channel matrix $\bm{\widetilde H}_{{\rm DL},l}^{[n]}[k]$ can then be formulated as
\begin{align}\label{eq_H_ttdu_k_dl} % eq 12
 \bm{\widetilde H}_{{\rm DL},l}^{[n]}[k] =& \sqrt{G_l} \alpha_l
 e^{\textsf{j} {2\pi \psi_{l,k} (n-1)T_{\rm sym}}} e^{-\textsf{j} {2\pi \left({\textstyle{k-1 \over K}}-{\textstyle{1 \over 2}}\right)f_s \tau_l}} \nonumber\\
 &\times \bm{\widetilde A}_{{\rm DL},l}[k] ,
\end{align}
 in which
\begin{equation}\label{eq_A_ttdu_dl_k} % eq 13
 \bm{\widetilde A}_{{\rm DL},l}[k]\! =\! \bm{A}_{{\rm DL},l}[k] \circ
 {\underbrace{\left( \bm{\bar a}_{\rm AC}({\widetilde \mu}_l^{\rm AC},{\widetilde \nu}_l^{\rm AC},k)
 \bm{\bar a}^{\rm H}_{\rm BS}({\widetilde \mu}_l^{\rm BS},{\widetilde \nu}_l^{\rm BS},k) \right)}_{\widetilde{\bm{\bar{A}}}_{{\rm DL},l}[k]}}^* .
\end{equation}
 By comparing $\widetilde{\bm{\bar{A}}}_{{\rm DL},l}[k]$ in (\ref{eq_A_ttdu_dl_k}) and $\bm{\bar{A}}_{{\rm DL},l}[k]$ in (\ref{eq_A_dl_k1}), we can find that if we can acquire the perfect angle information, the beam squint effect part can be perfectly eliminated, i.e., $\widetilde{\bm{\bar{A}}}_{{\rm DL},l}[k]\! =\! \bm{\bar{A}}_{{\rm DL},l}[k]$ and then $\bm{\widetilde A}_{{\rm DL},l}[k]\! =\! \bm{A}_{{\rm DL},l}$ when ${\widetilde \mu}_l^{\rm AC}\! =\! \mu_l^{\rm AC}$, ${\widetilde \nu}_l^{\rm AC}\! =\! \nu_l^{\rm AC}$, ${\widetilde \mu}_l^{\rm BS}\! =\! \mu_l^{\rm BS}$, and ${\widetilde \nu}_l^{\rm BS}\! =\! \nu_l^{\rm BS}$.
 Moreover, according to (\ref{eq_H_k_ul}) and (\ref{eq_A_ul_k}),
 the compensated UL spatial-frequency channel matrix $\bm{\widetilde H}_{{\rm UL},l}^{[m]}[k]$ has the
 similar expressions, which are omitted for simplicity.
\end{lemma}

 The ideal TTDU module provides a performance upper-bounds for the parameter estimation or data transmission, and we can design the sub-optimal GTTDU module adopted by our solution and the corresponding signal processing algorithms to approach these upper-bounds.
 The practical DL/UL spatial-frequency channel matrices compensated by the GTTDU module can be
 derived from (\ref{eq_H_ttdu_k_dl}) and (\ref{eq_A_ttdu_dl_k}). Specifically, all antenna groups
 for GTTDU module have the same size, i.e., ${\widetilde M}_{\rm BS}^{\rm h}\! \times\!
 {\widetilde M}_{\rm BS}^{\rm v}$ at BSs and ${\widetilde M}_{\rm AC}^{\rm h}\! \times\!
 {\widetilde M}_{\rm AC}^{\rm v}$ at aircraft, and the central antenna in each group can be
 regarded as the benchmark of antenna transmission delay for designing the corresponding TTDU.
 Moreover, to minimize the beam squint effect caused by antenna grouping as much as possible,
 the phase deviations of the rest antennas in one group can be compensated using the low-cost PSN,
 where the phase values at central carrier are treated as the benchmark for calculating these
 deviations. For convenience, the effective UL and DL channel matrices compensated by the GTTDU module
 can be also denoted as $\bm{\widetilde H}_{{\rm UL},l}^{[m]}[k]$ and $\bm{\widetilde H}_{{\rm DL},l}^{[n]}[k]$,
 respectively.

 At the initial channel estimation stage, we adopt the Orthogonal Frequency Division Multiple Access
 (OFDMA) to distinguish the pilot signals transmitted from different BSs and improve the accuracy of
 the estimated channel parameters. Hence, $K$ subcarriers can be equally assigned to $L$ BSs, where
 the alternating subcarrier index allocation with equal intervals is adopted and the ordered subcarrier
 index set assigned to the $l$th BS is ${\cal K}_l$ with $K_l = |{\cal K}_l|_c$. Moreover, the
 azimuth/elevation angles at BSs can be estimated in UL, while the rest of channel parameters are
 acquired in DL.

\begin{figure*}[!tp]
%\vspace{-5mm}
\begin{center}
 \includegraphics[width=1.6\columnwidth, keepaspectratio]{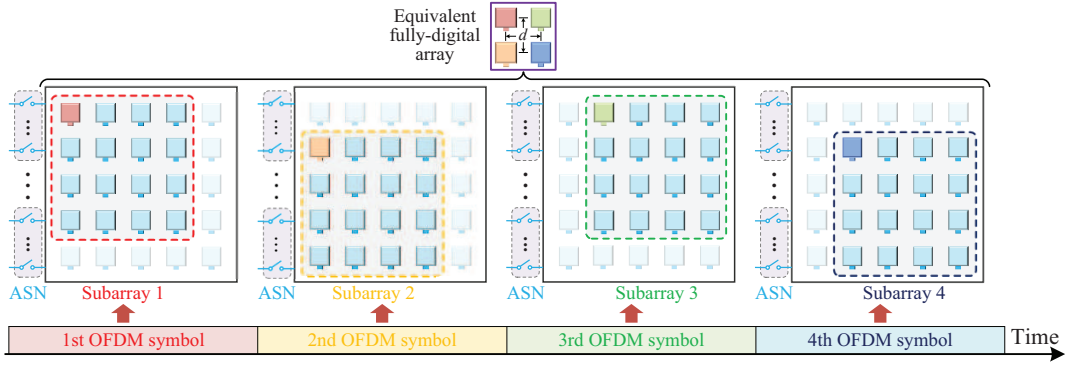}
\end{center}
\captionsetup{font = {footnotesize}, singlelinecheck = off, name = {Fig.}, labelsep = period} %, justification = raggedright
\caption{The schematic diagram of subarray selection at the initial angle estimation stage, where the different antenna connection patterns can be formed by controlling the ASN of the reconfigurable RF selection network. Taking the UPA of size $5\!\times\! 5$ as an example, this UPA can be partitioned into $4$ subarrays of size $4\! \times\! 4$, and the interval between each subarray is the width of one antenna. The same RF chain sequentially selects the corresponding subarrays in $4$ successive OFDM symbols to receive signals, and these received signals will be equivalent to the signals received by a low-dimensional fully-digital array  of size $2\! \times\! 2$ with the critical antenna spacing $d$.}
%\vspace{-6mm}
\label{FIG7}
\end{figure*}

\subsection{Fine Angle Estimation Based on Reconfigurable RF Selection Network}\label{S3.1}

\subsubsection{Fine Angle Estimation at BSs}\label{S3.1.1}
 Due to the insufficient valid observation caused by the limited number of RF chains
 at the BSs, it is necessary to accumulate multiple OFDM symbols in the time domain to estimate
 the angles. To mitigate the inter-carrier interference within one OFDM symbol caused by the
 large Doppler shifts, the acquired rough Doppler shift estimates are first utilized to compensate
 the transmitted signals, so that the compensated channels of multiple OFDM symbols can be
 slow time-varying. By transforming the different RF connection pattern of antenna array, we
 observe a fact that the received signals adopting different selected subarrays only differ
 by one envisaged phase value if the transceiver has the same configuration, and those
 regular phase differences can construct the array response vector of low-dimensional fully-digital array.
 Taking the UPA with size of $5\!\times\! 5$ in Fig.~\ref{FIG7} as an example, we can select $4$ subarrays
 of size $4\! \times\! 4$ in $4$ successive OFDM symbols to form the array response vector of equivalent
 fully-digital array with size of $2\! \times\! 2$ by controlling the reconfigurable RF selection
 network. Specifically, we intend to use $I_{\rm BS}$ OFDM symbols to estimate the angles at BSs,
 where each OFDM symbol adopts a dedicated RF connection pattern (i.e., the selected subarray).
 By employing the rough angle estimates at aircraft and BSs, the analog precoding and combining
 vectors, i.e., $\bm{p}_{{\rm RF},l}$ and $\bm{q}_{{\rm RF},l}^{[m]}$ for $1\! \le\! l\! \le\! L$,
 $1\! \le\! m\! \le\! I_{\rm BS}$, can be first designed. In terms of $\bm{p}_{{\rm RF},l}$,
 initialize $\bm{p}_{{\rm RF},l}$ as $\bm{p}_{{\rm RF},l}\! =\! \bm{0}_{N_{\rm AC}}$, and then
 let $[\bm{p}_{{\rm RF},l}]_{{\cal I}_{{\rm AC},l}}\! =\! \textstyle{1 \over \sqrt{M_{\rm AC}}}
 [\bm{a}_{\rm AC}({\widetilde \mu}_l^{\rm AC},{\widetilde \nu}_l^{\rm AC})]_{{\cal I}_{{\rm AC},l}}$.
 Here ${\cal I}_{{\rm AC},l}$ with $M_{\rm AC}\! =\! |{\cal I}_{{\rm AC},l}|_c$ denotes the
 antenna index of subarray assigned to the $l$th BS, since each subarray at aircraft only communicates
 with its corresponding BS as shown in Fig.~\ref{FIG4}(a). To design $\{\bm{q}_{{\rm RF},l}^{[m]}\}_{m=1}^{I_{\rm BS}}$,
 the UM-MIMO array at BS can be partitioned into $I_{\rm BS}\! =\! I_{\rm BS}^{\rm h}I_{\rm BS}^{\rm v}$
 smaller subarrays to yield the array response vector of equivalent low-dimensional fully-digital
 array with size of $I_{\rm BS}^{\rm h}\! \times\! I_{\rm BS}^{\rm v}$, where the sizes of
 these smaller subarrays are ${\bar M}_{\rm BS}^{\rm h}\! \times\! {\bar M}_{\rm BS}^{\rm v}$
 (${\bar M}_{\rm BS}^{\rm h}\! =\! N_{\rm BS}^{\rm h}\! -\! I_{\rm BS}^{\rm h}\! +\! 1$ and
 ${\bar M}_{\rm BS}^{\rm v}\! =\! N_{\rm BS}^{\rm v}\! -\! I_{\rm BS}^{\rm v}\! +\! 1$) and their
 number of antennas is ${\bar M}_{\rm BS}\! =\! {\bar M}_{\rm BS}^{\rm h}{\bar M}_{\rm BS}^{\rm v}$.
 Defining $m\! =\! (i_{\rm BS}^{\rm v}-1)I_{\rm BS}^{\rm h} \! +\! i_{\rm BS}^{\rm h}$ with
 $i_{\rm BS}^{\rm h}$ and $i_{\rm BS}^{\rm v}$ being the ($i_{\rm BS}^{\rm h},i_{\rm BS}^{\rm v}$)th
 subarray for $1\! \le\! i_{\rm BS}^{\rm h}\! \le\! I_{\rm BS}^{\rm h}$ and $1\! \le\!
 i_{\rm BS}^{\rm v}\! \le\! I_{\rm BS}^{\rm v}$, respectively, the antenna index of the selected
 $m$th subarray that corresponds to the $m$th OFDM symbol can be denoted by ${\cal I}_{\rm BS}^{[m]}$
 with ${\bar M}_{\rm BS}\! =\! |{\cal I}_{\rm BS}^{[m]}|_c$, so that $\bm{q}_{{\rm RF},l}^{[m]}$
 can be also initialized as $\bm{q}_{{\rm RF},l}^{[m]}\! =\! \bm{0}_{N_{\rm BS}}$, and then let
 $[\bm{q}_{{\rm RF},l}^{[m]}]_{{\cal I}_{\rm BS}^{[m]}}\! =\! \textstyle{1 \over \sqrt{{\bar M}_{\rm BS}}}
 [\bm{a}_{\rm BS}({\widetilde \mu}_l^{\rm BS},{\widetilde \nu}_l^{\rm BS})]_{{\cal I}_{\rm BS}^{[1]}}$
 for $1\! \le\! m\! \le\! I_{\rm BS}$.

 According to the UL transmission model in (\ref{eq_y_ul}), the received signal $y_{{\rm UL},l}^{[m]}[k_l]$
 at the $k_l$th subcarrier of the $m$th OFDM symbol transmitted by the $l$th BS can be expressed as
\begin{equation}\label{eq_y_ukl} % eq 14
 y_{{\rm UL},l}^{[m]}[k_l]\! =\! \sqrt{P_l} (\bm{q}_{{\rm RF},l}^{[m]})^{\rm H}
 \bm{\widetilde H}_{{\rm UL},l}^{'[m]}[k_l] \bm{p}_{{\rm RF},l} s_{{\rm UL},l}^{[m]}[k_l] + n_{{\rm UL},l}^{[m]}[k_l] ,
\end{equation}
 where $k_l\! \in\! {\cal K}_l$, $1\! \le\! m\! \le\! I_{\rm BS}$,
 $\bm{\widetilde H}_{{\rm UL},l}^{'[m]}[k_l]$ is the channel matrix compensated by GTTDU module and
 rough Doppler shift estimates, and $s_{{\rm UL},l}^{[m]}[k_l]$ and $n_{{\rm UL},l}^{[m]}[k_l]$ are
 the transmitted pilot signal and noise, respectively. By collecting the received signals at $K_l$
 subcarriers as $\bm{y}_{{\rm UL},l}^{[m]}\! \in\! \mathbb{C}^{K_l}$ and substituting the UL channel
 matrix in (\ref{eq_H_k_ul}) into $\bm{y}_{{\rm UL},l}^{[m]}$, we have
\begin{align}\label{eq_y_ul_vec1} % eq 15
 \bm{y}_{{\rm UL},l}^{[m]} =&
 \left[ y_{{\rm UL},l}^{[m]}[\{{\cal K}_l\}_1]\cdots y_{{\rm UL},l}^{[m]}[\{{\cal K}_l\}_{K_l}]\right]^{\rm T} \nonumber\\
 =& \sqrt{P_lG_l} \alpha_l (\bm{q}_{{\rm RF},l}^{[m]})^{\rm H} \bm{A}_{{\rm UL},l} \bm{p}_{{\rm RF},l} \bm{s}_{{\rm UL},l}^{[m]}
 \circ \bm{\widetilde y}_{{\rm UL},l}^{[m]} \nonumber\\
 &+ \bm{n}_{{\rm UL},l}^{[m]} ,
\end{align}
 where $\bm{s}_{{\rm UL},l}^{[m]}\! =\! \left[ s_{{\rm UL},l}^{[m]}[\{{\cal K}_l\}_1]\! \cdots\!
 s_{{\rm UL},l}^{[m]}[\{{\cal K}_l\}_{K_l}]\right]^{\rm T}\! \in\! \mathbb{C}^{K_l}$,
 $\bm{\widetilde y}_{{\rm UL},l}^{[m]}$ is the error vector including the residual beam squint caused by
 inaccurate prior information, and $\bm{n}_{{\rm UL},l}^{[m]}$ is the corresponding noise vector.
 Moreover, the same transmitted pilot signals are adopted for $I_{\rm BS}$ OFDM symbol,
 i.e., $s_{{\rm UL},l}[k_l]\! =\! s_{{\rm UL},l}^{[m]}[k_l]$, and accordingly,
 $\bm{s}_{{\rm UL},l}\! =\! \bm{s}_{{\rm UL},l}^{[m]}$ for $1\! \le\! m\! \le\! I_{\rm BS}$.
 Taking the transposition of $\{\bm{y}_{{\rm UL},l}^{[m]}\}_{m=1}^{I_{\rm BS}}$
 received from $I_{\rm BS}$ OFDM symbols, we can stack them as
 $\bm{Y}_{{\rm UL},l}\! =\! \left[ \bm{y}_{{\rm UL},l}^{[1]}\cdots \bm{y}_{{\rm UL},l}^{[I_{\rm BS}]} \right]^{\rm T}\! \in\! \mathbb{C}^{I_{\rm BS}\! \times\! K_l}$, i.e.,
\begin{align}\label{eq_Y_ul_angle1} % eq 16
 \bm{Y}_{{\rm UL},l} =& \sqrt{P_lG_l} \alpha_l \left( \bm{Q}_{{\rm RF},l}^{\rm H} \bm{A}_{{\rm UL},l} \bm{p}_{{\rm RF},l} \bm{s}_{{\rm UL},l}^{\rm T} \right)
 \circ \bm{\widetilde Y}_{{\rm UL},l} \nonumber\\
 &+ \bm{N}_{{\rm UL},l} ,
\end{align}
 where $\bm{Q}_{{\rm RF},l}\! =\! \left[ \bm{q}_{{\rm RF},l}^{[1]}\! \cdots\!
 \bm{q}_{{\rm RF},l}^{[I_{\rm BS}]} \right]\! \in\! \mathbb{C}^{N_{\rm BS}\!\times\! I_{\rm BS}}$
 and $\bm{\widetilde Y}_{{\rm UL},l}\! =\! \left[ \bm{\widetilde y}_{{\rm UL},l}^{[1]}\! \cdots\!
 \bm{\widetilde y}_{{\rm UL},l}^{[I_{\rm BS}]} \right]$ are the analog combining and residual
 beam squint matrices, respectively, and $\bm{N}_{{\rm UL},l}$ is the noise matrix.
 By utilizing this analog combining matrix $\bm{Q}_{{\rm RF},l}$, the array response
 vector of equivalent low-dimensional fully-digital array can be formed to estimate the
 angles at BSs using array signal processing techniques. To be specific, compared with
 $(\bm{q}_{{\rm RF},l}^{[1]})^{\rm H}\bm{a}_{\rm BS}(\mu_l^{\rm BS},\nu_l^{\rm BS})$ for
 $m\! =\! 1$ in (\ref{eq_y_ul_vec1}), $(\bm{q}_{{\rm RF},l}^{[m]})^{\rm H}\bm{a}_{\rm BS}(\mu_l^{\rm BS},\nu_l^{\rm BS})$
 is multiplied by an extra phase shift $e^{\textsf{j} {\left( {(i_{\rm BS}^{\rm h}-1)\mu_l^{\rm BS}\!
 +\! (i_{\rm BS}^{\rm v}-1)\nu_l^{\rm BS}} \right)}}$ for $m\! =\! (i_{\rm BS}^{\rm v}-1)I_{\rm BS}^{\rm h}
 \! +\! i_{\rm BS}^{\rm h}$ and $2\! \le\! m\! \le\! I_{\rm BS}$. Obviously, these regular phase
 shifts can constitute the effective array response vector of equivalent fully-digital array
 with size of $I_{\rm BS}^{\rm h}\! \times\! I_{\rm BS}^{\rm v}$, i.e., ${\bm{\bar{\bar a}}}_{\rm BS}
 (\mu_l^{\rm BS},\nu_l^{\rm BS})\! =\! \bm{a}_{\rm v}(\nu_l^{\rm BS},I_{\rm BS}^{\rm v})\!\otimes\!
 \bm{a}_{\rm h}(\mu_l^{\rm BS},I_{\rm BS}^{\rm h})\! \in\! \mathbb{C}^{I_{\rm BS}}$.
 Thus, the UL received signal matrix $\bm{Y}_{{\rm UL},l}$ in (\ref{eq_Y_ul_angle1})
 can be then rewritten as
\begin{equation}\label{eq_Y_ul_angle2} % eq 17
 \bm{Y}_{{\rm UL},l} = \gamma_{{\rm UL},l} \left( {\bm{\bar{\bar a}}}_{\rm BS}(\mu_l^{\rm BS},\nu_l^{\rm BS})
 \bm{s}_{{\rm UL},l}^{\rm T} \right) \circ \bm{\widetilde Y}_{{\rm UL},l} + \bm{N}_{{\rm UL},l} ,
\end{equation}
 where $\gamma_{{\rm UL},l}\! =\! \sqrt{P_lG_l} \alpha_l (\bm{q}_{{\rm RF},l}^{[1]})^{\rm H}
 \bm{A}_{{\rm UL},l} \bm{p}_{{\rm RF},l}$ is the beam-aligned effective channel gain.

 For the received signal model in (\ref{eq_Y_ul_angle2}), we propose a prior-aided iterative angle
 estimation algorithm as follows. At the first iteration, i.e., $i_{\rm BS}\! =\! 1$, the azimuth
 and elevation angles at the $l$th BS can be first estimated as ${\widehat \theta}_l^{(i_{\rm BS})}$
 and ${\widehat \varphi}_l^{(i_{\rm BS})}$, and the corresponding horizontally and vertically virtual
 angles are ${\widehat \mu}_l^{(i_{\rm BS})}$ and ${\widehat \nu}_l^{(i_{\rm BS})}$ for
 $1\! \le\! l\! \le\! L$ by applying the TDU-ESPRIT algorithm \cite{Liao_Tcom19,Haardt_SSD_TSP98} to
 the received signal matrix $\bm{Y}_{{\rm UL},l}$. Furthermore, to minimize the impact of
 $\bm{\widetilde Y}_{{\rm UL},l}$ on (\ref{eq_Y_ul_angle2}), more accurate angle estimates can be
 acquired by utilizing the estimated angles above to iteratively compensate $\bm{Y}_{{\rm UL},l}$
 at the subsequent iterations (i.e., $i_{\rm BS}\! \ge\! 2$). Specifically, for the $i_{\rm BS}$th iteration,
 according to the rough virtual angle estimates ${\widetilde \mu}_l^{\rm BS}$ and ${\widetilde \nu}_l^{\rm BS}$,
 and ${\widehat \mu}_l^{(i_{\rm BS}-1)}$ and ${\widehat \nu}_l^{(i_{\rm BS}-1)}$ estimated
 at the $(i_{\rm BS}\! -\! 1)$th iteration, we define the compensation matrix as
 $\bm{\widetilde Y}_{{\rm UL},l}^{(i_{\rm BS}-1)}\! =\! \left[ \bm{\widetilde y}_{{\rm UL},l}^{(i_{\rm BS}-1)}[\{{\cal K}_l\}_1]\! \cdots\! \bm{\widetilde y}_{{\rm UL},l}^{(i_{\rm BS}-1)}[\{{\cal K}_l\}_{K_l}] \right]$,
 whose the $k_l$th column $\bm{\widetilde y}_{{\rm UL},l}^{(i_{\rm BS}-1)}[k_l]\! \in\! \mathbb{C}^{I_{\rm BS}}$
 is given by
\begin{align}\label{eq_y_tilde_ul_pie} % eq 18
 &\bm{\widetilde y}_{{\rm UL},l}^{(i_{\rm BS}-1)}[k_l] \nonumber\\
 &= \left( \bm{\bar a}_{\rm v}({\widetilde \nu}_l^{\rm BS},I_{\rm BS}^{\rm v},k_l)
 \otimes \bm{\bar a}_{\rm h}({\widetilde \mu}_l^{\rm BS},I_{\rm BS}^{\rm h},k_l) \right)^* \nonumber\\
 &\quad \circ \left( \bm{\bar a}_{\rm v}({\widehat \nu}_l^{(i_{\rm BS}-1)},I_{\rm BS}^{\rm v},k_l)
 \otimes \bm{\bar a}_{\rm h}({\widehat \mu}_l^{(i_{\rm BS}-1)},I_{\rm BS}^{\rm h},k_l) \right) .
\end{align}
 After the compensation matrix $\bm{\widetilde Y}_{{\rm UL},l}^{(i_{\rm BS}-1)}$ processing,
 the processed matrix $\bm{Y}_{{\rm UL},l}^{(i_{\rm BS})}\! =\! \left(\bm{\widetilde Y}_{{\rm UL},l}^{(i_{\rm BS}-1)}\right)^{*} \circ \bm{Y}_{{\rm UL},l}$ can be written as
\begin{align}\label{eq_Y_ul_pie} % eq 19
 \bm{Y}_{{\rm UL},l}^{(i_{\rm BS})} &= \gamma_{{\rm UL},l} \left( {\bm{\bar{\bar a}}}_{\rm BS}(\mu_l^{\rm BS},\nu_l^{\rm BS})
 \bm{s}_{{\rm UL},l}^{\rm T} \right) \nonumber\\
 &\quad \circ \left( \bm{\widetilde Y}_{{\rm UL},l} \circ
 \left(\bm{\widetilde Y}_{{\rm UL},l}^{(i_{\rm BS}-1)}\right)^{*} \right) + \bm{N}_{{\rm UL},l}^{(i_{\rm BS})} ,
\end{align}
 where $\bm{N}_{{\rm UL},l}^{(i_{\rm BS})}$ is the processed noise matrix.
 By applying the TDU-ESPRIT algorithm to those matrices $\{ \bm{Y}_{{\rm UL},l}^{(i_{\rm BS})} \}_{l=1}^L$
 again, we can obtain the more accurate angle estimates until the maximum number of iterations
 $i_{\rm BS}^{\rm max}$ is reached, i.e., $i_{\rm BS}\! =\! i_{\rm BS}^{\rm max}$. Finally,
 the estimates of azimuth and elevation angles and the corresponding virtual angles at BSs can be
 denoted as ${\widehat \theta}_l^{\rm BS}\! =\! {\widehat \theta}_l^{(i_{\rm BS}^{\rm max})}$,
 ${\widehat \varphi}_l^{\rm BS}\! =\!  {\widehat \varphi}_l^{(i_{\rm BS}^{\rm max})}$,
 ${\widehat \mu}_l^{\rm BS}\! =\! {\widehat \mu}_l^{(i_{\rm BS}^{\rm max})}$, and
 ${\widehat \nu}_l^{\rm BS}\! =\! {\widehat \nu}_l^{(i_{\rm BS}^{\rm max})}$ for $1\! \le\! l\! \le\! L$.
 The proposed prior-aided iterative angle estimation algorithm above is summarized in \textbf{Algorithm~\ref{ALG1}},
 where the beam squint effect can be addressed well.

\SetAlCapFnt{\normalsize}
\SetAlCapNameFnt{\normalsize}
\SetAlFnt{\small}
\begin{algorithm}[tp!]
\caption{Proposed Prior-Aided Iterative Angle Estimation Algorithm}\label{ALG1}
\LinesNumbered
\KwIn{Rough virtual angle information $\{ {\widetilde \mu}_l^{\rm BS},{\widetilde \nu}_l^{\rm BS},{\widetilde \mu}_l^{\rm AC},{\widetilde \nu}_l^{\rm AC} \}$, transmitted pilot signal $\bm{s}_{{\rm UL},l}$, maximum iterations $i_{\rm BS}^{\rm max}$, and dimensional parameters $\{ N_{\rm AC},M_{\rm AC},{\bar M}_{\rm BS},I_{\rm BS},I_{\rm BS}^{\rm h},I_{\rm BS}^{\rm v},K_l \}$}
\KwOut{Estimated azimuth/elevation angles $\{ {\widehat \theta}_l^{\rm BS},{\widehat \varphi}_l^{\rm BS} \}$ and virtual angles $\{ {\widehat \mu}_l^{\rm BS},{\widehat \nu}_l^{\rm BS} \}$}
{\% Preliminary (subarray selection and signal transmission)}\\
{Determine antenna indices ${\cal I}_{{\rm AC},l}$ and ${\cal I}_{\rm BS}^{[1]}$}\;
{Initialize $\bm{p}_{{\rm RF},l}\! =\! \bm{0}_{N_{\rm AC}}$ and then let $[\bm{p}_{{\rm RF},l}]_{{\cal I}_{{\rm AC},l}}\! =\! \textstyle{1 \over \sqrt{M_{\rm AC}}} [\bm{a}_{\rm AC}({\widetilde \mu}_l^{\rm AC},{\widetilde \nu}_l^{\rm AC})]_{{\cal I}_{{\rm AC},l}}$}\;
\For{$m\! =\! 1,\! \cdots\!, I_{\rm BS}$}
{{Determine antenna index ${\cal I}_{\rm BS}^{[m]}$}\;
{Initialize $\bm{q}_{{\rm RF},l}^{[m]}\! =\! \bm{0}_{N_{\rm BS}}$ and then let
 $[\bm{q}_{{\rm RF},l}^{[m]}]_{{\cal I}_{\rm BS}^{[m]}}\! =\! \textstyle{1 \over \sqrt{{\bar M}_{\rm BS}}}
 [\bm{a}_{\rm BS}({\widetilde \mu}_l^{\rm BS},{\widetilde \nu}_l^{\rm BS})]_{{\cal I}_{\rm BS}^{[1]}}$}\;
 {Transmit pilot signal $\bm{s}_{{\rm UL},l}$ to obtain received signal vector $\bm{y}_{{\rm UL},l}^{[m]}$ in (\ref{eq_y_ul_vec1})}\;
}
{Stack as $\{\bm{y}_{{\rm UL},l}^{[m]}\}_{m=1}^{I_{\rm BS}}$ as $\bm{Y}_{{\rm UL},l}\! =\! \left[ \bm{y}_{{\rm UL},l}^{[1]}\cdots \bm{y}_{{\rm UL},l}^{[I_{\rm BS}]} \right]^{\rm T}$ in (\ref{eq_Y_ul_angle1}) and (\ref{eq_Y_ul_angle2})}\;
{\% Prior-aided iterative angle estimation}\\
\For{$i_{\rm BS}\! =\! 1,\! \cdots\!, i_{\rm BS}^{\rm max}$}
{ \eIf{$i_{\rm BS}\! =\! 1$}
{ {Apply TDU-ESPRIT algorithm to $\bm{Y}_{{\rm UL},l}$}\;
 {Obtain angle estimates of first iteration as $\{ {\widehat \theta}_l^{(i_{\rm BS})},{\widehat \varphi}_l^{(i_{\rm BS})} \}$ and $\{ {\widehat \mu}_l^{(i_{\rm BS})},{\widehat \nu}_l^{(i_{\rm BS})} \}$}\;
}
{ {Design compensation matrix $\bm{\widetilde Y}_{{\rm UL},l}^{(i_{\rm BS}-1)}$, whose $k_l$th column $\bm{\widetilde y}_{{\rm UL},l}^{(i_{\rm BS}-1)}[k_l]$ is shown in (\ref{eq_y_tilde_ul_pie})}\;
 {Obtain compensated matrix $\bm{Y}_{{\rm UL},l}^{(i_{\rm BS})}\! =\! \left(\bm{\widetilde Y}_{{\rm UL},l}^{(i_{\rm BS}-1)}\right)^{*} \circ \bm{Y}_{{\rm UL},l}$ in (\ref{eq_Y_ul_pie})}\;
 {Apply TDU-ESPRIT algorithm to $\bm{Y}_{{\rm UL},l}^{(i_{\rm BS})}$}\;
 {Obtain angle estimates of $i_{\rm BS}$th iteration as $\{ {\widehat \theta}_l^{(i_{\rm BS})},{\widehat \varphi}_l^{(i_{\rm BS})} \}$ and $\{ {\widehat \mu}_l^{(i_{\rm BS})},{\widehat \nu}_l^{(i_{\rm BS})} \}$}\;
}
}
{\bf Return}: ${\widehat \theta}_l^{\rm BS}\! =\! {\widehat \theta}_l^{(i_{\rm BS}^{\rm max})}$, ${\widehat \varphi}_l^{\rm BS}\! =\!  {\widehat \varphi}_l^{(i_{\rm BS}^{\rm max})}$, ${\widehat \mu}_l^{\rm BS}\! =\! {\widehat \mu}_l^{(i_{\rm BS}^{\rm max})}$, and ${\widehat \nu}_l^{\rm BS}\! =\! {\widehat \nu}_l^{(i_{\rm BS}^{\rm max})}$
\end{algorithm}

\begin{remark} % Remark 1
 Based on the analysis above, by controlling the connection patterns, the reconfigurable RF selection
 network can select the desired subarrays to obtain an equivalent low-dimensional fully-digital array, so that
 the robust array signal processing techniques can be utilized to obtain the accurate angle estimates.
 On the other hand, the size of each selected subarray, i.e., ${\bar M}_{\rm BS}^{\rm h}\! \times\!
 {\bar M}_{\rm BS}^{\rm v}$, is large enough. This indicates that  at the initial angle estimation stage,
 we can achieve the sufficient full-dimensional beamforming gain with the aid of rough angle estimates
 to effectively combat the severe path loss of long-distance THz links and improve the receive SNR.
\end{remark}

\newcounter{TempEqCnt}
\setcounter{TempEqCnt}{\value{equation}}
\setcounter{equation}{22}
\begin{figure*}[hb]
\hrulefill
\begin{align}\label{eq_y_dop_lk} % eq 23
 y_{{\rm do},l}^{[{\bar m}]}[k_l] &= \sqrt{P_l} \bm{w}_{{\rm RF},l}^{\rm H} \bm{\widetilde H}_{{\rm DL},l}^{'[{\bar m}]}[k_l]
 \bm{f}_{{\rm RF},l} s_{{\rm do},l}^{[{\bar m}]}[k_l] + n_{{\rm do},l}^{[{\bar m}]}[k_l] \nonumber \\
 &= \underbrace{\sqrt{P_lG_l} \alpha_l e^{\textsf{j} {\pi f_s \tau_l}}}_{\gamma_{{\rm do},l}}
 e^{\textsf{j} {2\pi \Delta{\widetilde \psi}_{l,k_l} ({\bar m} - 1) T_{\rm sym} }} \underbrace{\bm{w}_{{\rm RF},l}^{\rm H} \bm{\widetilde A}_{{\rm DL},l}[k_l] \bm{f}_{{\rm RF},l} e^{\textsf{j} {(k_l - 1) \mu_l^\tau}} s_{{\rm do},l}[k_l]}_{{\bar s}_{{\rm do},l}[k_l]} + n_{{\rm do},l}^{[{\bar m}]}[k_l] .
\end{align}
\end{figure*}
 
\subsubsection{Fine Angle Estimation at Aircraft}\label{S3.1.2}
 Due to the channel reciprocity of UL and DL, the acquisition of fine angle estimates at aircraft
 in DL is similar to the fine angle estimation at BSs. At this stage, instead of using the rough angle
 estimates, the fine angles estimated at BSs in Section~\ref{S3.1.1} can be used not only to design the analog precoding
 vectors at BSs for beam alignment with improved receive SNR, but also to refine the
 GTTDU modules at BSs. Specifically, we consider $I_{\rm AC}\! =\! I_{\rm AC}^{\rm h}I_{\rm AC}^{\rm v}$
 OFDM symbols to estimate the fine azimuth and elevation angles at aircraft, where the size of the
 equivalent low-dimensional fully-digital array is $I_{\rm AC}^{\rm h}\! \times\! I_{\rm AC}^{\rm v}$.
 Based on the estimated $\{ {\widehat \mu}_l^{\rm BS},{\widehat \nu}_l^{\rm BS} \}_{l=1}^L$,
 the analog precoding vector can be designed as $\bm{f}_{{\rm RF},l}\! =\!
 \bm{a}_{\rm BS}({\widehat \mu}_l^{\rm BS},{\widehat \nu}_l^{\rm BS})$ for $1\! \le\! l\! \le\! L$.
 By employing the reconfigurable RF selection network, the selected antenna index in the $n$th OFDM sysmbol
 at the $l$th aircraft subarray is denoted by ${\cal I}_{{\rm AC},l}^{[n]}$ with
 ${\bar M}_{\rm AC}\! =\! |{\cal I}_{{\rm AC},l}^{[n]}|_c$.
 Then, initialize the analog combining vector as $\bm{w}_{{\rm RF},l}^{[n]}\! =\! \bm{0}_{N_{\rm AC}}$,
 and then let $[\bm{w}_{{\rm RF},l}^{[n]}]_{{\cal I}_{{\rm AC},l}^{[n]}}\! =\! \textstyle{1 \over \sqrt{{\bar M}_{\rm AC}}}
 [\bm{a}_{\rm AC}({\widetilde \mu}_l^{\rm AC},{\widetilde \nu}_l^{\rm AC})]_{{\cal I}_{{\rm AC},l}^{[1]}}$,
 for $1\! \le\! n\! \le\! I_{\rm AC}$, $1\! \le\! l\! \le\! L$.

 According to the DL transmission in (\ref{eq_y_dl}), at the $l$th RF chain of aircraft,
 the received signal $y_{{\rm DL},l}^{[n]}[k_l]$ at the $k_l$th subcarrier of the $n$th
 OFDM symbol corresponding to the $l$th BS can be expressed as
\setcounter{equation}{19}
\begin{equation}\label{eq_y_dkl} % eq 20
 y_{{\rm DL},l}^{[n]}[k_l]\! =\! \sqrt{P_l} (\bm{w}_{{\rm RF},l}^{[n]})^{\rm H} \bm{\widetilde H}_{{\rm DL},l}^{'[n]}[k_l]
 \bm{f}_{{\rm RF},l} s_{{\rm DL},l}^{[n]}[k_l] + n_{{\rm DL},l}^{[n]}[k_l] ,
\end{equation}
 where $k_l\! \in\! {\cal K}_l$, $1\! \le\! n\! \le\! I_{\rm AC}$, $\bm{\widetilde H}_{{\rm DL},l}^{'[n]}[k_l]$
 is the compensated DL channel matrix, and $s_{{\rm DL},l}^{[n]}[k_l]$ and $n_{{\rm DL},l}^{[n]}[k_l]$
 are the transmitted pilot signal and noise, respectively. Considering the received signals at
 $K_l$ subcarriers of $I_{\rm AC}$ OFDM symbols, we can obtain the DL received signal matrix
 $\bm{Y}_{{\rm DL},l}\! \in\! \mathbb{C}^{I_{\rm AC}\! \times\! K_l}$ as
\begin{align}\label{eq_Y_dl_angle1} % eq 21
 &\bm{Y}_{{\rm DL},l} \nonumber\\
 &=\! \sqrt{P_lG_l} \alpha_l e^{\textsf{j} {\pi f_s \tau_l}}
 \big( \bm{\bar W}_{{\rm RF},l}^{\rm H} \bm{A}_{{\rm DL},l}
 \bm{f}_{{\rm RF},l} {\underbrace{\left( \bm{a}_{\tau}(\mu_l^\tau,K_l) \circ
 \bm{s}_{{\rm DL},l} \right)}_{\bm{\bar s}_{{\rm DL},l}}}^{\rm T} \big) \nonumber\\
 &\quad \circ \bm{\widetilde Y}_{{\rm DL},l}
 + \bm{N}_{{\rm DL},l} ,
\end{align}
 where $\bm{\bar W}_{{\rm RF},l}\! =\! \left[ \bm{w}_{{\rm RF},l}^{[1]}\! \cdots\!
 \bm{w}_{{\rm RF},l}^{[I_{\rm AC}]} \right]\! \in\! \mathbb{C}^{N_{\rm AC}\!\times\! I_{\rm AC}}$
 and $\bm{\widetilde Y}_{{\rm DL},l}$ are the analog combining and residual beam squint matrices, respectively,
 $\bm{s}_{{\rm DL},l}\! =\! \bm{s}_{{\rm DL},l}^{[n]} \! =\! \left[ s_{{\rm DL},l}^{[n]}[\{{\cal K}_l\}_1]\! \cdots\!
 s_{{\rm DL},l}^{[n]}[\{{\cal K}_l\}_{K_l}]\right]^{\rm T}\! \in\! \mathbb{C}^{K_l}$ for
 $1\! \le\! n\! \le\! I_{\rm AC}$, and $\bm{N}_{{\rm DL},l}$ is the corresponding noise matrix.
 In (\ref{eq_Y_dl_angle1}), the steering vector associated with path delay $\tau_l$ can be defined as
 $\bm{a}_{\tau}(\mu_l^\tau,K_l)\! =\! \left[ e^{\textsf{j} (\{{\cal K}_l\}_1\! -\! 1) \mu_l^\tau} ~
 e^{\textsf{j} (\{{\cal K}_l\}_2\! -\! 1) \mu_l^\tau} \cdots
 e^{\textsf{j} (\{{\cal K}_l\}_{K_l}\! -\! 1) \mu_l^\tau} \right]^{\rm T}$
 with $\mu_l^\tau\! =\! -2\pi f_s \tau_l/ K$ being the virtual delay.
 Similar to (\ref{eq_Y_ul_angle2}), $\bm{Y}_{{\rm DL},l}$ can be rewritten as
\begin{equation}\label{eq_Y_dl_angle2} % eq 22
 \bm{Y}_{{\rm DL},l} = \gamma_{{\rm DL},l} \left( {\bm{\bar{\bar a}}}_{\rm AC}(\mu_l^{\rm AC},\nu_l^{\rm AC})
 \bm{\bar s}_{{\rm DL},l}^{\rm T} \right) \circ \bm{\widetilde Y}_{{\rm DL},l} + \bm{N}_{{\rm DL},l} ,
\end{equation}
 where $\gamma_{{\rm DL},l}\! =\! \sqrt{P_lG_l} \alpha_l e^{\textsf{j} {\pi f_s \tau_l}}
 (\bm{w}_{{\rm RF},l}^{[1]})^{\rm H} \bm{A}_{{\rm DL},l} \bm{f}_{{\rm RF},l}$, and
 ${\bm{\bar{\bar a}}}_{\rm AC}(\mu_l^{\rm AC},\nu_l^{\rm AC})\! =\!
 \bm{a}_{\rm v}(\nu_l^{\rm AC},I_{\rm AC}^{\rm v})\!\otimes\!
 \bm{a}_{\rm h}(\mu_l^{\rm AC},I_{\rm AC}^{\rm h})\! \in\! \mathbb{C}^{I_{\rm AC}}$
 is the effective array response vector of equivalent low-dimensional fully-digital array at the
 $l$th subarray of aircraft.
 For the received signal model in (\ref{eq_Y_dl_angle2}), we can also utilize the proposed prior-aided iterative angle estimation algorithm in \textbf{Algorithm~\ref{ALG1}} to obtain the more accurate angle estimates.
 By replacing the input parameters
 $\{ {\widetilde \mu}_l^{\rm BS},{\widetilde \nu}_l^{\rm BS},\{ {\cal I}_{\rm BS}^{[m]} \}_{m=1}^{I_{\rm BS}},\bm{s}_{{\rm UL},l},{\bar M}_{\rm BS},I_{\rm BS},I_{\rm BS}^{\rm h},I_{\rm BS}^{\rm v},i_{\rm BS},i_{\rm BS}^{\rm max} \}$
 for BSs with the corresponding parameters
 $\{ {\widehat \mu}_l^{\rm BS},{\widehat \nu}_l^{\rm BS},\{ {\cal I}_{{\rm AC},l}^{[n]} \}_{n=1}^{I_{\rm AC}},\bm{s}_{{\rm DL},l},{\bar M}_{\rm AC},I_{\rm AC},I_{\rm AC}^{\rm h},I_{\rm AC}^{\rm v},i_{\rm AC},i_{\rm AC}^{\rm max} \}$
 for aircraft, the estimates of azimuth and elevation angles and the corresponding virtual angles at aircraft can be
 obtained as ${\widehat \theta}_l^{\rm AC}\! =\! {\widehat \theta}_l^{(i_{\rm AC}^{\rm max})}$,
 ${\widehat \varphi}_l^{\rm AC}\! =\!  {\widehat \varphi}_l^{(i_{\rm AC}^{\rm max})}$,
 ${\widehat \mu}_l^{\rm AC}\! =\! {\widehat \mu}_l^{(i_{\rm AC}^{\rm max})}$, and
 ${\widehat \nu}_l^{\rm AC}\! =\! {\widehat \nu}_l^{(i_{\rm AC}^{\rm max})}$ for $1\! \le\! l\! \le\! L$.

\subsection{Fine Doppler Shift Estimation under Doppler-Squint Effect}\label{S3.2}
 Based on the fine angle estimates above, the analog combining vectors of $L$ subarrays at aircraft
 are designed to achieve beam alignment, i.e., initialize $\bm{w}_{{\rm RF},l}$ as $\bm{w}_{{\rm RF},l}\!
 =\! \bm{0}_{N_{\rm AC}}$ and then let $[\bm{w}_{{\rm RF},l}]_{{\cal I}_{{\rm AC},l}}\! =\! \textstyle{1
 \over \sqrt{M_{\rm AC}}} [\bm{a}_{\rm AC}({\widehat \mu}_l^{\rm AC},{\widehat \nu}_l^{\rm AC})]_{{\cal I}_{{\rm AC},l}}$
 for $1\! \le\! l\! \le\! L$. The GTTDU module at aircraft can be also refined to further mitigate
 the delay-beam squint effects. Since the rough Doppler shift estimates are not precise enough for
 data transmission, we will use $N_{\rm do}$ OFDM symbols to estimate the fine Doppler shifts in DL,
 where how to solve the Doppler squint effect is also considered. To ensure the effective channels
 within multiple OFDM symbols to be quasi-static observed at the aircraft, the transmitters at BSs
 still need to perform rough Doppler shift pre-compensation on the transmit signals at this stage.

\begin{algorithm}[tp!]
\caption{Proposed Prior-Aided Iterative Doppler Shift Estimation Algorithm}\label{ALG2}
\LinesNumbered
\KwIn{Estimated virtual angles $\{ {\widehat \mu}_l^{\rm BS},{\widehat \nu}_l^{\rm BS},{\widehat \mu}_l^{\rm AC},{\widehat \nu}_l^{\rm AC} \}$, rough Doppler shift estimates $\{ {\widetilde \psi}_{l,k_l} \}_{k_l=1}^{K_l}$ and ${\widetilde \psi}_{z,l}$, transmitted pilot signal $\{ s_{{\rm do},l}[k_l] \}_{k_l=1}^{K_l}$, maximum iterations $i_{\rm do}^{\rm max}$, wavelength $\lambda_z$ at central carrier frequency, and dimensional parameters $\{ N_{\rm AC},M_{\rm AC},N_{\rm do},K_l \}$}
\KwOut{Doppler shift estimates ${\widehat \psi}_{z,l}$ at center frequency and $\{ {\widehat \psi}_{l,k} \}_{k=1}^{K}$ at all subcarriers}
{\% Preliminary (signal transmission and preprocessing)}\\
{Determine antenna index ${\cal I}_{{\rm AC},l}$ and initialize $\bm{w}_{{\rm RF},l}\! =\! \bm{0}_{N_{\rm AC}}$}\;
{Let $[\bm{w}_{{\rm RF},l}]_{{\cal I}_{{\rm AC},l}}\! =\! \textstyle{1
 \over \sqrt{M_{\rm AC}}} [\bm{a}_{\rm AC}({\widehat \mu}_l^{\rm AC},{\widehat \nu}_l^{\rm AC})]_{{\cal I}_{{\rm AC},l}}$ and $\bm{f}_{{\rm RF},l}\! =\! \bm{a}_{\rm BS}({\widehat \mu}_l^{\rm BS},{\widehat \nu}_l^{\rm BS})$}\;
\For{${\bar m}\! =\! 1,\! \cdots\!, N_{\rm do}$}
{ \For{$k_l\! =\! 1,\! \cdots\!, K_l$}
{ {Transmit pilot signal $s_{{\rm do},l}[k_l]$ to obtain received signal $y_{{\rm do},l}^{[{\bar m}]}[k_l]$ in (\ref{eq_y_dop_lk})}\;
{Remove compensated phase $e^{-\textsf{j} {2\pi {\widetilde \psi}_{l,k_l} ({\bar m} - 1) T_{\rm sym} }}$ of $y_{{\rm do},l}^{[{\bar m}]}[k_l]$ to obtain ${\bar y}_{{\rm do},l}^{[{\bar m}]}[k_l]$ in (\ref{eq_y_bar_dop_lk})};
}
}
{Gather $\{ \{ {\bar y}_{{\rm do},l}^{[{\bar m}]}[k_l] \}_{k_l=1}^{K_l} \}_{{\bar m}=1}^{N_{\rm do}}$ into $\bm{Y}_{{\rm do},l}$ in (\ref{eq_Y_tilde_dop_l})}\;
{\% Prior-aided iterative Doppler shift estimation}\\
{{\bf Initialize}: $i_{\rm do}\! =\! 0$ and $\underline{\widehat v}_l^{(0)}\! =\! {\widetilde \psi}_{z,l}\lambda_z$}\;
\While{$i_{\rm do}\! \le\! i_{\rm do}^{\rm max}$}
{ \eIf{$i_{\rm do}\! =\! 0$}
{ {Obtain estimate ${\widehat \psi}_{z,l}^{(0)}$ (for comparison in simulations) by applying TLS-ESPRIT algorithm to $\bm{Y}_{{\rm do},l}$};
}
{ {Design compensation matrix $\bm{\widetilde Y}_{{\rm do},l}(\underline{\widehat v}_l^{(i_{\rm do}-1)})$, whose (${\bar m},k_l$)th entry is ${\widetilde y}_{{\rm do},l}^{[{\bar m}]}[k_l](\underline{\widehat v}_l^{(i_{\rm do}-1)})$}\;
 {Obtain compensated matrix $\bm{Y}_{{\rm do},l}^{(i_{\rm do})}\! =\! \bm{\widetilde Y}_{{\rm do},l}^{*}(\underline{\widehat v}_l^{(i_{\rm do}-1)}) \circ \bm{Y}_{{\rm do},l}$ in (\ref{eq_Y_bar_dop_l})}\;
 {Apply TLS-ESPRIT algorithm to $\bm{Y}_{{\rm do},l}^{(i_{\rm do})}$}\;
 {Obtain Doppler shift estimate of $i_{\rm do}$th iteration as ${\widehat \psi}_{z,l}^{(i_{\rm do})}$ and calculate $\underline{\widehat v}_l^{(i_{\rm do})}\! =\! {\widehat \psi}_{z,l}^{(i_{\rm do})}\lambda_z$}\;
}
$i_{\rm do}\! =\! i_{\rm do}\! +\! 1$
}
{\bf Return}: ${\widehat \psi}_{z,l}\! =\! {\widehat \psi}_{z,l}^{(i_{\rm do}^{\rm max})}$ and extend it to all subcarriers $\{ {\widehat \psi}_{l,k} \}_{k=1}^{K}$
\end{algorithm}

 According to the compensated DL channel matrix
 $\bm{\widetilde H}_{{\rm DL},l}^{'[{\bar m}]}[k_l]$, the received signal $y_{{\rm do},l}^{[{\bar m}]}[k_l]$
 at the $k_l$th subcarrier of the ${\bar m}$th OFDM symbol observed from the $l$th aircraft RF chain
 can be expressed as (\ref{eq_y_dop_lk}) on the bottom of this page.
%\begin{align}\label % eq 23
% &y_{{\rm do},l}^{[{\bar m}]}[k_l] \nonumber \\
% &= \sqrt{P_l} \bm{w}_{{\rm RF},l}^{\rm H} \bm{\widetilde H}_{{\rm DL},l}^{'[{\bar m}]}[k_l]
% \bm{f}_{{\rm RF},l} s_{{\rm do},l}^{[{\bar m}]}[k_l] + n_{{\rm do},l}^{[{\bar m}]}[k_l] , \nonumber \\
% &= \underbrace{\sqrt{P_lG_l} \alpha_l e^{\textsf{j} {\pi f_s \tau_l}}}_{\gamma_{{\rm do},l}}
% e^{\textsf{j} {2\pi \Delta{\widetilde \psi}_{l,k_l} ({\bar m} - 1) T_{\rm sym} }} \nonumber \\
% &\times \underbrace{\bm{w}_{{\rm RF},l}^{\rm H} \bm{\widetilde A}_{{\rm DL},l}[k_l] \bm{f}_{{\rm RF},l} e^{\textsf{j} {(k_l - 1) \mu_l^\tau}} s_{{\rm do},l}[k_l]}_{{\bar s}_{{\rm do},l}[k_l]} + n_{{\rm do},l}^{[{\bar m}]}[k_l] ,
%\end{align}
 In (\ref{eq_y_dop_lk}), $k_l\! \in\! {\cal K}_l$, $1\! \le\! {\bar m}\! \le\! N_{\rm do}$,
 $\Delta{\widetilde \psi}_{l,k_l}\! =\! \psi_{l,k_l}\! -\! {\widetilde \psi}_{l,k_l}$
 is the residual Doppler shift after compensation with ${\widetilde \psi}_{l,k_l}$ being the rough
 Doppler shift estimates at the $k_l$th subcarrier, and $s_{{\rm do},l}[k_l]\! =\! s_{{\rm do},l}^{[{\bar m}]}[k_l]$
 for $1\! \le\! {\bar m}\! \le\! N_{\rm do}$ and $n_{{\rm do},l}^{[{\bar m}]}[k_l]$
 are the transmitted pilot signal and noise, respectively. Since $\Delta{\widetilde \psi}_{l,k_l}$
 is too small to effectively estimate fine Doppler shifts using the limited OFDM symbols, the compensated
 phase difference $e^{-\textsf{j} {2\pi {\widetilde \psi}_{l,k_l} ({\bar m} - 1) T_{\rm sym} }}$ of
 $y_{{\rm do},l}^{[{\bar m}]}[k_l]$ in (\ref{eq_y_dop_lk}) can be removed to obtain
\setcounter{equation}{23}
\begin{align}\label{eq_y_bar_dop_lk} % eq 24
 {\bar y}_{{\rm do},l}^{[{\bar m}]}[k_l] &= \gamma_{{\rm do},l} e^{\textsf{j} ({\bar m} - 1) \nu_l^\psi}
 {\bar s}_{{\rm do},l}[k_l] \nonumber \\
 &\times \underbrace{e^{\textsf{j} \textstyle{{2\pi f_s \underline{v}_l} \over c} ({\textstyle{{k_l\!-\!1} \over K}}\!-\!{\textstyle{1 \over 2}}) ({\bar m} - 1) T_{\rm sym}}}_{{\widetilde y}_{{\rm do},l}^{[{\bar m}]}[k_l](\underline{v}_l)} + {\bar n}_{{\rm do},l}^{[{\bar m}]}[k_l] ,
\end{align}
 where $\nu_l^\psi\! =\! 2\pi \psi_{z,l} T_{\rm sym}$ denotes the virtual Doppler shift,
 and ${\widetilde y}_{{\rm do},l}^{[{\bar m}]}[k_l]$ and
 ${\bar n}_{{\rm do},l}^{[{\bar m}]}[k_l]$ are the Doppler squint value and noise, respectively.
 Considering the signals at $K_l$ subcarriers of $N_{\rm do}$ OFDM symbols, we can acquire the
 received signal matrix $\bm{Y}_{{\rm do},l}\! \in\! \mathbb{C}^{N_{\rm do}\! \times\! K_l}$ as
\begin{equation}\label{eq_Y_tilde_dop_l} % eq 25
 \bm{Y}_{{\rm do},l} = \gamma_{{\rm do},l} \left( \bm{a}_{\psi}(\nu_l^\psi,N_{\rm do}) \bm{\bar s}_{{\rm do},l}^{\rm T} \right)
 \circ \bm{\widetilde Y}_{{\rm do},l}(\underline{v}_l) + \bm{N}_{{\rm do},l} ,
\end{equation}
 where $\bm{a}_{\psi}(\nu_l^\psi,N_{\rm do})\! =\! \left[ 1 ~ e^{\textsf{j} \nu_l^\psi}
 \cdots e^{\textsf{j} (N_{\rm do}\! -\! 1) \nu_l^\psi} \right]^{\rm T}\! \in\! \mathbb{C}^{N_{\rm do}}$
 denotes the steering vector associated with the Doppler shift $\psi_{z,l}$,
 $\bm{\bar s}_{{\rm do},l}\! =\! \left[ {\bar s}_{{\rm do},l}[\{{\cal K}_l\}_1]\! \cdots\!
 {\bar s}_{{\rm do},l}[\{{\cal K}_l\}_{K_l}]\right]^{\rm T}\! \in\! \mathbb{C}^{K_l}$,
 $\bm{\widetilde Y}_{{\rm do},l}(\underline{v}_l)$ with
 $[\bm{\widetilde Y}_{{\rm do},l}(\underline{v}_l)]_{{\bar m},k_l}\! =\!
 {\widetilde y}_{{\rm do},l}^{[{\bar m}]}[k_l](\underline{v}_l)$
 and $\bm{N}_{{\rm do},l}$ are the Doppler squint and noise matrices, respectively.

\setcounter{TempEqCnt}{\value{equation}}
\setcounter{equation}{26}
\begin{figure*}[ht]
\begin{align}\label{eq_y_del_lk} % eq 27
 y_{{\rm de},l}^{[{\bar n}]}[k_l] =& \sqrt{P_l} \bm{w}_{{\rm RF},l}^{\rm H}
 \bm{\widetilde H}_{{\rm DL},l}^{'[{\bar n}]}[k_l] \bm{f}_{{\rm RF},l} s_{{\rm de},l}^{[{\bar n}]}[k_l]
 + n_{{\rm de},l}^{[{\bar n}]}[k_l] \nonumber \\
 =& \underbrace{\sqrt{P_lG_l} \alpha_l e^{\textsf{j} {\pi f_s \tau_l}} \bm{w}_{{\rm RF},l}^{\rm H} \bm{A}_{{\rm DL},l} \bm{f}_{{\rm RF},l}}_{\gamma_{{\rm de},l}} e^{\textsf{j}{(k_l - 1) \mu_l^\tau}}
 \underbrace{e^{\textsf{j}{2\pi (\psi_{z,l}\! -\! {\widehat \psi}_{z,l}) ({\bar n}\! -\! 1) T_{\rm sym} }} s_{{\rm de},l}^{[{\bar n}]}}_{{\bar s}_{{\rm de},l}^{[{\bar n}]}} \cdot {\widetilde y}_{{\rm de},l}^{[{\bar n}]}[k_l] + n_{{\rm de},l}^{[{\bar n}]}[k_l] .
\end{align}
\hrulefill
\end{figure*}

 To attenuate the impact of Doppler squint matrix $\bm{\widetilde Y}_{{\rm do},l}(\underline{v}_l)$
 on (\ref{eq_Y_tilde_dop_l}), we propose the following prior-aided iterative Doppler shift estimation
 algorithm. Define the rough Doppler shift estimate at the central carrier frequency as
 ${\widetilde \psi}_{z,l}$, and the initially relative radial velocity is given by
 $\underline{\widehat v}_l^{(0)}\! =\! {\widetilde \psi}_{z,l}\lambda_z$. At the $i_{\rm do}$th iteration,
 by exploiting the acquired $\underline{\widehat v}_l^{(i_{\rm do}-1)}$ at the $(i_{\rm do}\! -\! 1)$th
 iteration, the compensation matrix can be designed as
 $\bm{\widetilde Y}_{{\rm do},l}(\underline{\widehat v}_l^{(i_{\rm do}-1)})$, and its (${\bar m},k_l$)th
 element is ${\widetilde y}_{{\rm do},l}^{[{\bar m}]}[k_l](\underline{\widehat v}_l^{(i_{\rm do}-1)})$,
 which can be acquired by replacing $\underline{v}_l$ of
 ${\widetilde y}_{{\rm do},l}^{[{\bar m}]}[k_l](\underline{v}_l)$ in (\ref{eq_y_bar_dop_lk}) with
 $\underline{\widehat v}_l^{(i_{\rm do}-1)}$. The compensated receive matrix
 $\bm{Y}_{{\rm do},l}^{(i_{\rm do})}\! =\! \bm{\widetilde Y}_{{\rm do},l}^{*}(\underline{\widehat v}_l^{(i_{\rm do}-1)}) \circ \bm{Y}_{{\rm do},l}$ can be then rewritten as
\setcounter{equation}{25}
\begin{align}\label{eq_Y_bar_dop_l} % eq 26
 \bm{Y}_{{\rm do},l}^{(i_{\rm do})} =& \gamma_{{\rm do},l} \left( \bm{a}_{\psi}(\nu_l^\psi,{\bar m}) \bm{\bar s}_{{\rm do},l}^{\rm T} \right) \nonumber \\
 &\circ \left( \bm{\widetilde Y}_{{\rm do},l}(\underline{v}_l) \circ
 \bm{\widetilde Y}_{{\rm do},l}^{*}(\underline{\widehat v}_l^{(i_{\rm do}-1)}) \right) + \bm{N}_{{\rm do},l}^{(i_{\rm do})} ,
\end{align}
 where $\bm{N}_{{\rm do},l}^{(i_{\rm do})}$ is the associated noise matrix. According to
 $\bm{Y}_{{\rm do},l}^{(i_{\rm do})}$ in (26), we can obtain the Doppler
 shift estimate at the center frequency of the $i_{\rm do}$th iteration, denoted by
 ${\widehat \psi}_{z,l}^{(i_{\rm do})}$, using Total Least Squares ESPRIT (TLS-ESPRIT) \cite{Roy_TSP89}.
 By employing this estimated ${\widehat \psi}_{z,l}^{(i_{\rm do})}$ to calculate the
 finely relative radial velocity, i.e., $\underline{\widehat v}_l^{(i_{\rm do})}\! =\!
 {\widehat \psi}_{z,l}^{(i_{\rm do})}\lambda_z$, we can design fine compensation matrix
 to further improve the accuracy of Doppler estimation. Finally, at the $i_{\rm do}^{\rm max}$th
 iteration, we can obtain the fine estimates of Doppler shift corresponding to $L$ BSs, i.e.,
 ${\widehat \psi}_{z,l}\! =\! {\widehat \psi}_{z,l}^{(i_{\rm do}^{\rm max})}$, which can
 be extended to all subcarriers $\{ {\widehat \psi}_{l,k} \}_{k=1}^{K}$ for $1\! \le\! l\! \le\! L$.
 The proposed prior-aided iterative Doppler shift estimation algorithm above is summarized in
 \textbf{Algorithm~\ref{ALG2}}, where the Doppler squint effect can be addressed well.

\subsection{Path Delay and Channel Gain Estimation}\label{S3.3}
 At the path delay estimation stage, the fine Doppler shift estimates above can be used to accomplish
 the fine Doppler compensation as shown in Fig.~\ref{FIG3}, and $N_{\rm de}$ OFDM symbols will be utilized
 to estimate the path delays in DL. Recall that $\bm{a}_{\tau}(\mu_l^\tau,K_l)\! =\! \left[ e^{\textsf{j}
 (\{{\cal K}_l\}_1\! -\! 1) \mu_l^\tau} ~ e^{\textsf{j} (\{{\cal K}_l\}_2\! -\! 1) \mu_l^\tau} \cdots
 e^{\textsf{j} (\{{\cal K}_l\}_{K_l}\! -\! 1) \mu_l^\tau} \right]^{\rm T}$ in (\ref{eq_Y_dl_angle1})
 denotes the steering vector associated with path delay $\tau_l$, and $\mu_l^\tau\! =\! -2\pi f_s \tau_l/ K$.
 The DL received signal $y_{{\rm de},l}^{[{\bar n}]}[k_l]$ at the $k_l$th subcarrier of the ${\bar n}$th
 OFDM symbol can be expressed as (\ref{eq_y_del_lk}) on the top of the next page.
%\begin{align}\label{eq_y_del_lk} % eq 27
% y_{{\rm de},l}^{[{\bar n}]}[k_l] =& \sqrt{P_l} \bm{w}_{{\rm RF},l}^{\rm H}
% \bm{\widetilde H}_{{\rm DL},l}^{'[{\bar n}]}[k_l] \bm{f}_{{\rm RF},l} s_{{\rm de},l}^{[{\bar n}]}[k_l]
% + n_{{\rm de},l}^{[{\bar n}]}[k_l] , \nonumber \\
% =& \underbrace{\sqrt{P_lG_l} \alpha_l e^{\textsf{j} {\pi f_s \tau_l}} \bm{w}_{{\rm RF},l}^{\rm H} \bm{A}_{{\rm DL},l} \bm{f}_{{\rm RF},l}}_{\gamma_{{\rm de},l}} e^{\textsf{j}{(k_l - 1) \mu_l^\tau}} \nonumber \\
% &\times \underbrace{e^{\textsf{j}{2\pi (\psi_{z,l}\! -\! {\widehat \psi}_{z,l}) ({\bar n}\! -\! 1) T_{\rm sym} }} s_{{\rm de},l}^{[{\bar n}]}}_{{\bar s}_{{\rm de},l}^{[{\bar n}]}} \cdot {\widetilde y}_{{\rm de},l}^{[{\bar n}]}[k_l] + n_{{\rm de},l}^{[{\bar n}]}[k_l] ,
%\end{align}
 In (\ref{eq_y_del_lk}), $k_l\! \in\! {\cal K}_l$, $1\! \le\! {\bar n}\! \le\! N_{\rm de}$, $s_{{\rm de},l}^{[{\bar n}]}\! =\! s_{{\rm de},l}^{[{\bar n}]}[k_l]$ for $k_l\! \in\! {\cal K}_l$ is the transmitted pilot signal\footnote{Note that we assume the same pilot signals are adopted by $K_l$ subcarriers, which maybe lead to the high PAPR in OFDM systems. Fortunately,
 we can utilize a predefined pseudo-random descrambling code spread at all subcarriers \cite{Liao_Tcom19}
 to reduce the high PAPR effectively.}, ${\widetilde y}_{{\rm de},l}^{[{\bar n}]}[k_l]$ is the error value including the residual beam-Doppler squint errors caused by the channel estimation error, and $n_{{\rm de},l}^{[{\bar n}]}[k_l]$ is the noise. By collecting all received signals at $K_l$ subcarriers into
 the vector $\bm{y}_{{\rm de},l}^{[{\bar n}]}\! \in\! \mathbb{C}^{K_l}$, we have
\setcounter{equation}{27}
\begin{align}\label{eq_y_del_l} % eq 28
 \bm{y}_{{\rm de},l}^{[{\bar n}]} &= \left[ y_{{\rm de},l}^{[{\bar n}]}[\{{\cal K}_l\}_1] \cdots
 y_{{\rm de},l}^{[{\bar n}]}[\{{\cal K}_l\}_{K_l}]\right]^{\rm T} \nonumber \\
 &= \gamma_{{\rm de},l} \bm{a}_{\tau}(\mu_l^\tau,K_l) {\bar s}_{{\rm de},l}^{[{\bar n}]}
 \circ \bm{\widetilde y}_{{\rm de},l}^{[{\bar n}]} + \bm{n}_{{\rm de},l}^{[{\bar n}]} ,
\end{align}
 where $\bm{\widetilde y}_{{\rm de},l}^{[{\bar n}]}\! =\! \left[ {\widetilde y}_{{\rm de},l}^{[{\bar n}]}[\{{\cal K}_l\}_1] \cdots {\widetilde y}_{{\rm de},l}^{[{\bar n}]}[\{{\cal K}_l\}_{K_l}]\right]^{\rm T}$ and $\bm{n}_{{\rm de},l}^{[{\bar n}]}$ denote the error and noise vector, respectively. Considering the received signals of $N_{\rm de}$ OFDM symbols,
 we can obtain the matrix $\bm{Y}_{{\rm de},l}\! =\! \left[ \bm{y}_{{\rm de},l}^{[1]} \cdots \bm{y}_{{\rm de},l}^{[N_{\rm de}]}\right]\! \in\! \mathbb{C}^{K_l\! \times\! N_{\rm de}}$ as
\begin{equation}\label{eq_Y_del_l} % eq 29
 \bm{Y}_{{\rm de},l} = \gamma_{{\rm de},l} \left( \bm{a}_{\tau}(\mu_l^\tau,K_l) \bm{\bar s}_{{\rm de},l}^{\rm T} \right)
 \circ \bm{\widetilde Y}_{{\rm de},l} + \bm{N}_{{\rm de},l} ,
\end{equation}
 where $\bm{\bar s}_{{\rm de},l}\! =\! \left[ {\bar s}_{{\rm de},l}^{[1]}\! \cdots\!
 {\bar s}_{{\rm de},l}^{[N_{\rm de}]}\right]^{\rm T}\! \in\! \mathbb{C}^{N_{\rm de}}$, and
 $\bm{\widetilde Y}_{{\rm de},l}\! =\! \left[ \bm{\widetilde y}_{{\rm de},l}^{[1]}\! \cdots\!
 \bm{\widetilde y}_{{\rm de},l}^{[N_{\rm de}]}\right]$ and $\bm{N}_{{\rm de},l}$ are the residual beam-Doppler squint and noise
 matrices, respectively. By exploiting the TLS-ESPRIT algorithm \cite{Roy_TSP89}, we can
 obtain the path delay estimates corresponding to $L$ BSs, i.e., $\{ {\widehat \tau}_l \}_{l=1}^L$.
 From (\ref{eq_Y_del_l}), we observe that the accuracy of path delay estimation
 depends on the angle and Doppler estimation accuracy, and this conclusion can be further
 verified by the simulation results in Section~\ref{S7}.

 To estimate the channel gains, we need to harness the received signal matrix $\bm{Y}_{{\rm de},l}$
 in (\ref{eq_Y_del_l}). Specifically, this matrix $\bm{Y}_{{\rm de},l}$ can be split into the
 equivalent channel gain ${\bar \alpha}_l$ and $\bm{\bar Y}_{{\rm de},l}$, i.e.,
 $\bm{Y}_{{\rm de},l}\! =\! {\bar \alpha}_l \bm{\bar Y}_{{\rm de},l}$, where
 ${\bar \alpha}_l\! =\! \sqrt{P_lG_l} \alpha_l$. Regardless of the residual beam-Doppler squint
 and noise matrices of $\bm{Y}_{{\rm de},l}$, we can then utilize the previously
 estimated dominant channel parameters, i.e., the azimuth/elevation angles at BSs
 and aircraft, Doppler shifts, and path delays, to reestablish the estimated matrix
 of $\bm{\bar Y}_{{\rm de},l}$ as $\bm{\widehat{\bar Y}}_{{\rm de},l}$.
 Finally, we can obtain the estimation of ${\bar \alpha}_l$, denoted by
 ${\widehat \alpha}_l$, as
\begin{equation}\label{eq_alpha_l} % eq 30
 {\widehat \alpha}_l = \frac{1}{N_{\rm de} K_l} \sum\limits_{{\bar n}=1}^{N_{\rm de}} \sum\limits_{k_l=1}^{K_l}
 \left[\bm{Y}_{{\rm de},l}\right]_{k_l,{\bar n}} \big/ \left[\bm{\widehat{\bar Y}}_{{\rm de},l}\right]_{k_l,{\bar n}} .
\end{equation}

\setcounter{TempEqCnt}{\value{equation}}
\setcounter{equation}{31}
\begin{figure*}[hb]
\hrulefill
\begin{align}\label{eq_y_l_data} % eq 32
 y_l^{[r]}[k] =& \underbrace{\sqrt{P_l} \bm{w}_{{\rm RF},l}^{\rm H} \bm{\widetilde H}_{{\rm DL},l}^{'[r]}[k] \bm{f}_{{\rm RF},l}}_{h_l^{[r]}[k]} s_{l}^{[r]}[k] + \underbrace{\bm{w}_{{\rm RF},l}^{\rm H} \sum\nolimits_{l'=1 \atop l'\ne l}^{L}{ \sqrt{P_{l'}}
 \bm{\widetilde H}_{{\rm DL},l'}^{'[r]}[k] \bm{f}_{{\rm RF},l'} s_{l'}^{[r]}[k] } + n_l^{[r]}[k]}_{z_l^{[r]}[k]} .
\end{align}
\end{figure*}

\section{Data-Aided Channel Tracking}\label{S4}

 In Section~\ref{S3}, we have acquired the estimates of dominant channel parameters,
 which will be used for the following data transmission. Although THz UM-MIMO-based aeronautical communication channels exhibit the fast time-varying fading characteristic caused by the large Doppler shifts, the variations of dominant channel parameters, including angles, delays, Doppler shifts, and channel gains, can be relatively smooth within very transitory duration time $T_{\rm sym}$. Hence, we regard the duration time of $N_{\rm C}$ OFDM symbols as a Time Interval (TI), and the channel parameters within this TI are assumed to be stationary. Note that after the rough or fine Doppler compensation, the channel related to each OFDM symbol within the same TI is still slowly changing due to the imperfect Doppler compensation.
 Hence, after a long period of accumulation, the channels can change obviously, which
 would drastically degrade the detection accuracy of received data. To improve the
 reliability and efficiency of data transmission, a DADD-based channel tracking algorithm
 is developed to track the beam-aligned effective channels in real-time, which would save
 numerous pilot overhead as the time-varying channels should be updated frequently.
 The proposed DADD-based method utilizes the channel correlation of two adjacent OFDM symbols,
 where the estimated channels in the previous symbol can be approximately regarded as the real-time channels
 of the next symbol to detect the data sequentially. Meanwhile, the powerful error correction capability of
 the channel coding (e.g., Turbo or LDPC codings) can correct part of the erroneous detected data
 to minimize error propagation during the decision-directed process.
 Note that at the data transmission stage, we consider $L$ BSs can simultaneously
 serve the aircraft using the same time-frequency resource to achieve the high
 spectrum efficiency, i.e., signals associated with different BSs can be distinguished
 in the spatial domain, rather than the OFDMA utilized for the initial channel estimation.
 The proposed DADD-based channel tracking algorithm is summarized in \textbf{Algorithm~\ref{ALG3}}.

 Specifically, considering the $r$th OFDM symbol with $r\! =\! (q\! -\! 1)N_{\rm C}\! +\! p$
 that corresponds to the $p$th OFDM symbol of the $q$th TI, the DL channel matrix
 $\bm{H}_{{\rm DL},l}^{[n]}[k]$ in (\ref{eq_H_k_dl_1}) can be rewritten as $\bm{H}_{{\rm DL},l}^{[r]}[k]$,
 which contains the channel parameters $G_l^{[q]}$, $\alpha_l^{[q]}$, $\psi_{z,l}^{[q]}$,
 $\underline{v}_l^{[q]}$, $\tau_l^{[q]}$, $\theta_l^{\rm AC}[q]$, $\varphi_l^{\rm AC}[q]$,
 $\theta_l^{\rm BS}[q]$, and $\varphi_l^{\rm BS}[q]$. Define the initial data sequence in the
 $r$th OFDM symbol at the $l$th BS as $\bm{x}_l^{[r]}$, and this sequence can be mapped to
 $K$ subcarriers via channel coding and modulation to obtain the transmitted signal vector, i.e.,
 $\bm{s}_l^{[r]}\! =\! \left[s_l^{[r]}[1]\! \cdots\! s_l^{[r]}[K]\right]^{\rm T}\! \in\! \mathbb{C}^{K}$.
 The DL baseband signal vector $\bm{y}^{[r]}[k]\! \in\! \mathbb{C}^{L}$ received by aircraft
 at the $k$th subcarrier of the $r$th OFDM symbol can be expressed as
\setcounter{equation}{30}
\begin{align}\label{eq_y_data} % eq 31
 \bm{y}^{[r]}[k] &= \left[ y_1^{[r]}[k] \cdots y_L^{[r]}[k] \right]^{\rm T} \nonumber\\
 &= \bm{W}_{\rm RF}^{\rm H} \left(\sum\limits_{l=1}^{L}{\sqrt{P_l}
 \bm{\widetilde H}_{{\rm DL},l}^{'[r]}[k] \bm{f}_{{\rm RF},l} s_{l}^{[r]}[k]} + \bm{n}^{[r]}[k] \right) ,
\end{align}
 where $1\! \le\! k\! \le\! K$, $\bm{W}_{\rm RF}\! =\! \left[ \bm{w}_{{\rm RF},1}\! \cdots\! \bm{w}_{{\rm RF},L}\right]$,
 and $\bm{n}^{[r]}[k]$ is the noise vector. In (\ref{eq_y_data}), the $l$th received signal
 $y_l^{[r]}[k]$ in $\bm{y}^{[r]}[k]$ corresponding to the transmitted signal of the $l$th BS is given by (\ref{eq_y_l_data}) on the bottom of this page.
%\begin{align}\label{eq_y_l_data} % eq 32
% y_l^{[r]}[k] =& \underbrace{\sqrt{P_l} \bm{w}_{{\rm RF},l}^{\rm H} \bm{\widetilde H}_{{\rm DL},l}^{'[r]}[k] \bm{f}_{{\rm RF},l}}_{h_l^{[r]}[k]} s_{l}^{[r]}[k] \nonumber\\
% +& \underbrace{\bm{w}_{{\rm RF},l}^{\rm H} \sum\nolimits_{l'=1 \atop l'\ne l}^{L}{ \sqrt{P_{l'}}
% \bm{\widetilde H}_{{\rm DL},l'}^{'[r]}[k] \bm{f}_{{\rm RF},l'} s_{l'}^{[r]}[k] }\! +\! n_l^{[r]}[k]}_{z_l^{[r]}[k]} ,
%\end{align}
 In (\ref{eq_y_l_data}), the second entry is the interference from other BSs, $n_l^{[r]}[k]$ is the combining noise, and $h_l^{[r]}[k]$ and $z_l^{[r]}[k]$ are the beam-aligned effective channel coefficient and interference plus noise, respectively. Note that the interference entry in (\ref{eq_y_l_data}) is regarded as the additional noise due to the small interference from other BSs caused by the large angle differences among different BSs and the extremely narrow beams formed by THz UM-MIMO array at aircraft. Thus, (\ref{eq_y_l_data}) can be rewritten as $y_l^{[r]}[k]\! =\! h_l^{[r]}[k] s_{l}^{[r]}[k]\! +\! z_l^{[r]}[k]$. The $K$ channel coefficients $\{ h_l^{[r]}[k] \}_{k=1}^{K}$ can form together the beam-aligned true effective channel vector $\bm{h}_l^{[r]}\! \in\! \mathbb{C}^{K}$ at the $k$th subcarrier in the $r$th OFDM symbol.

\begin{algorithm}[tp!]
\caption{Proposed DADD-Based Channel Tracking Algorithm}\label{ALG3}
\LinesNumbered
\KwIn{Estimated channel parameters $\{ {\widehat \theta}_l^{\rm BS},{\widehat \varphi}_l^{\rm BS},{\widehat \theta}_l^{\rm AC},{\widehat \varphi}_l^{\rm AC},{\widehat \psi}_{z,l},{\widehat \tau}_l,{\widehat \alpha}_l \}_{l=1}^{L}$, dimensional parameters $\{ K,L,N_{\rm C},{\widetilde K} \}$, and preset threshold ratio $\varepsilon$}
\KwOut{Estimated effective channel vector $\{ \bm{\widehat h}_l^{[r]} \}_{l=1}^{L}$ and detected data sequence $\{ \bm{\widehat x}_l^{[r]} \}_{l=1}^{L}$ for $r\! =\! 1,2,3,\! \cdots$}
{{\bf Initialize}: ${\widetilde{\cal K}}_l^{[0]}\! =\! \emptyset$ and ${\widehat h}_l^{[0]}[k]\! =\! {\widehat \alpha}_l e^{-\textsf{j} {2\pi \left({\textstyle{k-1 \over K}}-{\textstyle{1 \over 2}}\right)f_s {\widehat \tau}_l}} \bm{w}_{{\rm RF},l}^{\rm H} \bm{\widehat A}_{{\rm DL},l} \bm{f}_{{\rm RF},l}$ for $1\! \le\! k\! \le\! K$ and $1\! \le\! l\! \le\! L$}\;
\For{$q\! =\! 1,2,3,\! \cdots$}
{ \For{$p\! =\! 1,\! \cdots\! ,N_{\rm C}$}
{ $r\! =\! (q\! -\! 1)N_{\rm C}\! +\! p$\;
\eIf{$|{\widetilde{\cal K}}_l^{[r-1]}|_c\! \le\! {\widetilde K}$ {\rm for} $1\! \le\! l\! \le\! L$}
{ {Map initial data sequence $\{ \bm{x}_l^{[r]} \}_{l=1}^{L}$ to transmitted signal vector $\{ \bm{s}_l^{[r]} \}_{l=1}^{L}$}\;
{Obtain baseband signal vector $\{ \bm{y}^{[r]}[k] \}_{k=1}^{K}$ in (\ref{eq_y_data}), whose $l$th entry is $y_l^{[r]}[k]$ in (\ref{eq_y_l_data})}\;
{Design the digital combining matrix $\bm{W}_{\rm BB}^{[r]}[k]\! =\! \text{diag}({\widehat h}_1^{[r\!-\!1]}[k]\! \cdots\! {\widehat h}_L^{[r\!-\!1]}[k])$ for $1\! \le\! k\! \le\! K$}\;
{Obtain $\{ \bm{\widehat s}^{[r]}[k] \}_{k=1}^{K}$ in (\ref{eq_s_hat}) and extract $\{ \bm{\widehat s}_l^{[r]} \}_{l=1}^{L}$ to restore data sequence as $\{ \bm{\widehat x}_l^{[r]} \}_{l=1}^{L}$}\;
{Code and modulate $\{ \bm{\widehat x}_l^{[r]} \}_{l=1}^{L}$ again to yield $\{ \bm{\widetilde s}_l^{[r]} \}_{l=1}^{L}$ as pilot signal}\;
{Substitute ${\widetilde s}_l^{[r]}[k]$ into (\ref{eq_y_l_data}) to acquire ${\widehat h}_l^{[r]}[k]\! =\! y_l^{[r]}[k]/{\widetilde s}_l^{[r]}[k]$ for $1\! \le\! k\! \le\! K$ and $1\! \le\! l\! \le\! L$}\;
{Collect $\{ \bm{\widehat h}_l^{[r]} \}_{l=1}^{L}$ and initialize ${\widetilde{\cal K}}_l^{[r]}\! =\! \emptyset$ for $1\! \le\! l\! \le\! L$}\;
{ \For{$k\! =\! 1,\! \cdots\! ,K$ {\rm and} $l\! =\! 1,\! \cdots\! ,L$}
{Satisfy $\left| {\widehat h}_l^{[r]}[k] - {\widehat h}_l^{[r-1]}[k] \right|\! >\! \frac{\varepsilon}{K} \sum\nolimits_{k=1}^{K}{\big| {\widehat h}_l^{[r-1]}[k] \big|}$ in (\ref{eq_h_r_pi_hat}) and let ${\widetilde{\cal K}}_l^{[r]}\! =\! {\widetilde{\cal K}}_l^{[r]}\! \cup\! k$\;}
}
}
{{\bf Return}: $\{ \bm{\widehat h}_l^{[r]} \}_{l=1}^{L}$ and $\{ \bm{\widehat x}_l^{[r]} \}_{l=1}^{L}$ for $r\! =\! 1,2,3,\! \cdots$\;
{\bf Terminate} current algorithm and {\bf trigger off} pilot-aided channel tracking.}
}
}
\end{algorithm}

 Based on the estimated effective channel coefficient in the $(r\! -\! 1)$th OFDM symbol, denoted by
 ${\widehat h}_l^{[r\!-\!1]}[k]$ for $1\! \le\! l\! \le\! L$, we can design the digital combining matrix as
 $\bm{W}_{\rm BB}^{[r]}[k]\! =\! \text{diag}({\widehat h}_1^{[r\!-\!1]}[k]\! \cdots\! {\widehat h}_L^{[r\!-\!1]}[k])$.
 According to (\ref{eq_y_data}), the signal vector $\bm{s}^{[r]}[k]\! =\! \left[ s_{1}^{[r]}[k]\! \cdots\! s_{L}^{[r]}[k] \right]\! \in\! \mathbb{C}^{L}$
 can be estimated as
\setcounter{equation}{32}
\begin{equation}\label{eq_s_hat} % eq 33
 \bm{\widehat s}^{[r]}[k] = \left[ {\widehat s}_1^{[r]}[k] \cdots {\widehat s}_L^{[r]}[k] \right]^{\rm T}
 = \left(\bm{W}_{\rm BB}^{[r]}[k]\right)^{\rm H} \bm{y}^{[r]}[k] ,
\end{equation}
 where the $l$th entry of $\bm{\widehat s}^{[r]}[k]$ is ${\widehat s}_l^{[r]}[k]\! =\!
 {\textstyle{{{\widehat h}_l^{[r]}[k]} \over {{\widehat h}_l^{[r\!-\!1]}[k]}}}
 s_{l}^{[r]}[k]\! +\! {\textstyle{{z_l^{[r]}[k]} \over {{\widehat h}_l^{[r\!-\!1]}[k]}}}$.
 By extracting the received signal processed by the $l$th RF chain and gathering these signals
 at $K$ subcarriers, the estimation of transmitted signal vector $\bm{s}_l^{[r]}$
 can be denoted by $\bm{\widehat s}_l^{[r]}\! \in\! \mathbb{C}^{K}$. To track the effective
 channel of the current $r$th OFDM symbol, i.e., $\bm{h}_l^{[r]}$, this signal vector
 $\bm{\widehat s}_l^{[r]}$ can be demodulated and decoded as the detected data sequence
 $\bm{\widehat x}_l^{[r]}$ (i.e., the estimate of initial data sequence $\bm{x}_l^{[r]}$).
 This data sequence $\bm{\widehat x}_l^{[r]}$ can be then coded and modulated again to yield the
 transmitted signal vector $\bm{\widetilde s}_l^{[r]}$, which should be more accurate than the
 estimated $\bm{\widehat s}_l^{[r]}$ thanks to the error correction of channel coding.
 By considering $\bm{\widetilde s}_l^{[r]}$ as the pilot signal, we substitute its $k$th element,
 denoted by ${\widetilde s}_l^{[r]}[k]$, into the received signal $y_l^{[r]}[k]$ in (\ref{eq_y_l_data})
 to acquire the estimate of effective channel coefficient $h_l^{[r]}[k]$, i.e.,
 ${\widehat h}_l^{[r]}[k]\! =\! y_l^{[r]}[k]/{\widetilde s}_l^{[r]}[k]$.
 Finally, considering $K$ subcarriers, the estimated effective channel vector of the $r$th OFDM symbol is
 $\bm{\widehat h}_l^{[r]}\! \in\! \mathbb{C}^{K}$ for $1\! \le\! l\! \le\! L$.
 Accordingly, the digital combining matrix at the $k$th subcarrier in the $(r\!+\!1)$th OFDM symbol
 can be designed as $\bm{W}_{\rm BB}^{[r\!+\!1]}[k]\! =\! \text{diag}({\widehat h}_1^{[r]}[k]\!
 \cdots\! {\widehat h}_L^{[r]}[k])$, which is used to perform the subsequent channel equalization.
 Furthermore, by utilizing the previously estimated channel parameters at the initial channel estimation stage, the
 estimates of initial beam-aligned effective channel vectors $\{\bm{\widehat h}_l^{[0]}\}_{l\!=\!1}^L$
 can be obtained as ${\widehat h}_l^{[0]}[k]\! =\! {\widehat \alpha}_l e^{-\textsf{j}
 {2\pi \left({\textstyle{k-1 \over K}}-{\textstyle{1 \over 2}}\right)f_s {\widehat \tau}_l}}
 \bm{w}_{{\rm RF},l}^{\rm H} \bm{\widehat A}_{{\rm DL},l} \bm{f}_{{\rm RF},l}$ for
 $1\! \le\! k\! \le\! K$ and $1\! \le\! l\! \le\! L$, where $\bm{\widehat A}_{{\rm DL},l}$
 is the reconstructed DL array response matrix in (\ref{eq_A_dl_k1}) using the fine angle estimates.

 As the time goes on, the previously estimated channel parameters will not match the
 current effective channels. Therefore, the quality of the tracked effective
 channel vectors at the data-aided channel tracking stage should be monitored in
 real-time by exploiting the temporal correlation of two adjacent OFDM symbols. Specifically,
 for the estimated effective channel vector $\bm{\widehat h}_l^{[r]}$ in the $r$th OFDM symbol,
 its $k$th channel coefficient ${\widehat h}_l^{[r]}[k]$ can be regarded as a wrong coefficient
 if ${\widehat h}_l^{[r]}[k]$ satisfies
\begin{equation}\label{eq_h_r_pi_hat} % eq 34
 \left| {\widehat h}_l^{[r]}[k] - {\widehat h}_l^{[r-1]}[k] \right|
 > \frac{\varepsilon}{K} \sum\nolimits_{k=1}^{K}{\big| {\widehat h}_l^{[r-1]}[k] \big|} ,
\end{equation}
 where $\varepsilon$ is a preset threshold ratio. The indices of subcarriers involving erroneous
 coefficients composes a set ${\widetilde{\cal K}}_l^{[r]}$. Let ${\widetilde K}$ as the acceptable
 number of erroneous channel coefficients, the tracked effective channel vectors can be regarded
 as the invalid estimates if $|{\widetilde{\cal K}}_l^{[r]}|_c\! >\! {\widetilde K}$ for
 $1\! \le\! l\! \le\! L$, which will trigger off the pilot-aided channel tracking in Section~\ref{S5}.

\section{Pilot-Aided Channel Tracking}\label{S5}

 In this section, the previously estimated channel parameters in Section~\ref{S3} will be exploited as
 the prior information for facilitating the pilot-aided channel tracking. This is because according to
 the previous analysis, these channel parameters including angles, Doppler shifts are changing slowly and
 can usually not vary dramatically. Since previously estimated channel parameters can be more accurate
 than the rough estimates based on navigation information, the tracked channel parameters at this stage
 would be more accurate than those acquired at the initial channel estimation stage. The main process of
 the pilot-aided channel tracking is similar to the initial channel estimation in Section~\ref{S3}.
 The difference between them lies in that the azimuth and elevation angles at BSs and aircraft in this
 section are estimated by forming the array response vector of \emph{equivalent low-dimensional fully-digital
 sparse array}. By contrast, an equivalent fully-digital array with critical antenna spacing (i.e.,
 the half-wavelength antenna spacing) is considered in Section~\ref{S3}. The existing conclusions indicate
 that the usage of sparse array can improve the accuracy of angle estimation significantly, but these
 estimated angles would suffer from the angle ambiguity issue \cite{DavidTse05,Chuang_TAP15}. Fortunately,
 this angle ambiguity can be solved with the aid of the previously estimated angles. Due to space constraints,
 this section focuses on the pilot-aided angle tracking at BSs.

\begin{figure*}[!tp]
%\vspace{-5mm}
\begin{center}
 \includegraphics[width=1.6\columnwidth, keepaspectratio]{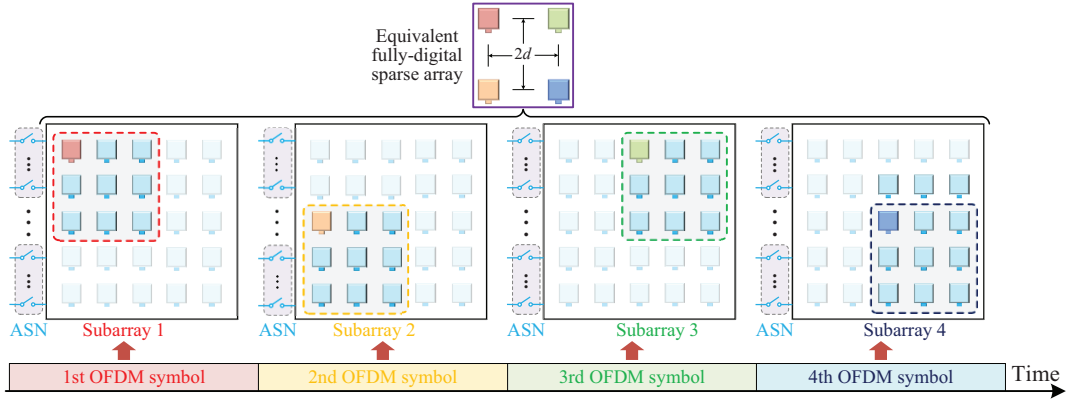}
\end{center}
\captionsetup{font = {footnotesize}, singlelinecheck = off, name = {Fig.}, labelsep = period} %, justification = raggedright
\caption{The schematic diagram of subarray selection based on different antenna connection patterns of the reconfigurable RF selection network at the angle tracking stage. Taking the UPA of size $5\!\times\! 5$ as an example, this UPA can be partitioned into $4$ subarrays of size $3\! \times\! 3$, and the interval between each subarray is the width of two antennas. The same RF chain sequentially selects the corresponding subarrays in $4$ successive OFDM symbols to receive signals, and these received signals will be equivalent to the signals received by a low-dimensional fully-digital sparse array of size $2\! \times\! 2$ with the sparse spacing $\varOmega\! =\! 2$.}
%\vspace{-6mm}
\label{FIG8}
\end{figure*}

 Specifically, $I'_{\rm BS}$ OFDM symbols are used to obtain the equivalent fully-digital sparse
 array of size $I_{\rm BS}^{'{\rm h}}\! \times\! I_{\rm BS}^{'{\rm v}}$ at BSs, where
 $I'_{\rm BS}\! =\! I_{\rm BS}^{'{\rm h}} I_{\rm BS}^{'{\rm v}}$ subarrays can be acquired
 by reconfiguring the dedicated connection pattern of the RF selection network.
 Define $\varOmega$ as the sparse antenna spacing relative to the critical
 antenna spacing $d$. The size of the selected subarray is
 ${\bar M}_{\rm BS}^{'{\rm h}}\! \times\! {\bar M}_{\rm BS}^{'{\rm v}}$ with ${\bar M}'_{\rm BS}\!
 =\! {\bar M}_{\rm BS}^{'{\rm h}} {\bar M}_{\rm BS}^{'{\rm v}}$ antenna elements, where
 ${\bar M}_{\rm BS}^{'{\rm h}}\! =\! N_{\rm BS}^{\rm h}\! -\! \varOmega(I_{\rm BS}^{'{\rm h}}\! -\! 1)$
 and ${\bar M}_{\rm BS}^{'{\rm v}}\! =\! N_{\rm BS}^{\rm v}\! -\! \varOmega(I_{\rm BS}^{'{\rm v}}\! -\! 1)$.
 Fig.~\ref{FIG8} depicts an example that the UPA with size of $5\!\times\! 5$ can be divided into $4$
 subarrays of size $3\! \times\! 3$, and these subarrays construct the array response vector of equivalent fully-digital
 sparse array of size $2\! \times\! 2$ with the sparse spacing $\varOmega\! =\! 2$.
 Similar to the fine angle estimation at BSs
 in Section~\ref{S3.1.1}, we can obtain the homologous UL received signal matrix $\bm{\bar Y}_{{\rm UL},l}\!
 \in\! \mathbb{C}^{I'_{\rm BS}\! \times\! K_l}$ in (\ref{eq_Y_ul_angle2}), where the
 effective array response vector of the sparse array can be expressed as
 ${\bm{\bar{\bar a}}}_{\rm BS}({\bar \mu}_l^{\rm BS},{\bar \nu}_l^{\rm BS})\! =\!
 \bm{a}_{\rm v}({\bar \nu}_l^{\rm BS},I_{\rm BS}^{'{\rm v}})\!\otimes\!
 \bm{a}_{\rm h}({\bar \mu}_l^{\rm BS},I_{\rm BS}^{'{\rm h}})\! \in\! \mathbb{C}^{I'_{\rm BS}}$
 with ${\bar \mu}_l^{\rm BS}\! =\! \varOmega\mu_l^{\rm BS}$ and
 ${\bar \nu}_l^{\rm BS}\! =\! \varOmega\nu_l^{\rm BS}$.
 By exploiting the proposed prior-aided iterative angle estimation in
 \textbf{Algorithm~\ref{ALG1}} as before, the estimates of ${\bar \mu}_l^{\rm BS}$ and
 ${\bar \nu}_l^{\rm BS}$ can be respectively obtained as $\widehat{\bar \mu}_l^{\rm BS}$ and
 $\widehat{\bar \nu}_l^{\rm BS}$ at each iteration.
 Note that $\widehat{\bar \mu}_l^{\rm BS}$ and $\widehat{\bar \nu}_l^{\rm BS}$
 suffer from the inherent angle ambiguity problem.
 To further address this angle ambiguity issue, we define an ordered
 index set ${\cal B}\! =\! \left\{ -1, -1\! +\! {\textstyle{1 \over \varOmega }},
 -1\! +\! {\textstyle{2 \over \varOmega }},\! \cdots\! ,1 \right\}$
 with $|{\cal B}|_c\! =\! 2\varOmega\! +\! 1$, and let $\widetilde{\bar \mu}_l^{\rm BS}\! =\!
 \widehat{\bar \mu}_l^{\rm BS}/\varOmega$ and $\widetilde{\bar \nu}_l^{\rm BS}\! =\!
 \widehat{\bar \nu}_l^{\rm BS}/\varOmega$. Thus, the estimates of virtual angles corresponding
 to $\widetilde{\bar \mu}_l^{\rm BS}$ and $\widetilde{\bar \nu}_l^{\rm BS}$,
 denoted by ${\widehat \mu}_l^{'\rm BS}$ and ${\widehat \nu}_l^{'\rm BS}$, should satisfy
 ${\widehat \mu}_l^{'\rm BS}\! =\! \widetilde{\bar \mu}_l^{\rm BS}\! +\! b_{\mu}^\star\pi$ and
 ${\widehat \mu}_l^{'\rm BS}\! =\! \widetilde{\bar \nu}_l^{\rm BS}\! +\! b_{\nu}^\star\pi$,
 where $b_{\mu}^\star\! \in\! {\cal B}$ and $b_{\nu}^\star\! \in\! {\cal B}$
 are the optimal indices. Due to the limited elements in ${\cal B}$, we adopt
 the exhaustive method to search for these optimal indices $b_{\mu}^\star$ and $b_{\nu}^\star$.
 The previously estimated ${\widehat \mu}_l^{\rm BS}$ and ${\widehat \nu}_l^{\rm BS}$
 in Section~\ref{S3.1.1} can be regarded as the prior information, i.e.,
 ${\widetilde \mu}_l^{\rm BS}\! =\! {\widehat \mu}_l^{\rm BS}$
 and ${\widetilde \nu}_l^{\rm BS}\! =\! {\widehat \nu}_l^{\rm BS}$, and
 $b_{\mu}^\star$ and $b_{\nu}^\star$ can be then obtained as
\begin{align} % eqs 43,44
 b_{\mu}^\star &= \arg \min\limits_{b_{\mu} \in {\cal B}}
 \left| \widetilde{\bar \mu}_l^{\rm BS} + b_{\mu}\pi - {\widetilde \mu}_l^{\rm BS} \right| , \label{eq_b_mu_star} \\
 b_{\nu}^\star &= \arg \min\limits_{b_{\nu} \in {\cal B}}
 \left| \widetilde{\bar \nu}_l^{\rm BS} + b_{\nu}\pi - {\widetilde \nu}_l^{\rm BS} \right| . \label{eq_b_nu_star}
\end{align}
 Based on the acquired estimates ${\widehat \mu}_l^{'\rm BS}$ and ${\widehat \nu}_l^{'\rm BS}$, we can calculate
 the updated estimates of azimuth and elevation angles at the $l$th BS as
 ${\widehat \theta}_l^{'\rm BS}$ and ${\widehat \varphi}_l^{'\rm BS}$, for $1\! \le\! l\! \le\! L$.
 The remaining steps are the same as those in Section~\ref{S3.1.1} except that the exhaustive search
 in (\ref{eq_b_mu_star}) and (\ref{eq_b_nu_star}) should be taken into account. Finally, we can obtain
 the fine estimates of azimuth and elevation angles at BSs, denoted by
 $\{ {\widehat \theta}_l^{\rm BS},{\widehat \varphi}_l^{\rm BS} \}_{l=1}^L$.
 In a similar way, the fine estimates of azimuth and elevation angles at aircraft can be also
 acquired as $\{ {\widehat \theta}_l^{\rm AC},{\widehat \varphi}_l^{\rm AC} \}_{l\!=\!1}^L$,
 where $I'_{\rm AC}\! =\! I_{\rm AC}^{'{\rm h}} I_{\rm AC}^{'{\rm v}}$ OFDM symbols are required.
 Moreover, with the help of the previously estimated Doppler shifts, the updated Doppler shift
 estimates $\{ {\widehat \psi}_{z,l} \}_{l=1}^{L}$ via the pilot-aided channel tracking will be
 more accurate than those estimated at the initial channel estimation stage, and so do the
 estimates of path delays $\{ {\widehat \tau}_l \}_{l=1}^L$ and channel gains
 $\{ {\bar \alpha}_l \}_{l=1}^L$. As shown in Fig.~\ref{FIG3}, the updated beam-aligned
 effective channels can be then used for the following data transmission, and the tracked
 channel parameters will be regarded as the prior information for the next pilot-aided channel tracking.

\begin{figure*}[!tp]
%\vspace{-5mm}
\begin{center}
 \includegraphics[width=1.5\columnwidth,keepaspectratio]{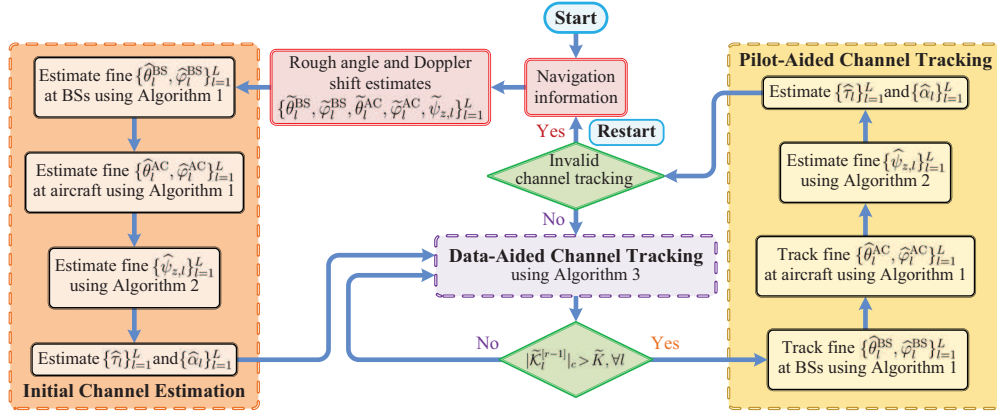}
\end{center}
 \captionsetup{font = {footnotesize}, singlelinecheck = off, justification = raggedright, name = {Fig.}, labelsep = period}%
 \caption{Flow diagram of the proposed channel estimation and tracking solution.}
 \label{FIG9}
%\vspace{-6mm}
\end{figure*}

 In order to intuitively describe the relationship among different channel estimation and tracking stages above, the block diagram of the proposed channel estimation and tracking solution is illustrated in Fig.~\ref{FIG9}.

\section{Performance Analysis}\label{S6}

\subsection{CRLBs of Channel Parameters}\label{S6.1}
 According to the effective received signal models in Section~\ref{S3}, we will investigate the CRLBs
 of the dominant channel parameters, i.e., azimuth/elevation angles at aerial BSs and aircraft,
 Doppler shifts, and path delays. Note that practical triple squint effects of aeronautical
 THz UM-MIMO channels would weaken the accuracy of channel parameter estimation, and these negative effects are not
 considered in deriving the CRLBs. So these CRLBs serve as the lower-bound of parameter estimation.

\setcounter{TempEqCnt}{\value{equation}}
\setcounter{equation}{37}
\begin{figure*}[hb]
\hrulefill
\begin{align} % eqs 38,40
\ln p(\bm{\bar Y}_{{\rm UL},l};\bm{\eta}_l) =& -{\bar I}_{\rm BS}K_l\ln(\pi\sigma_n^2)
 -\frac{1}{\sigma_n^2} \sum\limits_{{k_l}=1}^{K_l}\big( \left[ \bm{\bar y}_{{\rm UL},l}[\{{\cal K}_l\}_{k_l}]
 - \gamma_{{\rm UL},l} {\bm{\bar{\bar a}}}_{\rm BS}({\bar \mu}_l^{\rm BS},{\bar \nu}_l^{\rm BS})
 s_{{\rm UL},l}[\{{\cal K}_l\}_{k_l}] \right]^{\rm H} \nonumber \\
 \ & \times \left[ \bm{\bar y}_{{\rm UL},l}[\{{\cal K}_l\}_{k_l}] -
 \gamma_{{\rm UL},l} {\bm{\bar{\bar a}}}_{\rm BS}({\bar \mu}_l^{\rm BS},{\bar \nu}_l^{\rm BS})
 s_{{\rm UL},l}[\{{\cal K}_l\}_{k_l}] \right]\big) .\label{eq_lnp_ul_bar_CRLB}\\
 \mathrm{CRLB}_{\bm{\xi}_l^{\rm BS}} =&\,\bm{G}^{-1}(\bm{\eta}_l) = \frac{\sigma_n^2}{2\left|\gamma_{{\rm UL},l}\right|^2}
 \left\{ \sum\limits_{{k_l}=1}^{K_l} {\Re \left\{ \bm{B}_{{\rm BS},{k_l}}^{\rm H} \bm{\varGamma}_{\rm BS}^{\rm H}
 \left( \bm{I}_{{\bar I}_{\rm BS}} - \bm{\varPhi}_{\rm BS} \right) \bm{\varGamma}_{\rm BS} \bm{B}_{{\rm BS},{k_l}} \right\}}
 \right\}^{-1} . \tag{40} \label{eq_CRLB_xi}
\end{align}
\end{figure*}

\subsubsection{CRLBs of Angle Estimation at BSs and Aircraft}\label{S6.1.1}
 To investigate the performance at both the initial angle estimation stage and the following angle tracking stage,
 we consider the received signal model corresponding to the equivalent fully-digital sparse array
 with size of ${\bar I}_{\rm BS}^{\rm h}\! \times\! {\bar I}_{\rm BS}^{\rm v}$, where the
 sparse spacing is $\varOmega\! \ge\! 1$. Based on the expression of (\ref{eq_Y_ul_angle2}),
 the effective received signal model without considering the triple squint effects,
 denoted by $\bm{\bar Y}_{{\rm UL},l}\! =\! \left[ \bm{\bar y}_{{\rm UL},l}[\{{\cal K}_l\}_1]\!
 \cdots\! \bm{\bar y}_{{\rm UL},l}[\{{\cal K}_l\}_{K_l}] \right]\! \in\! \mathbb{C}^{{\bar I}_{\rm BS}\!
 \times\! K_l}$, can be written as
\setcounter{equation}{36}
\begin{equation}\label{eq_Y_ul_bar_CRLB} % eq 37
 \bm{\bar Y}_{{\rm UL},l} = \gamma_{{\rm UL},l} {\bm{\bar{\bar a}}}_{\rm BS}({\bar \mu}_l^{\rm BS},{\bar \nu}_l^{\rm BS})
 \bm{s}_{{\rm UL},l}^{\rm T} + \bm{\bar N}_{{\rm UL},l} ,
\end{equation}
 where $1\! \le\! l\! \le\! L$, ${\bar I}_{\rm BS}\! =\! {\bar I}_{\rm BS}^{\rm h} {\bar I}_{\rm BS}^{\rm v}$,
 ${\bm{\bar{\bar a}}}_{\rm BS}({\bar \mu}_l^{\rm BS},{\bar \nu}_l^{\rm BS})\! =\!
 \bm{a}_{\rm v}({\bar \nu}_l^{\rm BS},{\bar I}_{\rm BS}^{\rm v})\!\otimes\!
 \bm{a}_{\rm h}({\bar \mu}_l^{\rm BS},{\bar I}_{\rm BS}^{\rm h})\! \in\! \mathbb{C}^{{\bar I}_{\rm BS}}$
 with ${\bar \mu}_l^{\rm BS}\! =\! \varOmega\mu_l^{\rm BS}\! =\!
 \varOmega\pi\sin(\theta_l^{\rm BS})\cos(\varphi_l^{\rm BS})$ and
 ${\bar \nu}_l^{\rm BS}\! =\! \varOmega\nu_l^{\rm BS}\! =\!
 \varOmega\pi\sin(\varphi_l^{\rm BS})$, and $\bm{\bar N}_{{\rm UL},l}$
 is the noise matrix with its entry following ${\cal CN}\! (0,\sigma_n^2)$.
 The likelihood function of $\bm{\bar Y}_{{\rm UL},l}$ is $p(\bm{\bar Y}_{{\rm UL},l};\bm{\eta}_l)$, and the corresponding the log-likelihood function can be expressed as (\ref{eq_lnp_ul_bar_CRLB}) on the bottom of this page by defining $\bm{\eta}_l\! =\! [\alpha_l,(\bm{\xi}_l^{\rm BS})^{\rm T}]^{\rm T}$ with
 $\bm{\xi}_l^{\rm BS}\! =\! [{\bar \nu}_l^{\rm BS},{\bar \mu}_l^{\rm BS}]^{\rm T}$.
%\begin{align}\label{eq_lnp_ul_bar_CRLB} % eq 38
% &\ln p(\bm{\bar Y}_{{\rm UL},l};\bm{\eta}_l) \nonumber\\
% &= -{\bar I}_{\rm BS}K_l\ln(\pi\sigma_n^2)-\frac{1}{\sigma_n^2} \sum\nolimits_{{k_l}=1}^{K_l}\Big( \Big[ \bm{\bar y}_{{\rm UL},l}[\{{\cal K}_l\}_{k_l}] \nonumber\\
% &\quad - \gamma_{{\rm UL},l} {\bm{\bar{\bar a}}}_{\rm BS}({\bar \mu}_l^{\rm BS},{\bar \nu}_l^{\rm BS})
% s_{{\rm UL},l}[\{{\cal K}_l\}_{k_l}] \Big]^{\rm H} \Big[ \bm{\bar y}_{{\rm UL},l}[\{{\cal K}_l\}_{k_l}] \nonumber \\
% &\quad - \gamma_{{\rm UL},l} {\bm{\bar{\bar a}}}_{\rm BS}({\bar \mu}_l^{\rm BS},{\bar \nu}_l^{\rm BS})
% s_{{\rm UL},l}[\{{\cal K}_l\}_{k_l}] \Big]\Big) ,
%\end{align}
% In (\ref{eq_lnp_ul_bar_CRLB}), $p(\bm{\bar Y}_{{\rm UL},l};\bm{\eta}_l)$ is the corresponding likelihood function.
 Thus, the ($i,\!j$)th entry of Fisher Information Matrix (FIM), denoted by
 $[\bm{G}(\bm{\eta}_l)]_{i,j}$, is given by
\setcounter{equation}{38}
\begin{equation}\label{eq_Fisher_CRLB} % eq 39
 [\bm{G}(\bm{\eta}_l)]_{i,j} = -\mathbb{E}\left( \frac{\partial^2\ln p(\bm{\bar Y}_{{\rm UL},l};\bm{\eta}_l)}
 {\partial[\bm{\eta}_l]_i \partial[\bm{\eta}_l]_j}  \right).
\end{equation}

%\setcounter{TempEqCnt}{\value{equation}}
%\setcounter{equation}{39}
%\begin{figure*}[hb]
%\begin{align}\label{eq_CRLB_xi} % eq 40
% \mathrm{CRLB}_{\bm{\xi}_l^{\rm BS}} = \bm{G}^{-1}(\bm{\eta}_l) = \frac{\sigma_n^2}{2\left|\gamma_{{\rm UL},l}\right|^2}
% \left\{ \sum\limits_{{k_l}=1}^{K_l} {\Re \left\{ \bm{B}_{{\rm BS},{k_l}}^{\rm H} \bm{\varGamma}_{\rm BS}^{\rm H}
% \left( \bm{I}_{{\bar I}_{\rm BS}} - \bm{\varPhi}_{\rm BS} \right) \bm{\varGamma}_{\rm BS} \bm{B}_{{\rm BS},{k_l}} \right\}}
% \right\}^{-1} ,
%\end{align}
%\end{figure*}

 According to the results in \cite{Stoica_TSP89,Paulraj_TSP98}, the CRLB of $\bm{\xi}_l^{\rm BS}$
 consisting of the virtual angles ${\bar \mu}_l^{\rm BS}$ and ${\bar \nu}_l^{\rm BS}$ can be expressed as (\ref{eq_CRLB_xi}) on the bottom of this page.
%\begin{align}\label{eq_CRLB_xi} % eq 40
% \mathrm{CRLB}_{\bm{\xi}_l^{\rm BS}} = \bm{G}^{-1}(\bm{\eta}_l) = \frac{\sigma_n^2}{2\left|\gamma_{{\rm UL},l}\right|^2}
% \left\{ \sum\limits_{{k_l}=1}^{K_l} {\Re \left\{ \bm{B}_{{\rm BS},{k_l}}^{\rm H} \bm{\varGamma}_{\rm BS}^{\rm H}
% \left( \bm{I}_{{\bar I}_{\rm BS}} - \bm{\varPhi}_{\rm BS} \right) \bm{\varGamma}_{\rm BS} \bm{B}_{{\rm BS},{k_l}} \right\}}
% \right\}^{-1} ,
%\end{align}
 In (\ref{eq_CRLB_xi}), $\bm{B}_{{\rm BS},{k_l}}\! =\! \bm{I}_2\!\otimes\! s_{{\rm UL},l}[\{{\cal K}_l\}_{k_l}]$,
 $\bm{\varGamma}_{\rm BS}\! =\! \left[\bm{a}_{\rm v}({\bar \nu}_l^{\rm BS},{\bar I}_{\rm BS}^{\rm v})\!\otimes\!
 {\textstyle{{\partial\bm{a}_{\rm h}({\bar \mu}_l^{\rm BS},{\bar I}_{\rm BS}^{\rm h})} \over
 {\partial{\bar \mu}_l^{\rm BS}}}},{\textstyle{{\partial\bm{a}_{\rm v}({\bar \nu}_l^{\rm BS},{\bar I}_{\rm BS}^{\rm v})}
 \over {\partial{\bar \nu}_l^{\rm BS}}}}\!\otimes\!\bm{a}_{\rm h}({\bar \mu}_l^{\rm BS},{\bar I}_{\rm BS}^{\rm h}) \right]$,
 and the projection operator $\bm{\varPhi}_{\rm BS}\! =\! {\bm{\bar{\bar a}}}_{\rm BS}({\bar \mu}_l^{\rm BS},{\bar \nu}_l^{\rm BS})$ $\times\!
 \big( {\bm{\bar{\bar a}}}_{\rm BS}^{\rm H}({\bar \mu}_l^{\rm BS},{\bar \nu}_l^{\rm BS})
 {\bm{\bar{\bar a}}}_{\rm BS}({\bar \mu}_l^{\rm BS},{\bar \nu}_l^{\rm BS}) \big)^{-\!1}
 {\bm{\bar{\bar a}}}_{\rm BS}^{\rm H}({\bar \mu}_l^{\rm BS},{\bar \nu}_l^{\rm BS})$.

 To obtain the CRLBs of azimuth and elevation angles, we define the transformation relationship
 between the virtual angles and the corresponding physical angles as
\setcounter{equation}{40}
\begin{equation}\label{eq_J_xi} % eq 41
 \bm{J}(\bm{\xi}_l^{\rm BS}) = \left[ \begin{array}{*{20}{c}} \varphi _l^{\rm BS}\\
 \theta_l^{\rm BS} \end{array} \right] = \left[ \begin{array}{*{20}{c}}
 \arcsin \left({\textstyle{{\bar \nu_l^{\rm BS}} \over {\varOmega \pi }}}\right)\\
 \arcsin \left({\textstyle{{\bar \mu_l^{\rm BS}} \over {\varOmega \pi \cos (\varphi _l^{\rm BS})}}}\right) \end{array} \right] .
\end{equation}
 Based on the transformation of vector parameter CRLB in \cite{Kay_SSP93},
 defining $\partial\bm{J}(\bm{\xi}_l^{\rm BS})/\partial\bm{\xi}_l^{\rm BS}$
 as the Jacobian matrix, the CRLBs of azimuth angle
 $\theta_l^{\rm BS}$ and elevation angle $\varphi_l^{\rm BS}$,
 denoted by $\mathrm{CRLB}_{\theta_l^{\rm BS}}(\varOmega)$ and
 $\mathrm{CRLB}_{\varphi_l^{\rm BS}}(\varOmega)$, can be then formulated as (\ref{eq_CRLB_phi}) and (\ref{eq_CRLB_theta}), respectively, on the top of the next page.
%\begin{align} % eqs 42,43
% \mathrm{CRLB}_{\varphi_l^{\rm BS}}(\varOmega) =& \left[ {\textstyle{{\partial\bm{J}(\bm{\xi}_l^{\rm BS})}
% \over {\partial\bm{\xi}_l^{\rm BS}}}} \mathrm{CRLB}_{\bm{\xi}_l^{\rm BS}}
% {\textstyle{{\partial\bm{J}(\bm{\xi}_l^{\rm BS})^{\rm T}} \over {\partial\bm{\xi}_l^{\rm BS}}}} \right]_{1,1}
% = {\textstyle{ \left[ \mathrm{CRLB}_{\bm{\xi}_l^{\rm BS}} \right]_{1,1} \over
% {\varOmega^2 \left( \pi^2 - (\nu_l^{\rm BS})^2 \right) }}}, \label{eq_CRLB_phi} \\%[-1mm]
% \mathrm{CRLB}_{\theta_l^{\rm BS}}(\varOmega) =& \left[ {\textstyle{{\partial\bm{J}(\bm{\xi}_l^{\rm BS})}
% \over {\partial\bm{\xi}_l^{\rm BS}}}} \mathrm{CRLB}_{\bm{\xi}_l^{\rm BS}}
% {\textstyle{{\partial\bm{J}(\bm{\xi}_l^{\rm BS})^{\rm T}} \over {\partial\bm{\xi}_l^{\rm BS}}}} \right]_{2,2}
% = {\textstyle{ \left[ \mathrm{CRLB}_{\bm{\xi}_l^{\rm BS}} \right]_{2,2} \over
% {\varOmega^2 \left( \pi^2\cos^2(\varphi _l^{\rm BS}) - (\mu_l^{\rm BS})^2 \right) }}}, \label{eq_CRLB_theta}
%\end{align}
 Finally, the CRLBs of angles at BSs can be obtained as
 $\mathrm{CRLB}_{\theta^{\rm BS}}(\varOmega)\! =\!
 {\textstyle{1 \over L}}\sum\nolimits_{l=1}^{L}{\mathrm{CRLB}_{\theta_l^{\rm BS}}}(\varOmega)$
 and $\mathrm{CRLB}_{\varphi^{\rm BS}}(\varOmega)\! =\!
 {\textstyle{1 \over L}}\sum\nolimits_{l=1}^{L}{\mathrm{CRLB}_{\varphi_l^{\rm BS}}}(\varOmega)$, respectively.
 Furthermore, the CRLBs of angles at aircraft, i.e., $\mathrm{CRLB}_{\theta^{\rm AC}}(\varOmega)$
 and $\mathrm{CRLB}_{\varphi^{\rm AC}}(\varOmega)$, can be also acquired in a similar way,
 where the detailed derivations are omitted due to space constraints.

\setcounter{TempEqCnt}{\value{equation}}
\setcounter{equation}{41}
\begin{figure*}[ht]
\begin{align} % eqs 42,43,44,45
 \mathrm{CRLB}_{\varphi_l^{\rm BS}}(\varOmega) &= \left[ {\textstyle{{\partial\bm{J}(\bm{\xi}_l^{\rm BS})}
 \over {\partial\bm{\xi}_l^{\rm BS}}}} \mathrm{CRLB}_{\bm{\xi}_l^{\rm BS}}
 {\textstyle{{\partial\bm{J}(\bm{\xi}_l^{\rm BS})^{\rm T}} \over {\partial\bm{\xi}_l^{\rm BS}}}} \right]_{1,1}
 = {\textstyle{ \left[ \mathrm{CRLB}_{\bm{\xi}_l^{\rm BS}} \right]_{1,1} \over
 {\varOmega^2 \left( \pi^2 - (\nu_l^{\rm BS})^2 \right) }}}, \label{eq_CRLB_phi} \\
 \mathrm{CRLB}_{\theta_l^{\rm BS}}(\varOmega) &= \left[ {\textstyle{{\partial\bm{J}(\bm{\xi}_l^{\rm BS})}
 \over {\partial\bm{\xi}_l^{\rm BS}}}} \mathrm{CRLB}_{\bm{\xi}_l^{\rm BS}}
 {\textstyle{{\partial\bm{J}(\bm{\xi}_l^{\rm BS})^{\rm T}} \over {\partial\bm{\xi}_l^{\rm BS}}}} \right]_{2,2}
 = {\textstyle{ \left[ \mathrm{CRLB}_{\bm{\xi}_l^{\rm BS}} \right]_{2,2} \over
 {\varOmega^2 \left( \pi^2\cos^2(\varphi _l^{\rm BS}) - (\mu_l^{\rm BS})^2 \right) }}}. \label{eq_CRLB_theta} \\
 \mathrm{CRLB}_{\nu_l^\psi} &= \frac{\sigma_n^2}{2\left|\gamma_{{\rm do},l}\right|^2}
 \left\{ \sum\limits_{{k_l}=1}^{K_l} {\Re \left\{ \left|{\bar s}_{{\rm do},l}[\{{\cal K}_l\}_{k_l}]\right|^2
 \left( {\textstyle{{\partial\bm{a}_{\psi}(\nu_l^\psi,N_{\rm do})} \over {\partial\nu_l^\psi}}} \right)^{\rm H}
 \left( \bm{I}_{N_{\rm do}} - \bm{\varPhi}_{\rm Do} \right)
 {\textstyle{{\partial\bm{a}_{\psi}(\nu_l^\psi,N_{\rm do})} \over {\partial\nu_l^\psi}}} \right\}}
 \right\}^{-1} , \label{eq_CRLB_nu_psi} \\
 \mathrm{CRLB}_{\mu_l^\tau} &= \frac{\sigma_n^2}{2\left|\gamma_{{\rm de},l}\right|^2}
 \left\{ \sum\limits_{n=1}^{N_{\rm de}} {\Re \left\{ \left|{\bar s}_{{\rm de},l}^{[n]}\right|^2
 \left( {\textstyle{{\partial\bm{a}_{\tau}(\mu_l^\tau,K_l)} \over {\partial\mu_l^\tau}}} \right)^{\rm H}
 \left( \bm{I}_{N_{\rm de}} - \bm{\varPhi}_{\rm De} \right)
 {\textstyle{{\partial\bm{a}_{\tau}(\mu_l^\tau,K_l)} \over {\partial\mu_l^\tau}}} \right\}}
 \right\}^{-1} . \label{eq_CRLB_mu_tau}
\end{align}
\hrulefill
\end{figure*}

\begin{remark} % Remark 3
 According to (\ref{eq_CRLB_xi}), if the system configuration parameters of the transceiver are
 the same except for different sparse spacing $\varOmega$, the CRLB of
 $\bm{\xi}_l^{\rm BS}$, i.e., $\mathrm{CRLB}_{\bm{\xi}_l^{\rm BS}}$, is a constant.
 Therefore, we can observe from (\ref{eq_CRLB_phi}) and (\ref{eq_CRLB_theta}) that
 $\mathrm{CRLB}_{\theta_l^{\rm BS}}(1)$ and $\mathrm{CRLB}_{\varphi_l^{\rm BS}}(1)$ for $\varOmega\! =\! 1$
 are the $\varOmega^2$ times as much as $\mathrm{CRLB}_{\theta_l^{\rm BS}}(\varOmega)$
 and $\mathrm{CRLB}_{\varphi_l^{\rm BS}}(\varOmega)$ for $\varOmega\! >\! 1$, respectively.
 In other words, compared with the array with critical antenna spacing, the CRLB of sparse array
 with sparse spacing $\varOmega\! >\! 1$ can achieve the about $20\lg \varOmega\,{\rm dB}$
 Mean Square Error (MSE) performance gain, which theoretically testifies the improved accuracy of
 angle estimation using sparse array.
\end{remark}

\subsubsection{CRLBs of Doppler Shift and Path Delay Estimation}\label{S6.1.2}
 Similar to the CRLB derivations of angle estimation, according to (\ref{eq_Y_tilde_dop_l}) and (\ref{eq_Y_del_l}),
 the CRLBs of virtual Doppler $\nu_l^\psi$ and virtual delay $\mu_l^\tau$ can be obtained directly as (\ref{eq_CRLB_nu_psi}) and (\ref{eq_CRLB_mu_tau}), respectively, on the top of this page.
%\begin{align} % eqs 44,45
% \mathrm{CRLB}_{\nu_l^\psi} =& \frac{\sigma_n^2}{2\left|\gamma_{{\rm do},l}\right|^2}
% \left\{ \sum\limits_{{k_l}=1}^{K_l} {\Re \left\{ \left|{\bar s}_{{\rm do},l}[\{{\cal K}_l\}_{k_l}]\right|^2
% \left( {\textstyle{{\partial\bm{a}_{\psi}(\nu_l^\psi,N_{\rm do})} \over {\partial\nu_l^\psi}}} \right)^{\rm H}
% \left( \bm{I}_{N_{\rm do}} - \bm{\varPhi}_{\rm Do} \right)
% {\textstyle{{\partial\bm{a}_{\psi}(\nu_l^\psi,N_{\rm do})} \over {\partial\nu_l^\psi}}} \right\}}
% \right\}^{-1} , \label{eq_CRLB_nu_psi} \\[-1mm]
% \mathrm{CRLB}_{\mu_l^\tau} =& \frac{\sigma_n^2}{2\left|\gamma_{{\rm de},l}\right|^2}
% \left\{ \sum\limits_{n=1}^{N_{\rm de}} {\Re \left\{ \left|{\bar s}_{{\rm de},l}^{[n]}\right|^2
% \left( {\textstyle{{\partial\bm{a}_{\tau}(\mu_l^\tau,K_l)} \over {\partial\mu_l^\tau}}} \right)^{\rm H}
% \left( \bm{I}_{N_{\rm de}} - \bm{\varPhi}_{\rm De} \right)
% {\textstyle{{\partial\bm{a}_{\tau}(\mu_l^\tau,K_l)} \over {\partial\mu_l^\tau}}} \right\}}
% \right\}^{-1} , \label{eq_CRLB_mu_tau}
%\end{align}
 In (\ref{eq_CRLB_nu_psi}) and (\ref{eq_CRLB_mu_tau}), the projection operators $\bm{\varPhi}_{\rm Do}$ and $\bm{\varPhi}_{\rm De}$
 have the similar form to $\bm{\varPhi}_{\rm BS}$.
 By exploiting the transformation of parameter CRLB \cite{Kay_SSP93},
 the CRLBs of Doppler shift $\psi_{z,l}$ and the normalized delay
 ${\bar \tau_l}\! =\! f_s\tau_l$ can be then expressed as
 $\mathrm{CRLB}_{\psi_{z,l}}\! =\! {\textstyle{ {\mathrm{CRLB}_{\nu_l^\psi}}
 \over {\left( 2\pi T_{\rm sym} \right)^2 }}}$ and
 $\mathrm{CRLB}_{\bar \tau_l}\! =\! {\textstyle{ {K^2\mathrm{CRLB}_{\mu_l^\tau}}
 \over {\left( 2\pi \right)^2 }}}$, respectively.
 Finally, the CRLBs of Doppler shift and the normalized delay for $L$ BSs can be acquired as
 $\mathrm{CRLB}_{\psi_{z}}\! =\!
 {\textstyle{1 \over L}}\sum\nolimits_{l=1}^{L}\mathrm{CRLB}_{\psi_{z,l}}$
 and $\mathrm{CRLB}_{\bar \tau}\! =\!
 {\textstyle{1 \over L}}\sum\nolimits_{l=1}^{L}\mathrm{CRLB}_{\bar \tau_l}$, respectively.

\subsection{Computational Complexity}\label{S6.2}
 The computational complexity of the proposed channel estimation and tracking scheme
 mainly consists of two portions. The first one is to estimate and track the channel parameters,
 including the acquisition of azimuth/elevation angles at BSs and aircraft, Doppler shifts,
 and path delays using TDU-ESPRIT and TLS-ESPRIT algorithms. Since a mass of trivial computations
 with small computational complexity can be ignored, we focus on the dominant calculation steps
 involving numerous complex multiplications. For the estimation and tracking of angles
 at BSs and aircraft, their total computational complexity is $\textsf{O}\left(2LI_{\rm BS}K_l\!
 +\! 2LI_{\rm AC}K_l\! +\!2LI'_{\rm BS}K_l\! +\! 2LI'_{\rm AC}K_l\right)$, where
 $\textsf{O}(N)$ stands for ``on the order of $N$''. The computational complexity of Doppler
 shift and path delay estimation is $\textsf{O}\left(8LN_{\rm Do}^2K_l\! +\! 8LK_l^2N_{\rm De} \right)$.
 The second part is the data-aided channel tracking, and its computational complexity
 consists of the reestablishment of initial beam-aligned effective channel vectors and
 the tracking of subsequent effective channel vectors, i.e., $\textsf{O}\left(L(N_{\rm AC}\!
 +\! N_{\rm BS}\! +\! 3K)\right)$ and $\textsf{O}\left(LK\right)$, respectively.
 It can be seen from the above analysis that although the THz UM-MIMO arrays employing
 tens of thousands of antennas are equipped at BSs and aircraft, the computational complexity
 of the proposed solution is in polynomial time, since the effective low-dimensional signals at
 the receiver are utilized to estimate and track the aeronautical THz UM-MIMO channels.
 The state-of-the-art Digital Signal Processing (DSP) hardware devices, such as the latest Field Programmable Gate Array (FPGA), are capable of the operations with the order of trillions of Floating-Point Operations Per Second (FLOPS), which can be used for the proposed solution in THz UM-MIMO-based aeronautical communications with the acceptable processing time.

\section{Numerical Evaluation}\label{S7} % Simulation Results

\subsection{Simulation Setup}\label{S7.1}

\begin{figure*}[!tp]
%\vspace{-5mm}
\begin{center}
 \includegraphics[width=1.7\columnwidth, keepaspectratio]{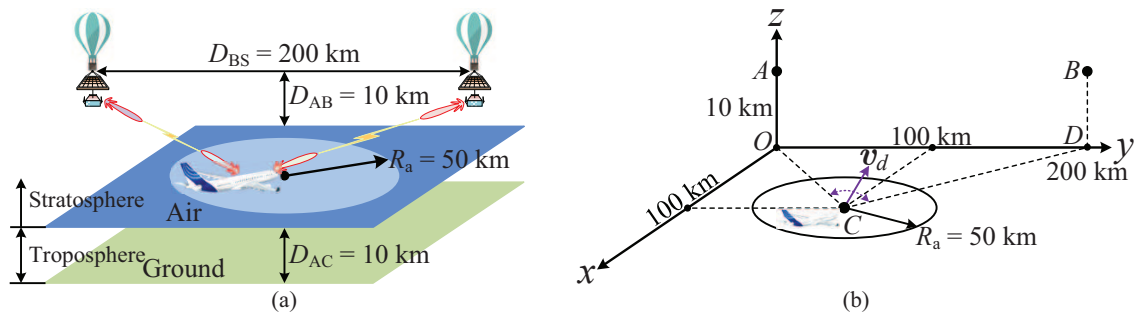}
\end{center}
\captionsetup{font = {footnotesize}, singlelinecheck = off, justification = raggedright, name = {Fig.}, labelsep = period}%
\caption{(a) Schematic diagram of simulation scenario, and (b) the corresponding spatial coordinate representation.}
%\vspace{-6mm}
\label{FIG10}
\end{figure*}

 In this section, we evaluate the performance of the proposed channel estimation and tracking
 scheme for THz UM-MIMO-based aeronautical communications, where the simulation scenario
 considered can be shown in Fig.~\ref{FIG10}. Without loss of generality, we set the reference altitudes of $L\! =\! 2$ suspended aerial BSs and an aircraft in Fig.~\ref{FIG10}(a) are $20$ kilometer (${\rm km}$) and $D_{\rm AC}\! =\! 10\, {\rm km}$ (at the top of the troposphere or the bottom of the stratosphere), respectively, and thus, the vertical distance between
 the aircraft and BSs is $D_{\rm AB}\! =\! 10\, {\rm km}$.
 The distance between two BSs is
 $D_{\rm BS}\! =\! 200\, {\rm km}$. In addition, we can abstract a spatial coordinate system as
 Fig.~\ref{FIG10}(b) from this real scenario, where point $O$ is the origin of coordinates, and
 the coordinates of points $A$, $B$, and $C$ are $(0,\!0,\!D_{\rm AB})$, $(0,\!D_{\rm BS},\!D_{\rm AB})$,
 and $(D_{\rm BS}/2,\!D_{\rm BS}/2,0)$, respectively. The position coordinate of the aircraft randomly
 appears in a horizontal circular plane with $C$ as the center and $R_{\rm a}\! =\! 50\, {\rm km}$
 as the radius, and the horizontal direction of aircraft $\bm{v}_d$ with flight speed
 $v_{\rm AC}\! =\! 200$ meter per second ($m/s$) falls in the intersection angle OCD.
 In order to simplify the simulation scenario, we consider that the altitude changes of aerial BSs and aircraft are reflected in the angle change over time.

 In simulations, the central carrier frequency is $f_z\! =\! 0.1\, {\rm THz}$ with system bandwidth
 $f_s\! = \! 1\, {\rm GHz}$, the horizontal/vertical antenna numbers of all subarrays at BSs and aircraft
 are $N_{\rm BS}^{\rm h}\! =\! N_{\rm BS}^{\rm v}\! =\! M_{\rm AC}^{\rm h}\! =\! M_{\rm AC}^{\rm v}\! =\!
 200$, and the horizontal and vertical numbers of subarrays at aircraft are
 ${\widetilde I}_{\rm AC}^{\rm h}\! =\! 1$ and ${\widetilde I}_{\rm AC}^{\rm v}\! =\! 2$, respectively,
 while the dimensions of the selected equivalent fully-digital (sparse) array are $I_{\rm BS}^{\rm h}\! =\!
 I_{\rm BS}^{\rm v}\! =\! I_{\rm AC}^{\rm h}\! =\! I_{\rm AC}^{\rm v}\! =\! 5$ ($I_{\rm BS}^{'{\rm h}}\!
 =\! I_{\rm BS}^{'{\rm v}}\! =\! I_{\rm AC}^{'{\rm h}}\! =\! I_{\rm AC}^{'{\rm v}}\! =\! 5)$. The
 numbers of antennas in each antenna group used for the GTTDU modules at BSs and aircraft are
 ${\widetilde M}_{\rm BS}^{\rm h}\! =\! {\widetilde M}_{\rm BS}^{\rm v}\! =\!
 {\widetilde M}_{\rm AC}^{\rm h}\! =\! {\widetilde M}_{\rm AC}^{\rm v}\! =\! 5$.
 Moreover, the number of OFDM symbols used to estimate and track the Doppler shifts and path delays
 are $N_{\rm do}\! =\! 6$ and $N_{\rm de}\! =\! 10$, respectively. The number of subcarriers is set
 to $K\! =\! 2048$ with the length of Cyclic Prefix (CP) being $N_{\rm cp}\! =\! 128$, and perfect
 frame synchronization and reliable delay compensation are assumed. The channel parameters are listed as follows. The azimuth and
 elevation angles at BSs and aircraft $\{ \theta_l^{\rm BS},\! \varphi_l^{\rm BS},\! \theta_l^{\rm AC},\!
 \varphi_l^{\rm AC} \}_{l=1}^L$ are generated from $\left[ -\pi/3, \pi/3 \right]$ randomly. Note that
 due to the long distance between the adjacent BSs, $\{ \theta_l^{\rm AC},\! \varphi_l^{\rm AC} \}_{l=1}^L$
 corresponding to different BSs have the large gaps, and these angles can be set based
 on the position of aircraft in Fig.~\ref{FIG10}(b). The Doppler shifts $\{ \psi_{z,l} \}_{l=1}^L$ can
 be set based on $\bm{v}_d$ and the relationship between spatial coordinates of the BSs and aircraft.
 The path delay $\tau_l$ follows uniform distribution ${\cal U}[0, N_{\rm cp}T_s]$ and each of channel
 gains $\alpha_l$ is generated according to ${\cal CN}(0,1)$, i.e., $\sigma_\alpha^2\! =\! 1$, for
 $1\! \le\! l\! \le\! L$. The rough estimates of azimuth/elevation angles at
 BSs and aircraft $\{ {\widetilde \theta}_l^{\rm BS}, {\widetilde \varphi}_l^{\rm BS},
 {\widetilde \theta}_l^{\rm AC}, {\widetilde \varphi}_l^{\rm AC} \}_{l=1}^{L}$
 can be randomly selected from the range of these true angles with offset $\pm\, 5^\circ$,
 while the rough Doppler shift estimate ${\widetilde \psi}_{z,l}$ can be randomly selected from the range of
 the true $\psi_{z,l}$ with offset $\pm\, 0.01\psi_{z,l}$ for $1\! \le\! l\! \le\! L$. Furthermore,
 to describe the fast time-varying fading channels, we define the relationship of these channel
 parameters between the $q$th and $(q\! +\! 1)$th TIs as $x^{[q\!+\!1]}\! =\! x^{[q]}\! +\!
 s_{\rm pm}\rho_x N_{\rm C} T_{\rm sym}$, where $x$ represents the channel parameter coming from $\alpha_l$,
 $\tau_l$, $\psi_{z,l}$, $\theta_l^{\rm AC}$, $\varphi_l^{\rm AC}$, $\theta_l^{\rm BS}$, or
 $\varphi_l^{\rm BS}$. Here, $s_{\rm pm}$ denotes a binary variable selected from $1$ or $-1$ randomly,
 $N_{\rm C}\! =\! 70$, $T_{\rm sym}\! =\! (N_{\rm cp}\! +\! K)T_s\! =\! 2.176$ Microseconds (${\rm \mu s}$), and the duration time of one TI is $T_{\rm TI}\! =\! N_{\rm C}T_{\rm sym}\! =\! 152.32\,{\rm \mu s}$,
 while $\rho_x$ is the rate
 of change associated with $x$. We consider $\rho_{\alpha_l}\! =\! \alpha_l^{(1)}\!/2$, $\rho_{\tau_l}\!
 =\! \tau_l^{(1)}\!/2$, $\rho_{\psi_{z,l}}\! =\! 0.01\psi_{z,l}^{(1)}$, $\rho_\theta^{\rm AC}\! =\!
 \rho_\varphi^{\rm AC}\! =\! \pi/4$, and $\rho_\theta^{\rm BS}\! =\! \rho_\varphi^{\rm BS}\! =\! \pi/12$.
 Note that the maximum value of angle changing during one TI can be approximately calculated as $\frac{\pi}{4}\! \times\! T_{\rm TI}\! \approx\! 0.0069^\circ$, which is extremely small, so that the assumption about TI is reasonable.
 For the data-aided channel tracking, $\varepsilon\! =\! 0.2$ and ${\widetilde K}\! =\! K/2$. Note that
 the relationship between transmit power $P_l$ and large-scale fading gain $G_l$ is complementary.
 Without loss of generality, assume that $P_lG_l\! =\! 1$ through the transmit power compensation.
 Therefore, to facilitate the simulation evaluation, we define $\sigma_\alpha^2/\sigma_n^2$ with $\sigma_n^2$
 being the noise variance as the transmitted SNR of UL and DL throughout our simulations.

\subsection{Simulation Results}\label{S7.2}

\begin{figure*}[!tp]
%\vspace{-5mm}
\captionsetup{singlelinecheck = off, justification = raggedright, font={footnotesize}, name = {Fig.}, labelsep = period}%
\captionsetup[subfigure]{singlelinecheck = on, justification = raggedright, font={footnotesize}}
\centering
\subfigure{%[The number of path $L = 3$]
\label{FIG11(a)}
%\hspace{-3.0mm}
\includegraphics[width=3.3in]{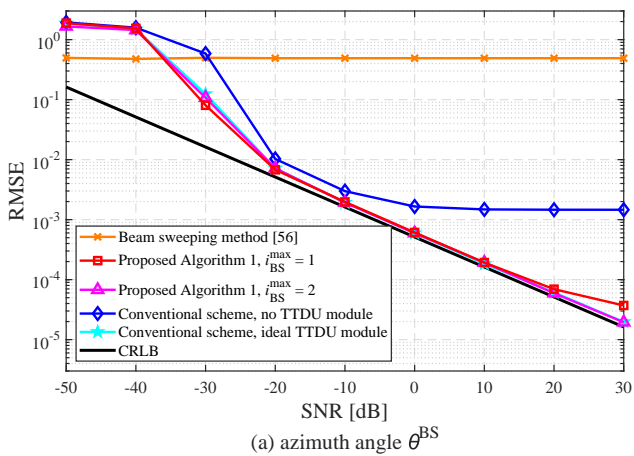}
}
\subfigure{%[The number of path $L = 5$]
\label{FIG11(b)}
%\hspace{6mm}
\includegraphics[width=3.3in]{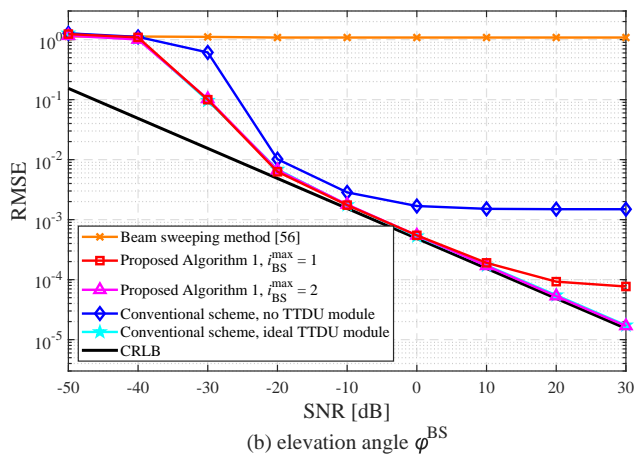}
}
\setlength{\abovecaptionskip}{-0.05mm}
\caption{RMSE comparison of $\{\theta^{\rm BS},\,\varphi^{\rm BS}\}$ at the initial angle estimation stage:
 (a)~azimuth angle $\theta^{\rm BS}$; and (b)~elevation angle $\varphi^{\rm BS}$.}
%\vspace{-4mm}
\label{FIG11}
\end{figure*}

\begin{figure*}[!tp]
%\vspace{-1mm}
\captionsetup{singlelinecheck = off, justification = raggedright, font={footnotesize}, name = {Fig.}, labelsep = period}%
\captionsetup[subfigure]{singlelinecheck = on, justification = raggedright, font={footnotesize}}
\centering
\subfigure{%[The number of path $L = 3$]
\label{FIG12(a)}
%\hspace{-3.0mm}
\includegraphics[width=3.3in]{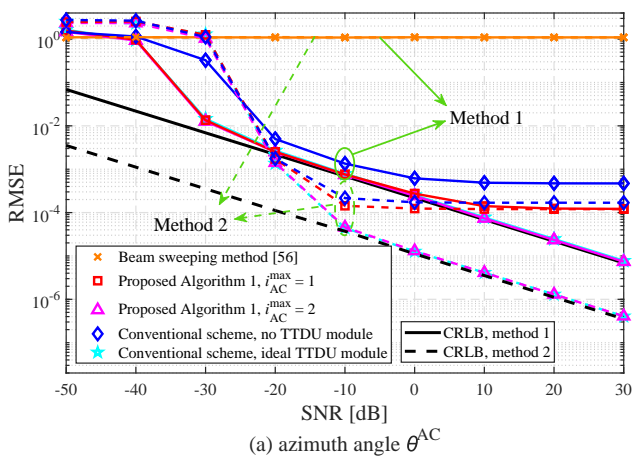}
}
\subfigure{%[The number of path $L = 5$]
\label{FIG12(b)}
%\hspace{6mm}
\includegraphics[width=3.3in]{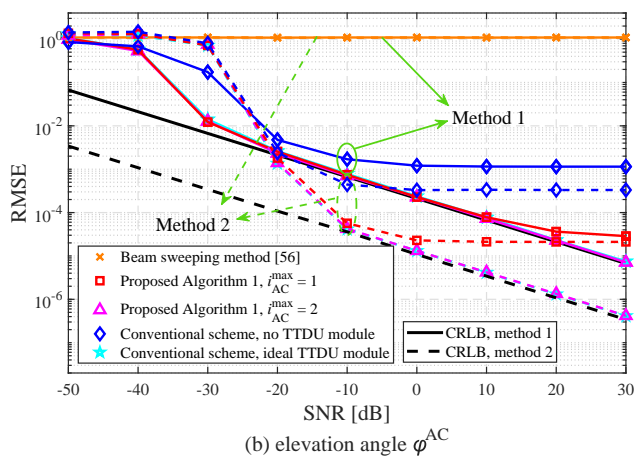}
}
\setlength{\abovecaptionskip}{-0.05mm}
\caption{RMSE comparison of $\{\theta^{\rm AC},\,\varphi^{\rm AC}\}$ at the initial angle estimation stage:
 (a)~azimuth angle $\theta^{\rm AC}$; and (b)~elevation angle $\varphi^{\rm AC}$.}
%\vspace{-6mm}
\label{FIG12}
\end{figure*}

 First the performance of the initial channel estimation is evaluated using the Root-MSE
 (RMSE) metric given by ${\mathrm{RMSE}}_{\bm{x}}\! =\! \sqrt{ \mathbb{E}\left({\textstyle{1 \over L}}
 \| \bm{x}\! -\! \bm{\widehat x} \|_2^2\right) }$, where $\bm{x}\! \in\! \mathbb{R}^{L}$ and
 $\bm{\widehat x}$ represent the true and the estimated channel parameter vectors, and
 $[\bm{x}]_l$ comes from the parameters $\theta_l^{\rm BS}$, $\varphi_l^{\rm BS}$, $\theta_l^{\rm AC}$,
 $\varphi_l^{\rm AC}$, $\psi_{z,l}$, or $\tau_l$.
 For the angle estimation at the BSs and aircraft, the state-of-the-art channel estimation and tracking schemes \cite{Gong_JSAC19,ZhangS_Tcom19M,ZhangS_Tcom19N,Qin_TVT18,Cheng_WCL19,GaoXY_TVT17,GaoFF_TSP19} are not suitable for the THz UM-MIMO based aeronautical communication channels with fast time-varying fading characteristics. Hence, we consider the beam sweeping method with severe beam squint effect in IEEE standards 802.11ad \cite{IEEE802.11ad} as one of the benchmarks, where its sweeping ranges are $\pm\, 5^\circ$ around the corresponding rough angle estimates acquired by BSs and aircraft.

 Fig.~\ref{FIG11} compares the RMSE performance of the proposed fine angle estimation for
 $\{ \theta_l^{\rm BS},\,\varphi_l^{\rm BS} \}_{l=1}^L$ at the initial channel estimation
 stage, where different processing methods are investigated. In Fig.~\ref{FIG11},
 the labels ``no TTDU module'' and ``ideal TTDU module'' indicate the transceiver
 adopting ideal TTDU module and without considering TTDU module, respectively. The label
 ``conventional scheme'' indicates directly applying the conventional TDU-ESPRIT algorithm to
 estimate angles as those used in existing mmWave systems \cite{Liao_Tcom19}, while
 $i_{\rm BS}^{\rm max}\! =\! 1$ and $i_{\rm BS}^{\rm max}\! =\! 2$ indicate the maximum
 iterations in the proposed \textbf{Algorithm~\ref{ALG1}}. From Fig.~\ref{FIG11}, it can
 be seen that the RMSE curves of ``proposed algorithm 1 with $i_{\rm BS}^{\rm max}\! =\! 2$''
 and ``conventional scheme'' using ``ideal TTDU module'' almost overlap, and they are very
 close to the CRLBs of azimuth and elevation angles at high SNR. The proposed
 \textbf{Algorithm~\ref{ALG1}} just needs $i_{\rm BS}^{\rm max}\! =\! 2$ iterations to
 achieve the performance upper-bound that uses ideal TTDU module without beam squint effect.
 If the beam squint effect is not well handled as ``conventional scheme'' with ``no TTDU module'',
 its performance of angle estimation will suffer from the obvious RMSE floor at medium-to-high SNR.
 Note that the angle estimation performance of beam sweeping method is very poor due to the limited training overhead in the fast time-varying channels.
 Moreover, due to the inaccurately rough angle estimates acquired, ``proposed algorithm 1 with
 $i_{\rm BS}^{\rm max}\! =\! 1$'' only using GTTDU module for compensation at transceiver still
 suffers from the RMSE floor at high SNR, while ``proposed algorithm 1 with $i_{\rm BS}^{\rm max}\!
 =\! 2$'' can further attenuate this beam squint error by finely compensating the received
 signal matrix $\bm{Y}_{{\rm UL},l}$ with the compensation matrix $\bm{\widetilde Y}_{{\rm UL},l}^{(1)}$.

\begin{figure*}[!tp]
%\vspace{-5mm}
\captionsetup{singlelinecheck = off, justification = raggedright, font={footnotesize}, name = {Fig.}, labelsep = period}
\begin{minipage}[t]{0.49\linewidth}
\centering
\includegraphics[width=3.3in]{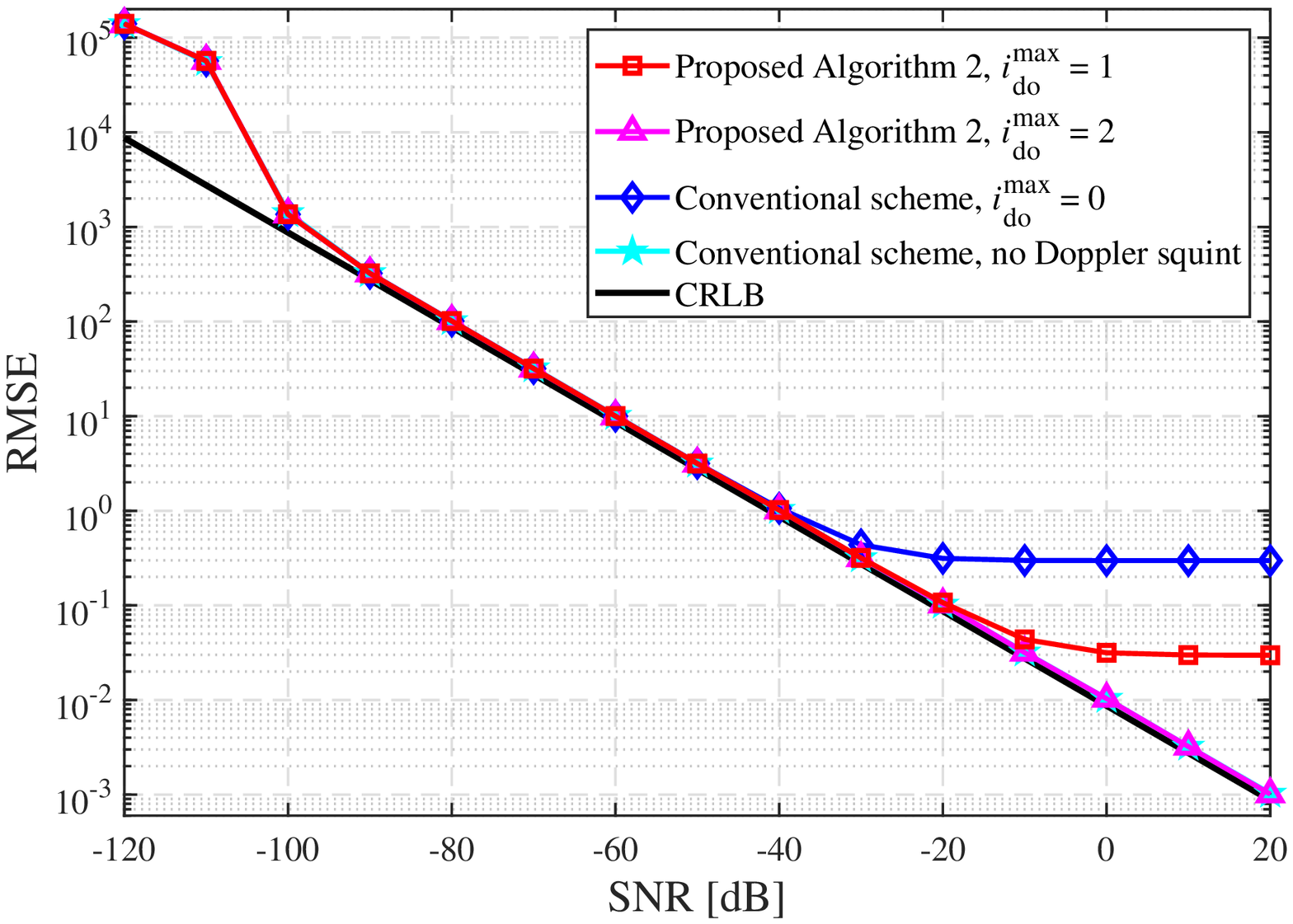}
\hspace{-5.5mm}
\caption{RMSE comparison of Doppler shift $\psi_z$ estimation.}
\label{FIG13}
\end{minipage}
\hfill
\begin{minipage}[t]{0.49\linewidth}
\centering
\includegraphics[width=3.3in]{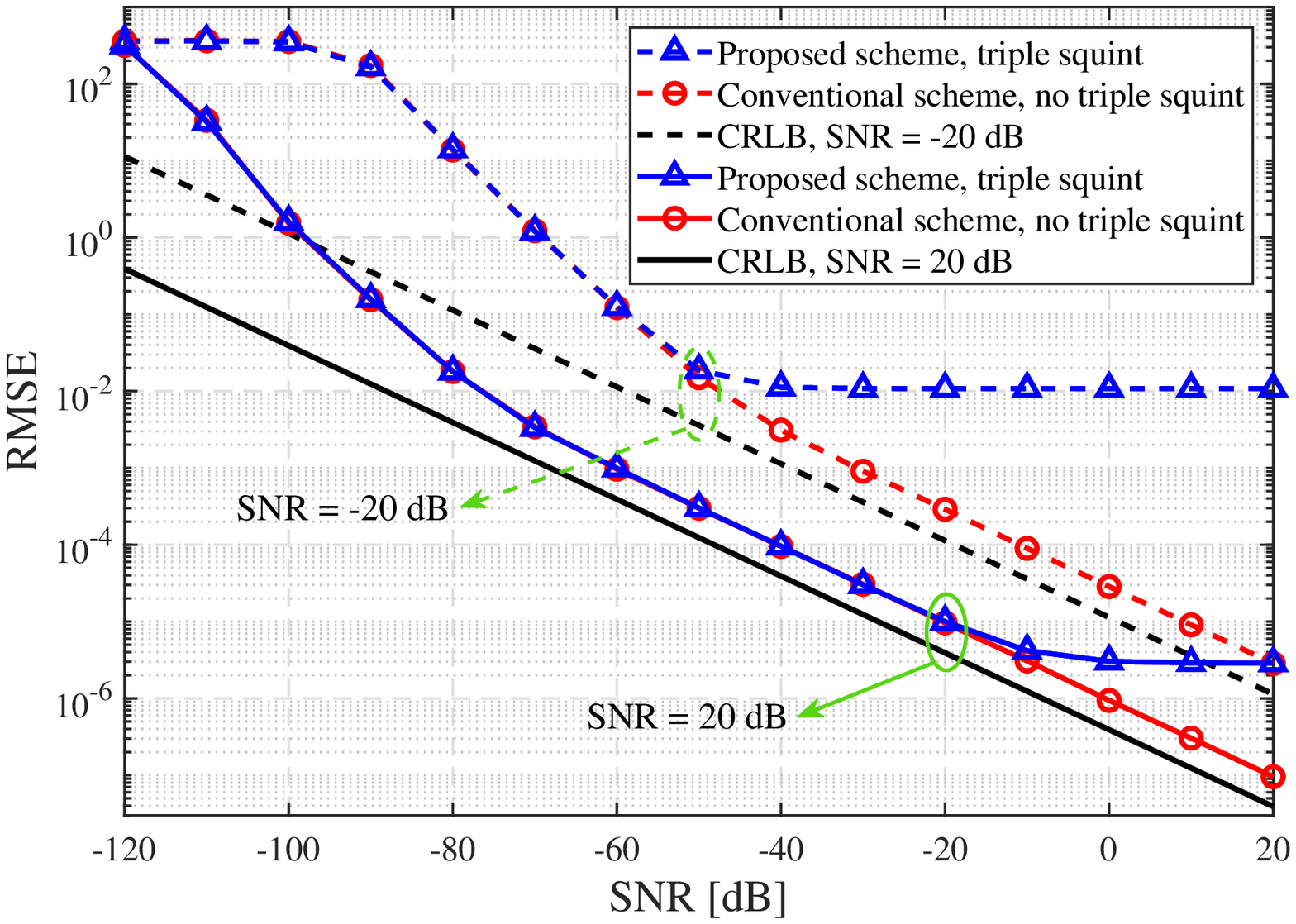}
\hspace{-0.5mm}
\caption{RMSE comparison of the normalized delay ${\bar \tau}$ estimation.}
\label{FIG14}
\end{minipage}
%\vspace{-6mm}
\end{figure*}

\begin{figure*}[!tp]
%\vspace{-5mm}
\captionsetup{singlelinecheck = off, justification = raggedright, font={footnotesize}, name = {Fig.}, labelsep = period}
\begin{minipage}[t]{0.49\linewidth}
\centering
\includegraphics[width=3.3in]{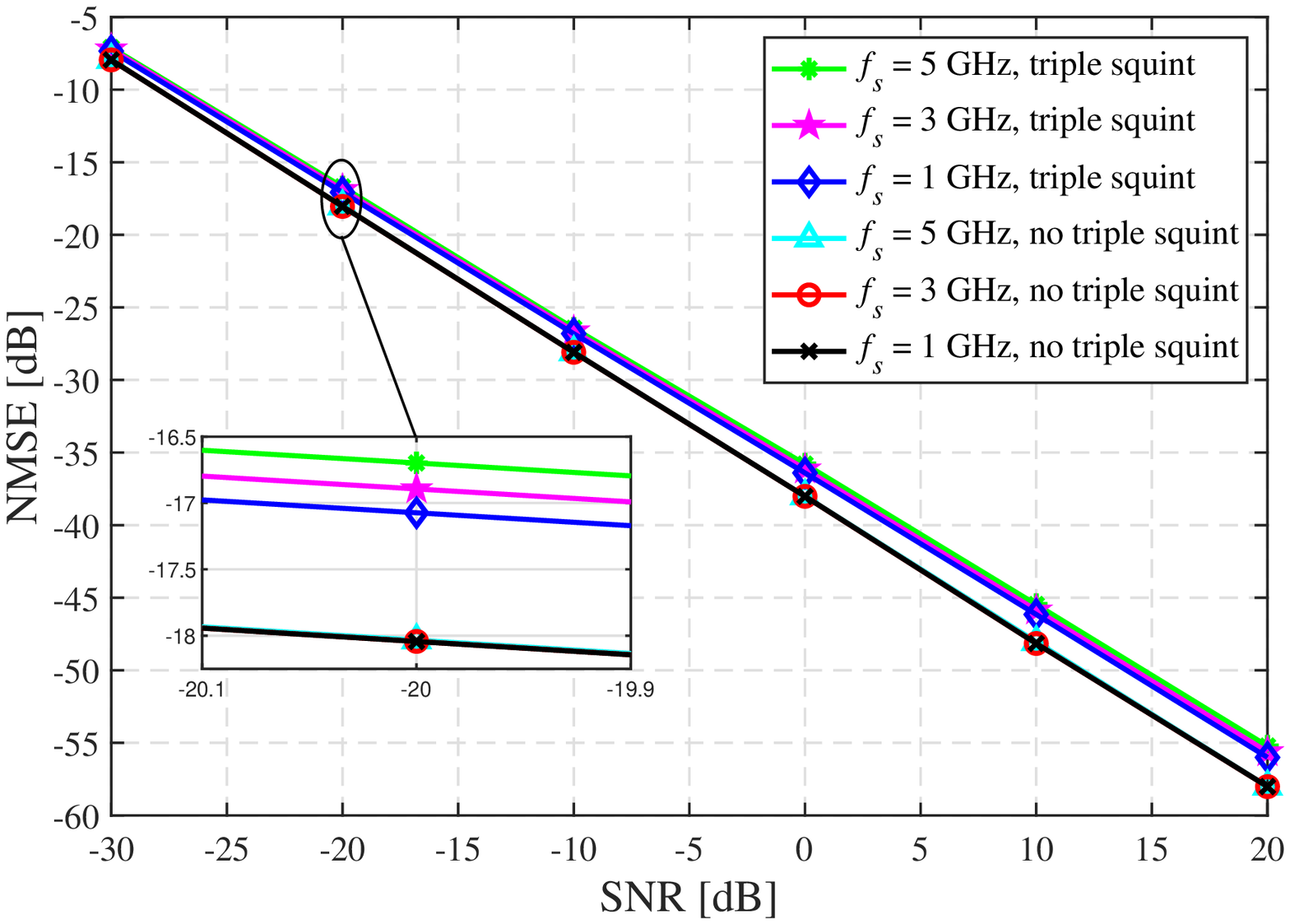}
\hspace{-7.0mm}
\caption{NMSE comparison with different bandwidths.}
\label{FIG15}
\end{minipage}
\hfill
\begin{minipage}[t]{0.49\linewidth}
\centering
\includegraphics[width=3.3in]{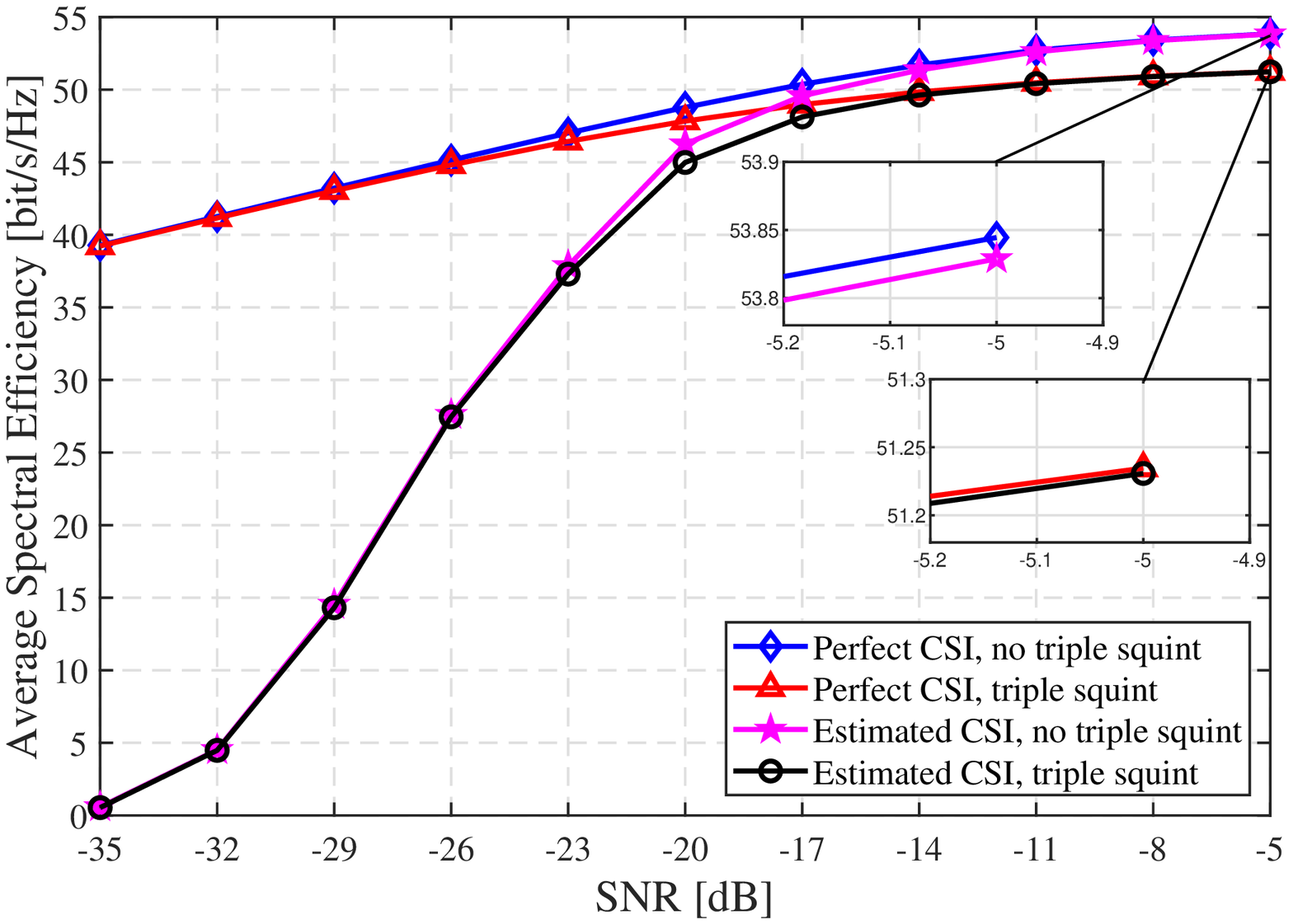}
\hspace{-1.3mm}
\caption{ASE comparison with different CSI.}
\label{FIG16}
\end{minipage}
%\vspace{-6mm}
\end{figure*}

 Fig.~\ref{FIG12} investigates the RMSE performance of the proposed fine angle estimation for
 $\{ \theta_l^{\rm AC},\,\varphi_l^{\rm AC} \}_{l=1}^L$ at the initial channel estimation
 stage. The accurate angle estimation of $\{ \theta_l^{\rm AC},\,\varphi_l^{\rm AC} \}_{l=1}^L$ relies
 on the fine estimates of $\{ \theta_l^{\rm BS},\,\varphi_l^{\rm BS} \}_{l=1}^L$ in Fig.~\ref{FIG11}.
 To investigate the impact of the estimated $\{ \theta_l^{\rm BS},\,\varphi_l^{\rm BS} \}_{l=1}^L$
 on the estimation of $\{ \theta_l^{\rm AC},\,\varphi_l^{\rm AC} \}_{l=1}^L$, we consider
 ``Method 1'' and ``Method 2''. ``Method 1'' adopts $\{ \theta_l^{\rm BS},\,\varphi_l^{\rm BS}
 \}_{l=1}^L$ estimated at BSs for the fixed ${\rm SNR}\! =\! -20\,{\rm dB}$, while ``Method 2''
 adopts the $\{ \theta_l^{\rm BS},\,\varphi_l^{\rm BS} \}_{l=1}^L$ estimated at BSs for the same
 SNRs with those of the angle estimation at aircraft\footnote{It's worth noting
 that to ensure the rationality of CRLB at
 low SNRs for ``Method 2'', the rough angle estimates $\{ {\widetilde \theta}_l^{\rm BS},\,{\widetilde
 \varphi}_l^{\rm BS} \}_{l=1}^{L}$ rather than the estimated angle $\{ {\widehat \theta}_l^{\rm BS},\,{\widehat
 \varphi}_l^{\rm BS} \}_{l=1}^{L}$ are considered as the beam-aligned angles at BSs when
 ${\rm SNR}\! \le\! -20\,{\rm dB}$.}. From Fig.~\ref{FIG12}, similar conclusions to those observed for
 Fig.~\ref{FIG11} can be obtained. Moreover, it can be observed that the ``Method 2'' can obtain more accurate
 angle estimation than that of ``Method 1'' when SNR is larger than $-20\,{\rm dB}$. For the curves labeled as
 ``proposed algorithm 1 with $i_{\rm AC}^{\rm max}\! =\! 1$'', ``proposed algorithm 1 with
 $i_{\rm AC}^{\rm max}\! =\! 2$'' and ``CRLB'', the improvement of RMSE performance are more than
 $12\,{\rm dB}$ when ${\rm SNR}\! \ge\! -10\,{\rm dB}$. This is because ``Method 2'' employs more accurate
 angles estimated at BSs in high SNR region to obtain the larger beam alignment gain than ``Method 1''.

\setcounter{TempEqCnt}{\value{equation}}
\setcounter{equation}{45}
\begin{figure*}[hb]
\hrulefill
\begin{equation}\label{eq_NMSE_H_dl} % eq 46
 {\mathrm{NMSE}}_{\bm{H}_{\rm DL}^{[2]}} = \mathbb{E}\left( \frac{1}{L}
 \sum\nolimits_{l=1}^{L} \left({ \sum\nolimits_{k=1}^{K} { \left\|\bm{H}_{{\rm DL},l}^{[2]}[k]
 - \bm{\widehat H}_{{\rm DL},l}^{[2]}[k]\right\|_F^2 } } \big/
 \sum\nolimits_{k=1}^{K} { {\left\|\bm{H}_{{\rm DL},l}^{[2]}[k]\right\|_F^2} } \right) \right) .
\end{equation}
\end{figure*}

 Fig.~\ref{FIG13} compares the RMSE performance of the proposed fine Doppler estimation for
 $\{ \psi_{z,l} \}_{l=1}^L $ at the initial channel estimation stage with different processing
 methods, where the angles at BSs and aircraft are estimated at the fixed ${\rm SNR}\! =\! -20\,{\rm dB}$.
 Note that the label ``no Doppler squint'' denotes the channel model without Doppler squint effect,
 and the label ``proposed algorithm 2 with $i_{\rm do}^{\rm max}\! =\! 0$'' indicates that the TLS-ESPRIT
 algorithm is applied directly to $\bm{Y}_{{\rm do},l}$ for obtaining the estimate
 ${\widehat \psi}_{z,l}^{(0)}$ in \textbf{Algorithm~\ref{ALG2}}.
 From Fig.~\ref{FIG13}, we observe that the THz UM-MIMO array can provide a large beam alignment gain
 and greatly improve the receive SNR for Doppler shift estimation, so that the RMSE curves are
 close to CRLB at very low SNR, even ${\rm SNR}\! =\! -100\,{\rm dB}$. Additionally,
 ``proposed algorithm 2 with $i_{\rm do}^{\rm max}\! =\! 0$'' and ``proposed algorithm 2 with
 $i_{\rm do}^{\rm max}\! =\! 1$'' will encounter the RMSE floors at high SNR, while the curve labeled as
 ``proposed algorithm 2 with $i_{\rm do}^{\rm max}\! =\! 2$'' almost overlap with ``conventional scheme'' with ``no
 beam squint'' when ${\rm SNR}\! >\! -100\,{\rm dB}$.

 Fig.~\ref{FIG14} compares the RMSE performance of the proposed path delay estimation for the normalized
 $\{ {\bar \tau}_l \}_{l=1}^L$ at the initial channel estimation stage, where the angles and Doppler shifts
 are estimated at fixed ${\rm SNR}\! =\! -20\,{\rm dB}$ and ${\rm SNR}\! =\! 20\,{\rm dB}$, respectively.
 Note that the labels ``triple squint'' and ``no triple squint'' indicate the channel model considering
 and not considering the practical triple squint effects, respectively. Clearly, when the triple squint
 effects are considered, the higher angle and Doppler estimation accuracy at ${\rm SNR}\! =\! 20\,{\rm dB}$
 will attenuate the impact of triple squint effects to acquire more accurate path delay estimation than
 that estimated at ${\rm SNR}\! =\! -20\,{\rm dB}$. Note that the errors of the previously estimated angles
 $\{ {\widehat \theta}_l^{\rm BS},\,{\widehat \varphi}_l^{\rm BS},\,{\widehat \theta}_l^{\rm AC},\,{\widehat
 \varphi}_l^{\rm AC} \}_{l=1}^L$ and Doppler shifts $\{ {\widehat \psi}_{z,l} \}_{l=1}^L$ impact on the estimation
 of $\{ {\bar \tau}_l \}_{l=1}^L$, which leads to the RMSE floors of the normalized delay estimation at high SNR.

\begin{figure*}[!tp]
%\vspace{-5mm}
\captionsetup{font={footnotesize}, name = {Fig.}, labelsep = period} % singlelinecheck = off, justification = raggedright,
\captionsetup[subfigure]{singlelinecheck = on, justification = raggedright, font={footnotesize}}
\centering
\subfigure{%[The number of path $L = 3$]
\label{FIG17(a)}
\hspace{-0.4mm}
\includegraphics[width=3.2in]{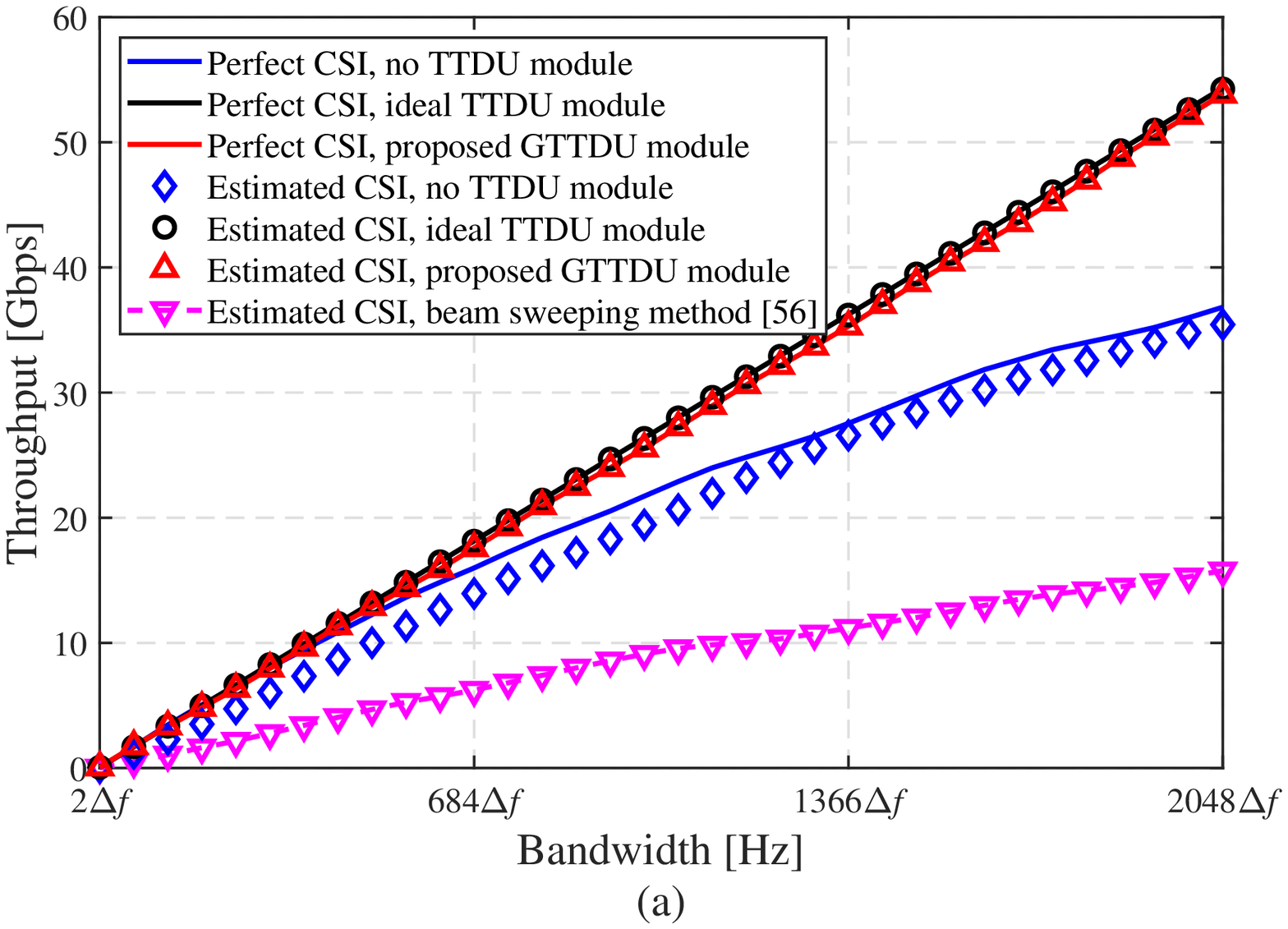}
}
\subfigure{%[The number of path $L = 5$]
\label{FIG17(b)}
%\hspace{0.1mm}
\includegraphics[width=3.2in]{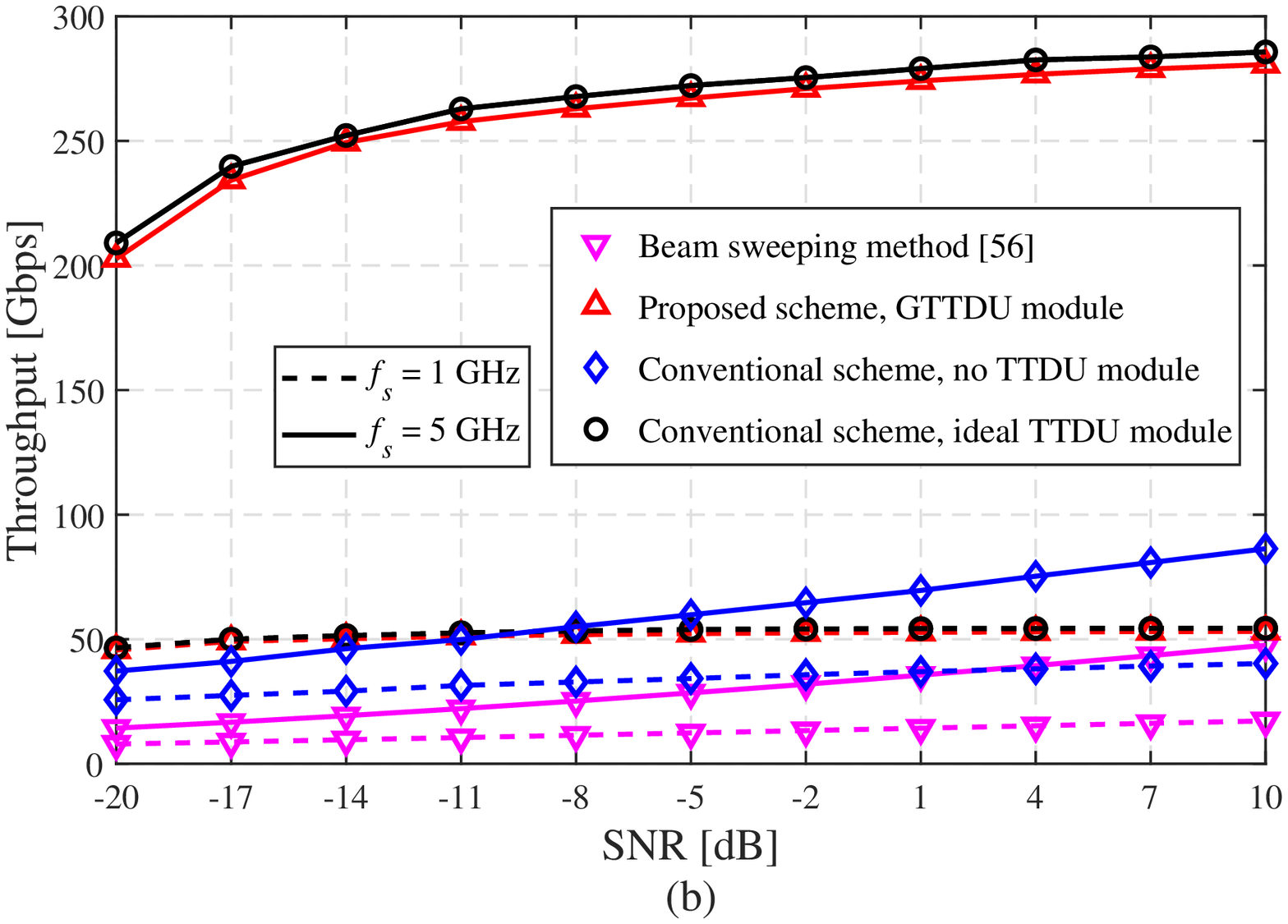}
}
\setlength{\abovecaptionskip}{-0.05mm}
\caption{Throughput performance comparison of THz UM-MIMO system adopting
 different TTDU modules: (a)~maximum bandwidth is $f_s\! =\! 1\,{\rm GHz}$ with perfect
 and the estimated CSI at ${\rm SNR}\! =\! 10\,{\rm dB}$; and (b)~bandwidth $f_s\! =\! 1\,{\rm GHz}$
 and $5\,{\rm GHz}$ with the estimated CSI.}
%\vspace{-4mm}
\label{FIG17}
\end{figure*}

 According to the estimated channel parameters, the Normalized-MSE (NMSE) metric
 \cite{Liao_Tcom19} for the initial channel estimation can be expressed as
 (\ref{eq_NMSE_H_dl}) on the bottom of this page.
%\begin{equation}\label{eq_NMSE_H_dl} % eq 46
% {\mathrm{NMSE}}_{\bm{H}_{\rm DL}^{[2]}} = \mathbb{E}\left( \frac{1}{L}
% \sum\limits_{l=1}^{L} \left({ \sum\limits_{k=1}^{K} { \left\|\bm{H}_{{\rm DL},l}^{[2]}[k]
% - \bm{\widehat H}_{{\rm DL},l}^{[2]}[k]\right\|_F^2 } } \big/
% \sum\limits_{k=1}^{K} { {\left\|\bm{H}_{{\rm DL},l}^{[2]}[k]\right\|_F^2} } \right) \right) ,
%\end{equation}
 In (\ref{eq_NMSE_H_dl}), $\bm{H}_{{\rm DL},l}^{[2]}[k]$ and $\bm{\widehat H}_{{\rm DL},l}^{[2]}[k]$ denote the DL
 spatial-frequency channel matrix at the $k$th subcarrier of the 2nd OFDM symbol (considering the
 impact of Doppler shifts) in (\ref{eq_H_k_dl_1}) and the reestablished channel matrix based on
 the estimated channel parameters, respectively. Fig.~\ref{FIG15} compares the NMSE performance
 at the initial channel estimation stage for different system bandwidths $f_s\! =\! \{ 1,3,5 \}\,{\rm GHz}$.
 From Fig.~\ref{FIG15}, we can observe that the channel estimation performance of the proposed
 solution under triple squint effects is very close to that of the proposed solution without triple
 squint effects, where the NMSE performance gap between them is about $1\,{\rm dB}$ at ${\rm SNR}\!
 =\! -20\,{\rm dB}$. Furthermore, the results of Fig.~\ref{FIG15} show that compared with the system
 bandwidth $f_s\! =\! 1\,{\rm GHz}$, the NMSE performance of the proposed solution using the larger
 bandwidth $f_s\! =\! 5\,{\rm GHz}$ does not deteriorate significantly.

\begin{figure}[!tp]
%\vspace{-1mm}
\begin{center}
 \includegraphics[width=0.9\columnwidth, keepaspectratio]{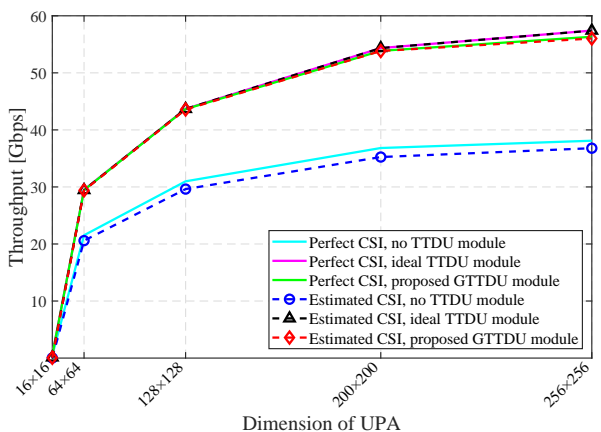}
\end{center}
\setlength{\abovecaptionskip}{-0.05mm}
\captionsetup{font = {footnotesize}, singlelinecheck = off, name = {Fig.}, labelsep = period} %, justification = raggedright
\caption{Throughput performance comparison of THz UM-MIMO system adopting
 different dimensions of UPA at ${\rm SNR}\! =\! 10\,{\rm dB}$.}
%\vspace{-6mm}
\label{FIG18}
\end{figure}

 Moreover, we consider the Average Spectral Efficiency (ASE) performance metric
 \cite{Liao_Tcom19,Wang_CL17} at the data transmission stage, defined as
 ${\mathrm{ASE}}\! =\! \sum\nolimits_{l=1}^{L} \left( {\textstyle{1 \over K}}
 \sum\nolimits_{k=1}^{K} \log_2 \left( 1 + {| h_l^{[2]}[k] |^2}/{|
 \mathbb{E}(z_l^{[2]}[k]) |^2} \right) \right)$, where $h_l^{[2]}[k]$ and
 $z_l^{[2]}[k]$ are the beam-aligned effective channel coefficient and interference
 plus noise at the $k$th subcarrier of the 2nd OFDM symbol, respectively.
 Fig.~\ref{FIG16} compares the ASE performance of the proposed solution with different CSI,
 where the perfect CSI known at both the BSs and aircraft is adopted as the performance
 upper bound. It can be observed from Fig.~\ref{FIG16} that the ASE performance
 using the estimated CSI almost attains the performance upper bound when
 ${\rm SNR}\! \ge\! -14\,{\rm dB}$ whether or not the triple squint effects are considered.
 In addition, since the practicable GTTDU module still has residual beam alignment error
 caused by beam squint effect, the ASE performance gain achieved by our solution
 with triple squint effects is 2.5 [bit/s/Hz] lower than the other one at high SNR.

\begin{figure*}[!tp]
%\vspace{-5mm}
\captionsetup{font={footnotesize}, name = {Fig.}, labelsep = period} % singlelinecheck = off, justification = raggedright,
\captionsetup[subfigure]{singlelinecheck = on, justification = raggedright, font={footnotesize}}
\centering
\subfigure{%[The number of path $L = 3$]
\label{FIG19(a)}
\hspace{-0.4mm}
\includegraphics[width=3.3in]{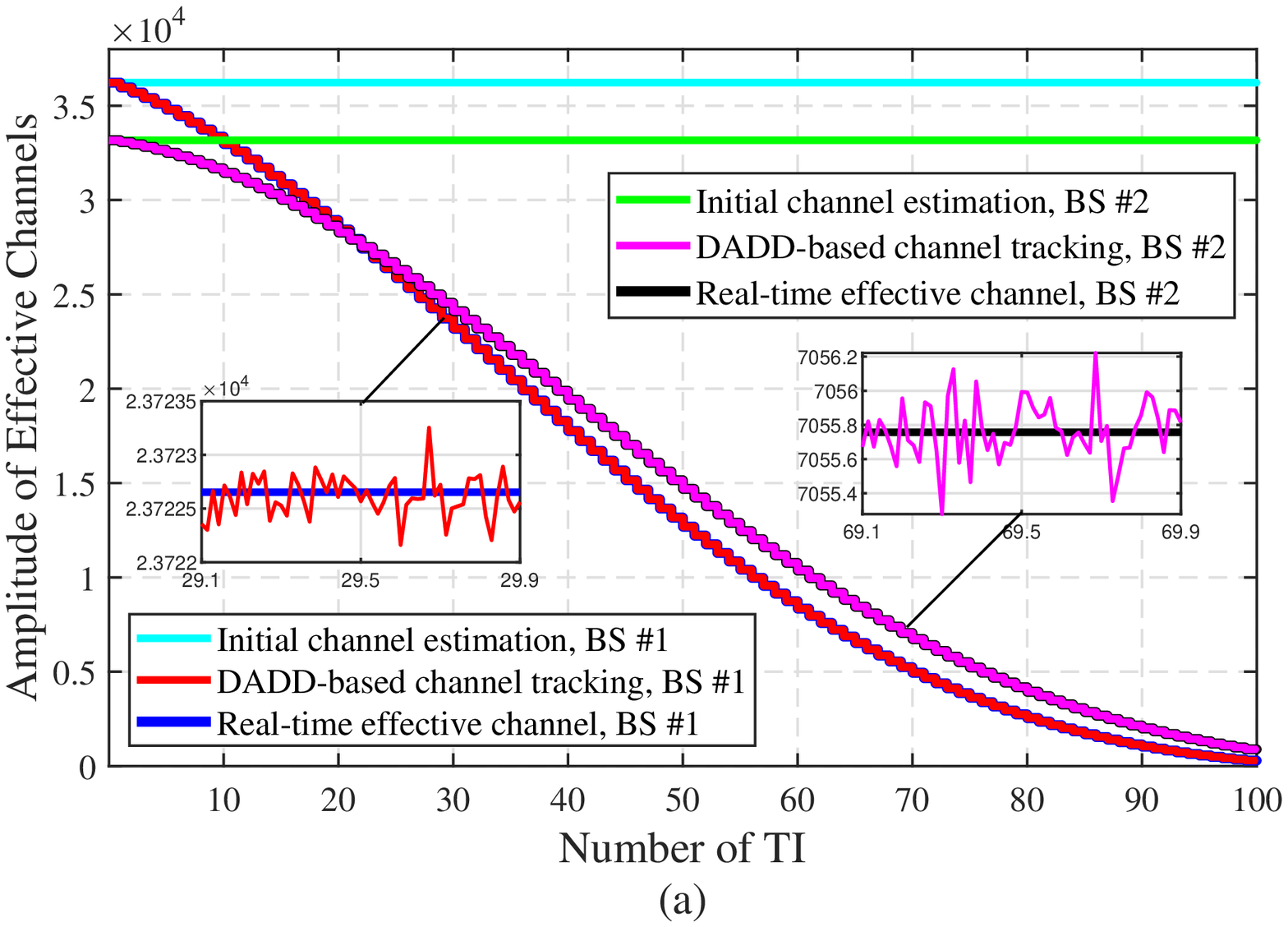}
}
\subfigure{%[The number of path $L = 5$]
\label{FIG19(b)}
%\hspace{0.1mm}
\includegraphics[width=3.3in]{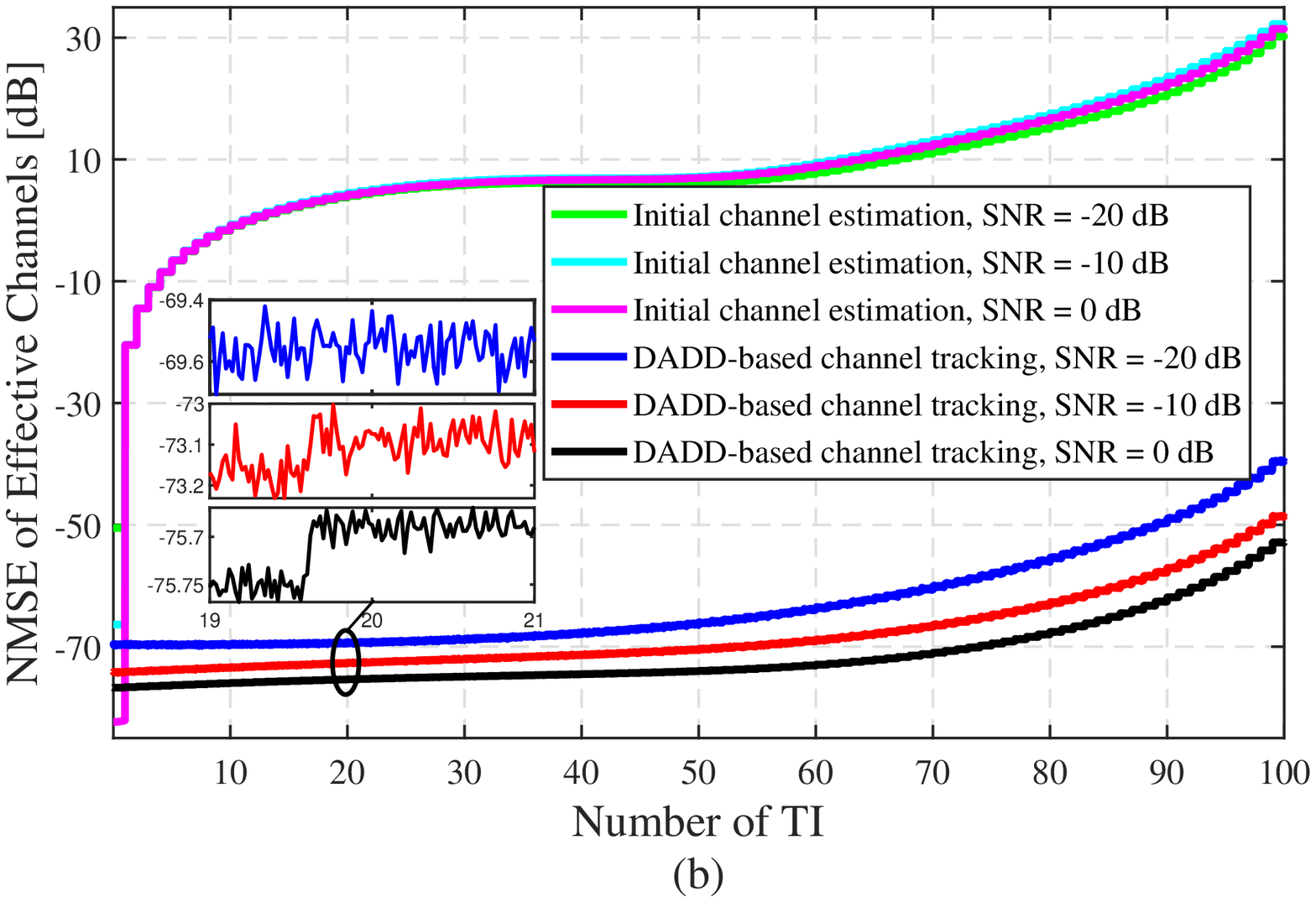}
}
\setlength{\abovecaptionskip}{-0.05mm}
\caption{Performance comparison of the proposed DADD-based channel tracking: (a)~amplitude of effective channels;
 and (b)~NMSE of effective channels.}
%\vspace{-4mm}
\label{FIG19}
\end{figure*}

\begin{figure*}[!tp]
%\vspace{-1mm}
\captionsetup{singlelinecheck = off, justification = raggedright, font={footnotesize}, name = {Fig.}, labelsep = period} %
\captionsetup[subfigure]{singlelinecheck = on, justification = raggedright, font={footnotesize}}
\centering
\subfigure{%[The number of path $L = 3$]
\label{FIG20(a)}
%\hspace{-2.0mm}
\includegraphics[width=3.3in]{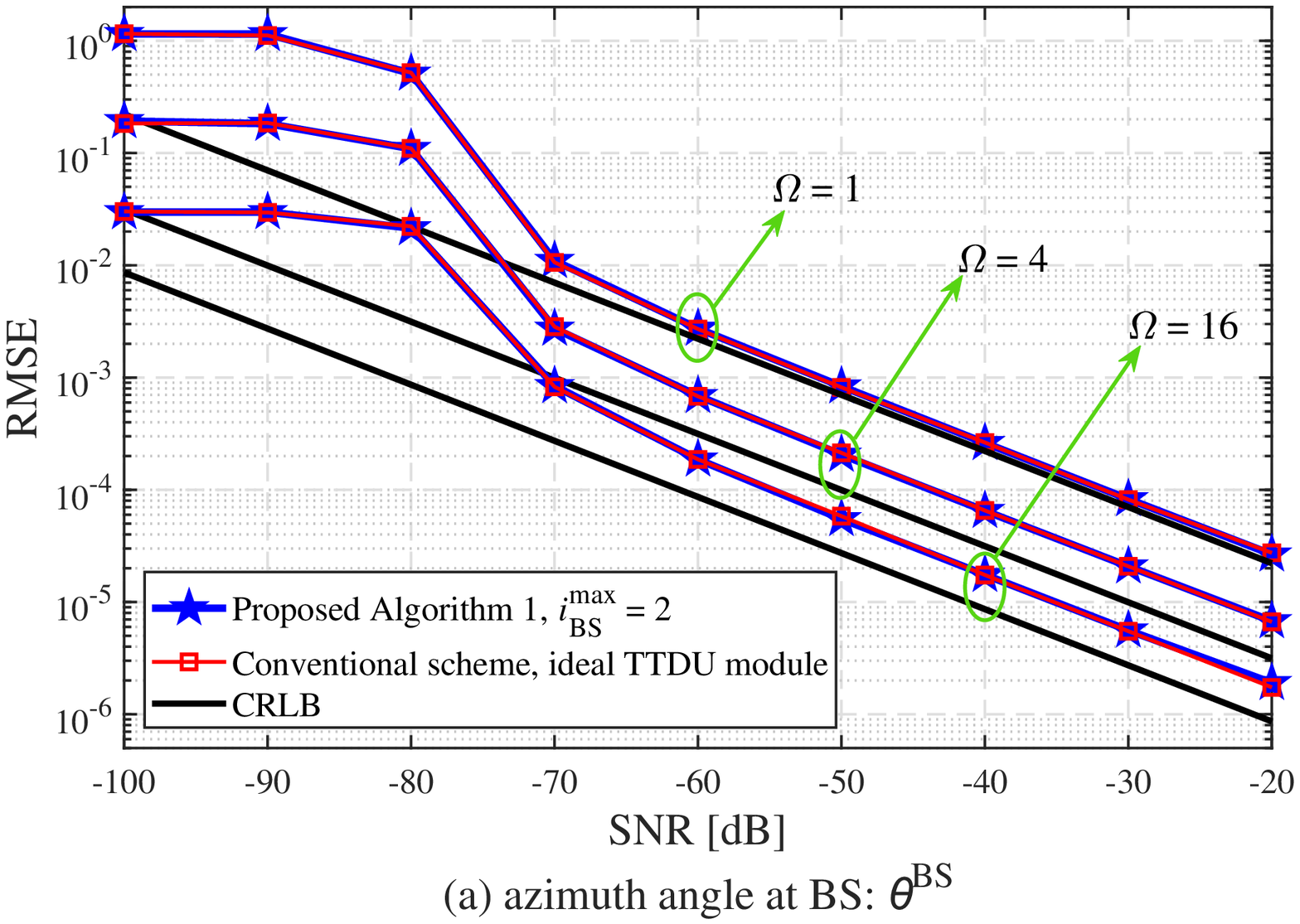}
}
\subfigure{%[The number of path $L = 5$]
\label{FIG20(b)}
%\hspace{5mm}
\includegraphics[width=3.3in]{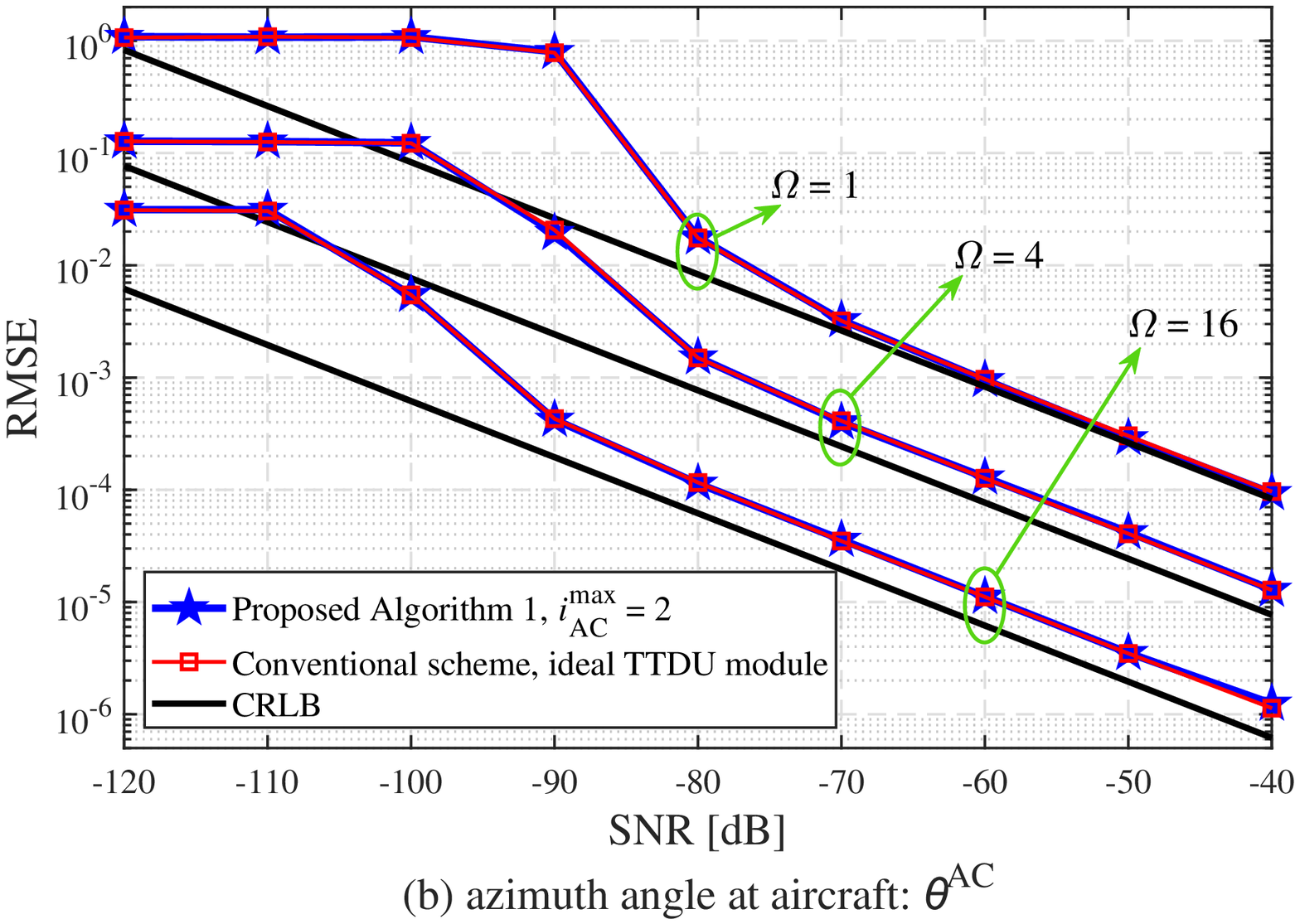}
}
\setlength{\abovecaptionskip}{-0.05mm}
\caption{RMSE performance at pilot-aided angle tracking stage:
 (a)~azimuth angle $\theta^{\rm BS}$ at BS; and (b)~azimuth angle $\theta^{\rm AC}$ at aircraft.}
%\vspace{-6mm}
\label{FIG20}
\end{figure*}

 Fig.~\ref{FIG17} compares the throughput performance of THz UM-MIMO system adopting
 different TTDU modules, where the transceivers using ideal TTDU module, the proposed
 GTTDU module, and without TTDU module are considered. Note that $\Delta f$ denotes the
 frequency spacing between adjacent subcarriers, typically, $\Delta f\! \approx\! 0.488$
 Megahertz (MHz) for $f_s\! =\! 1\,{\rm GHz}$ and $K\! =\! 2048$. In Fig.~\ref{FIG17}(a),
 for maximum bandwidth $f_s\! =\! 1\,{\rm GHz}$, an obvious throughput ceiling can be
 observed in ``beam sweeping method'' and conventional scheme with ``no TTDU module'' as
 the bandwidth increases, in other words, the severe beam squint effect will restrict
 the throughput of THz UM-MIMO systems. On the contrary, the throughputs adopting the
 proposed GTTDU module and ideal TTDU module present a linear growth with the increase
 of bandwidth. For the estimated CSI at $f_s\! =\! 2048\Delta f$, the throughput
 improvements of more than $15$ Gigabit per second (${\rm Gbps}$) and $35\,{\rm Gbps}$
 can be acquired by both ``ideal TTDU module'' and ``proposed GTTDU module'' compared
 with the throughput of ``no TTDU module'' and beam sweeping method in \cite{IEEE802.11ad},
 respectively. Furthermore, it can be also observed from Fig.~\ref{FIG17}(b) that
 the increase of throughput in the THz UM-MIMO system with severe beam squint effect
 is extremely limited when the bandwidth is increased to $f_s\! =\! 5\,{\rm GHz}$.

 Fig.~\ref{FIG18} compares the throughput performance of THz UM-MIMO system adopting different dimensions of UPA at ${\rm SNR}\! =\! 10\,{\rm dB}$, where bandwidth $f_s\! =\! 1\,{\rm GHz}$ and the same transmit power are considered.
 From Fig.~\ref{FIG18}, it can be observed that the usage of regular UPA with size of $16\! \times\! 16$ in the ultra-long-distance THz aeronautical communications cannot establish an efficient communication link, which causes the degraded throughput performance.
 Due to the pencil-like beams and less interference, the system throughput will be improved significantly as the dimension of UPA equipped at the transceiver increases.
 However, the increase of array dimension leads to more obvious beam squint effect, which inhibits the improvement of throughput performance in turn (observed from the curves labeled as ``no TTDU module'').
 For the transceiver equipped with UM-MIMO array of size $256\! \times\! 256$, the throughput adopting the proposed GTTDU module is closed to the throughput of ``ideal TTDU module'', and it can achieve throughput improvement more than $55\,{\rm Gbps}$ compared with that of transceiver using UPA of size $16\! \times\! 16$.
 Therefore, it is necessary to use UM-MIMO array in aeronautical communications to cater for the high data rate requirements of hundreds of users in the cabin.

 Next, the performance of the proposed DADD-based channel tracking algorithm is
 evaluated according to the metrics of effective channels' amplitude and NMSE, where the NMSE
 of effective channels for the $r$th OFDM symbol is given by ${\mathrm{NMSE}}_{\bm{h}^{[r]}}\!
 =\! \mathbb{E}\left( {\textstyle{1 \over L}} \sum\nolimits_{l=1}^{L} \left( { \|\bm{h}_l^{[r]}
 \! -\! \bm{\widehat h}_l^{[r]}\|_2^2 /{\|\bm{h}_l^{[r]}\|_2^2} } \right) \right)$. For the
 data-aided channel tracking scheme, Fig.~\ref{FIG19} compares the effective channels'
 amplitude performance (at ${\rm SNR}\! =\! -20\,{\rm dB}$) and NMSE performance
 (at ${\rm SNR}\! =\! -20,\,-10,$ and $0\,{\rm dB}$) for the different numbers of TI.
 Here, the Turbo coding and QPSK modulation are considered
 during the data transmission. From Fig.~\ref{FIG19(a)}, we can observe that the amplitude
 of effective channels decreases rapidly as time goes by, where the proposed DADD-based
 channel tracking method can track the amplitude changes of true effective channels in
 real-time. This decreasing amplitudes mean that the gains of beam alignment becomes small.
 Also observe in Fig.~\ref{FIG19(b)} that the NMSE performance of the proposed DADD-based
 channel tracking method slowly worsens as the number of TI increases, while the NMSE of the
 initial channel estimation without tracking will deteriorate rapidly after several TIs.

 Fig.~\ref{FIG20} investigates the RMSE performance of the proposed pilot-aided angle tracking
 scheme against different sparse spacing $\varOmega\! =\! 1$, $\varOmega\! =\! 4$, and
 $\varOmega\! =\! 16$. Here the angle tracking at aircraft adopts the angles $\{ {\widehat
 \theta}_l^{\rm BS},\,{\widehat \varphi}_l^{\rm BS} \}_{l=1}^L$ estimated at BSs using the
 fixed ${\rm SNR}\! =\! -60\,{\rm dB}$. Note that the RMSE curves of the elevation angles
 $\varphi^{\rm BS}$ and $\varphi^{\rm AC}$ are omitted due to the similar performance to
 the azimuth angles. From Fig.~\ref{FIG20}, it can be observed that the usage of sparse
 array can significantly improve the accuracy of angle estimation, and these results testify
 that the improved RMSE performance is consistent with the conclusion in \emph{Remark $2$},
 i.e., the proposed solution using the sparse array with sparse spacing $\varOmega$ can
 achieve about $20\lg \varOmega\,{\rm dB}$ performance gain.

\section{Conclusions}\label{S8}
 We have proposed an effective channel estimation and tracking scheme for THz UM-MIMO-based
 aeronautical communications in SAGIN, which can solve the unique triple delay-beam-Doppler
 squint effects not considered in the sub-6 GHz or mmWave systems. The proposed solution includes
 the initial channel estimation, data-aided channel tracking, and pilot-aided channel tracking.
 Specifically, based on the rough angle estimates acquired from navigation information, the initial
 THz UM-MIMO link can be established, where the delay-beam squint effects at transceiver can be
 significantly mitigated by employing the proposed GTTDU module. By exploiting the
 proposed prior-aided iterative angle estimation algorithm, the fine azimuth/elevation angles
 can be estimated based on the equivalent low-dimensional fully-digital array. These estimated
 angles can be used not only to achieve a highly accurate beam alignment, but also to refine the
 GTTDU module at the transceiver for further eliminating the delay-beam squint effects. The Doppler
 shifts can be subsequently estimated using the proposed prior-aided iterative Doppler shift
 estimation algorithm. On this basis, path delays and channel gains can be estimated accurately,
 where Doppler squint effect can be attenuated vastly via fine compensation process. At the data
 transmission stage, a DADD-based channel tracking algorithm is developed to track the beam-aligned
 effective channels. When the data-aided channel tracking is invalid,
 the pilot-aided channel tracking is proposed to re-estimate the angles at transceiver using an
 equivalent fully-digital sparse array, where the angle ambiguity issue derived from sparse array can
 be addressed based on the previously estimated angles. Finally, the CRLBs of dominant channel
 parameters and the simulation results evaluate the effectiveness of the proposed solution for
 THz UM-MIMO-based aeronautical communications.

 It is worth mentioning that the proposed solution in this paper still has some improvements in the following aspects.
 First, the proposed Rotman lens-based GTTDU module of transceiver in Fig.~\ref{FIG6} can be further researched.
 Second, the signal frame structure (e.g., the length of OFDM symbols, CP length) in THz communications can be also optimized based on the parameter configurations of specific scenarios. Third, some new data-aided channel tracking methods with the lower computational complexity can be considered in Section~IV, such as uniformly-spaced pilot interpolation in the frequency domain. Fourth, according to the specific communication scenarios, the transmit power at the transceiver can be also further optimized to improve the spectrum efficiency of systems and reduce the bit error rate.

\setcounter{TempEqCnt}{\value{equation}}
\setcounter{equation}{47}
\begin{figure*}[hb]
\hrulefill
\begin{align} % eqs 48,49,50,51
 [\bm{H}_{{\rm DL},l}^{(t)}(f)]_{n_{\rm AC},n_{\rm BS}} &= \sqrt{G_l} \alpha_l e^{\textsf{j} {2\pi\psi_l t}} e^{-\textsf{j} {2\pi f \tau_l}} e^{\textsf{j} {\textstyle{2d \over {\lambda_c}}}
 {\left( {(n_{\rm AC}^{\rm h}-1)\mu_l^{\rm AC} + (n_{\rm AC}^{\rm v}-1)\nu_l^{\rm AC}} \right)}}
 e^{-\textsf{j} {\textstyle{2d \over {\lambda_c}}}
 {\left( {(n_{\rm BS}^{\rm h}-1)\mu_l^{\rm BS} + (n_{\rm BS}^{\rm v}-1)\nu_l^{\rm BS}} \right)}} . \label{eq_H_f_dl_mn}\\[1mm]
 [\bm{H}_{{\rm DL},l}^{(t)}[k]]_{n_{\rm AC},n_{\rm BS}} &= \sqrt{G_l} \alpha_le^{\textsf{j} {2\pi \psi_{l,k} t}}
 e^{-\textsf{j} {2\pi \left({\textstyle{k-1 \over K}}-{\textstyle{1 \over 2}}\right)f_s \tau_l}}
 \!\left[ \bm{a}_{\rm AC}(\mu_l^{\rm AC},\nu_l^{\rm AC},k) \right]_{n_{\rm AC}}
 \!\left[ \bm{a}_{\rm BS}^*(\mu_l^{\rm BS},\nu_l^{\rm BS},k) \right]_{n_{\rm BS}} . \label{eq_H_k_dl_mn}\\[1mm]
\left[ \bm{a}_{\rm AC}(\mu_l^{\rm AC},\nu_l^{\rm AC},k) \right]_{n_{\rm AC}} &= e^{\textsf{j} {\left( {(n_{\rm AC}^{\rm h}-1)\mu_l^{\rm AC} + (n_{\rm AC}^{\rm v}-1)\nu_l^{\rm AC}} \right)}}
 e^{\textsf{j} \left({\textstyle{{k-1} \over K}}-{\textstyle{1 \over 2}}\right){\textstyle{f_s \over f_z}}
 {\left( {(n_{\rm AC}^{\rm h}-1)\mu_l^{\rm AC} + (n_{\rm AC}^{\rm v}-1)\nu_l^{\rm AC}} \right)}}
 , \label{eq_a_AC_n_AC_k}\\[1mm]
 \left[ \bm{a}_{\rm BS}(\mu_l^{\rm BS},\nu_l^{\rm BS},k) \right]_{n_{\rm BS}} &= e^{\textsf{j} {\left( {(n_{\rm BS}^{\rm h}-1)\mu_l^{\rm BS} + (n_{\rm BS}^{\rm v}-1)\nu_l^{\rm BS}} \right)}}
 e^{\textsf{j} \left({\textstyle{{k-1} \over K}}-{\textstyle{1 \over 2}}\right){\textstyle{f_s \over f_z}}
 {\left( {(n_{\rm BS}^{\rm h}-1)\mu_l^{\rm BS} + (n_{\rm BS}^{\rm v}-1)\nu_l^{\rm BS}} \right)}}
 . \label{eq_a_BS_n_BS_k}
\end{align}
\end{figure*}

 For future work, our proposed THz UM-MIMO-based aeronautical communication solution can be also suitable for the long distance communications or backhaul in SAGIN such as the space information network consisting of aircrafts/UAVs, aerial BSs, and LEO/MEO/GEO satellites, or the air-ground communication links between the high-altitude terrestrial stations and the LEO satellite systems. Potential research directions in the THz UM-MIMO-based aeronautical communications include more specific and universal THz UM-MIMO channel modeling under LoS path scenario \cite{HanC_CM18}, long-distance air-ground communication scheme design, low-complexity signal transmission and tracking methods for the large bandwidth and high dynamic environment, THz transceiver design using more practical hardware components (e.g., TTDU module, high-frequency switch \cite{Ghaleb_MWCL'17}, and low-energy antenna array \cite{Akyildiz_PC14}), advanced DSP module design supporting ultra-high data rate with the order of Tbps, modulation and coding design at the physical layer \cite{Akyildiz_WC14}, as well as the deployment and power optimization of aerial BSs at the network and transport layer.

\begin{appendices}

\section{Derivation of DL Channel Matrix $\bm{H}_{{\rm DL},l}^{[n]}[k]$}

 By taking the Fourier transform of (\ref{eq_H_tau_dl_mn}) with respect to $\tau$, the frequency response of
 $[\bm{\bar H}_{{\rm DL},l}^{(t)}(\tau)]_{n_{\rm AC},n_{\rm BS}}$ is given by
\setcounter{equation}{46}
\begin{align}\label{eq_H_fc_dl_mn} % eq 47
 [\bm{\bar H}_{{\rm DL},l}^{(t)}(f_c)]_{n_{\rm AC},n_{\rm BS}} =& \sqrt{G_l}
 \alpha_l e^{\textsf{j} {2\pi\psi_l t}} e^{-\textsf{j} {2\pi f_c \tau_l}}
 e^{-\textsf{j} {2\pi f_c \tau_l^{[n_{\rm AC}]}}} \nonumber\\
 &\times e^{-\textsf{j} {2\pi f_c \tau_l^{[n_{\rm BS}]}}} ,
\end{align}
 where the large-scale fading gain $G_l$ can be modeled as $G_l\! =\! \lambda_c^2 / (4\pi D_l)^2$ based on the free-space path loss of Friis' formula with $D_l$ being the communication distance between the aircraft and the $l$th BS.
 Considering the large system bandwidth $f_s$, the carrier frequency can be expressed as
 $f_c\! =\! f_z\! +\! f$, where $f$ denotes the baseband frequency satisfying $-{f_s}/2\! \le\! f\! \le\! {f_s}/2$
 and the wavelength corresponding to the central carrier frequency $f_z$ is $\lambda_z$. After
 the down-conversion and focussing on the baseband frequency, we can obtain the ($n_{\rm AC},n_{\rm BS}$)th
 element of the DL baseband channel matrix $\bm{H}_{{\rm DL},l}^{(t)}(f)$
 in the spatial-frequency domain \cite{GaoFF_TSP18,GaoFF_TSP19,LinXC_CL17}, i.e.,
 (\ref{eq_H_f_dl_mn}) on the bottom of this page.
%\begin{equation}\label{eq_H_f_dl_mn} % eq 48
% [\bm{H}_{{\rm DL},l}^{(t)}(f)]_{n_{\rm AC},n_{\rm BS}} = \sqrt{G_l} \alpha_l e^{\textsf{j} {2\pi\psi_l t}} e^{-\textsf{j} {2\pi f \tau_l}} e^{\textsf{j} {\textstyle{2d \over {\lambda_c}}}
% {\left( {(n_{\rm AC}^{\rm h}-1)\mu_l^{\rm AC} + (n_{\rm AC}^{\rm v}-1)\nu_l^{\rm AC}} \right)}}
% e^{-\textsf{j} {\textstyle{2d \over {\lambda_c}}}
% {\left( {(n_{\rm BS}^{\rm h}-1)\mu_l^{\rm BS} + (n_{\rm BS}^{\rm v}-1)\nu_l^{\rm BS}} \right)}} .
%\end{equation}

 Due to the large bandwidth in THz UM-MIMO, the carrier frequencies and wavelengths at different
 subcarriers are different, so the frequency-dependent Doppler shift at the $k$th subcarrier is given by
 $\psi_{l,k}\! =\! \psi_{z,l}\! +\! {\textstyle{\underline{v}_l \over c}}({\textstyle{{k\!-\!1} \over K}}\!-\!{\textstyle{1 \over 2}})f_s$ with $\psi_{z,l} = \underline{v}_l/\lambda_z$. Let the antenna spacing $d\! =\! \lambda_z/2$, the baseband frequency
 response in (\ref{eq_H_f_dl_mn}) can be further expressed as the spatial-frequency channel
 coefficient at the $k$th subcarrier, i.e., (\ref{eq_H_k_dl_mn}) on the bottom of this page.
%\begin{equation}\label{eq_H_k_dl_mn} % eq 49
% [\bm{H}_{{\rm DL},l}^{(t)}[k]]_{n_{\rm AC},n_{\rm BS}}\! =\! \sqrt{G_l} \alpha_le^{\textsf{j} {2\pi \psi_{l,k} t}}
% e^{-\textsf{j} {2\pi \left({\textstyle{k-1 \over K}}-{\textstyle{1 \over 2}}\right)f_s \tau_l}}
% \!\left[ \bm{a}_{\rm AC}(\mu_l^{\rm AC},\nu_l^{\rm AC},k) \right]_{n_{\rm AC}}
% \!\left[ \bm{a}_{\rm BS}^*(\mu_l^{\rm BS},\nu_l^{\rm BS},k) \right]_{n_{\rm BS}} ,
%\end{equation}
 In (\ref{eq_H_k_dl_mn}), $\bm{a}_{\rm AC}(\mu_l^{\rm AC},\nu_l^{\rm AC},k)\! \in\! \mathbb{C}^{N_{\rm AC}}$ and
 $\bm{a}_{\rm BS}(\mu_l^{\rm BS},\nu_l^{\rm BS},k)\! \in\! \mathbb{C}^{N_{\rm BS}}$
 are the array response vectors associated with the $k$th subcarrier at aircraft and the $l$th BS, respectively,
 and $\left[ \bm{a}_{\rm AC}(\mu_l^{\rm AC},\nu_l^{\rm AC},k) \right]_{n_{\rm AC}}$ and $\left[ \bm{a}_{\rm BS}(\mu_l^{\rm BS},\nu_l^{\rm BS},k) \right]_{n_{\rm BS}}$ can be expressed as (\ref{eq_a_AC_n_AC_k}) and (\ref{eq_a_BS_n_BS_k}), respectively, on the bottom of the previous page.
%\begin{align} % eqs 50,51
% \left[ \bm{a}_{\rm AC}(\mu_l^{\rm AC},\nu_l^{\rm AC},k) \right]_{n_{\rm AC}} =&\
% e^{\textsf{j} {\left( {(n_{\rm AC}^{\rm h}-1)\mu_l^{\rm AC} + (n_{\rm AC}^{\rm v}-1)\nu_l^{\rm AC}} \right)}}
% e^{\textsf{j} \left({\textstyle{{k-1} \over K}}-{\textstyle{1 \over 2}}\right){\textstyle{f_s \over f_z}}
% {\left( {(n_{\rm AC}^{\rm h}-1)\mu_l^{\rm AC} + (n_{\rm AC}^{\rm v}-1)\nu_l^{\rm AC}} \right)}}
% , \label{eq_a_AC_n_AC_k}\\[-2.0mm] % \underbrace{}_{\rm Beam\,\,squint}
% \left[ \bm{a}_{\rm BS}(\mu_l^{\rm BS},\nu_l^{\rm BS},k) \right]_{n_{\rm BS}} =&\
% e^{\textsf{j} {\left( {(n_{\rm BS}^{\rm h}-1)\mu_l^{\rm BS} + (n_{\rm BS}^{\rm v}-1)\nu_l^{\rm BS}} \right)}}
% e^{\textsf{j} \left({\textstyle{{k-1} \over K}}-{\textstyle{1 \over 2}}\right){\textstyle{f_s \over f_z}}
% {\left( {(n_{\rm BS}^{\rm h}-1)\mu_l^{\rm BS} + (n_{\rm BS}^{\rm v}-1)\nu_l^{\rm BS}} \right)}}
% . \label{eq_a_BS_n_BS_k} % \underbrace{}_{\rm Beam\,\,squint}
%\end{align}

\setcounter{TempEqCnt}{\value{equation}}
\setcounter{equation}{52}
\begin{figure*}[ht]
\begin{align} % eqs 53,55,56,58,60
 \bm{A}_{{\rm DL},l}[k] &= \bm{a}_{\rm AC}(\mu_l^{\rm AC},\nu_l^{\rm AC},k)
 \bm{a}_{\rm BS}^{\rm H}(\mu_l^{\rm BS},\nu_l^{\rm BS},k) \nonumber \\
 &= \left( \bm{a}_{\rm AC}(\mu_l^{\rm AC},\nu_l^{\rm AC}) \circ
 \bm{\bar a}_{\rm AC}(\mu_l^{\rm AC},\nu_l^{\rm AC},k) \right)\!
 \left( \bm{a}_{\rm BS}(\mu_l^{\rm BS},\nu_l^{\rm BS}) \circ
 \bm{\bar a}_{\rm BS}(\mu_l^{\rm BS},\nu_l^{\rm BS},k) \right)^{\rm H} \nonumber \\
 &{\mathop=\limits^{(a)}} \left( \bm{a}_{\rm AC}(\mu_l^{\rm AC},\nu_l^{\rm AC})
 \bm{a}^{\rm H}_{\rm BS}(\mu_l^{\rm BS},\nu_l^{\rm BS}) \right)
 \circ \left( \bm{\bar a}_{\rm AC}(\mu_l^{\rm AC},\nu_l^{\rm AC},k)
 \bm{\bar a}^{\rm H}_{\rm BS}(\mu_l^{\rm BS},\nu_l^{\rm BS},k) \right). \label{eq_A_dl_k_2}\\[2mm]
 {\widetilde \tau}_l^{[n_{\rm AC}]} &= \left((n_{\rm AC}^{\rm h} - 1)d
 \sin({\widetilde \theta}_l^{\rm AC})\cos({\widetilde \varphi}_l^{\rm AC})
 + (n_{\rm AC}^{\rm v} - 1)d\sin({\widetilde \varphi}_l^{\rm AC})\right)/c , \tag{55} \label{eq_tau_tidle_AC}\\[2mm]
 {\widetilde \tau}_l^{[n_{\rm BS}]} &= -\left((n_{\rm BS}^{\rm h} - 1)d
 \sin({\widetilde \theta}_l^{\rm BS})\cos({\widetilde \varphi}_l^{\rm BS})
 + (n_{\rm BS}^{\rm v} - 1)d\sin({\widetilde \varphi}_l^{\rm BS})\right)/c . \tag{56} \label{eq_tau_tidle_BS}\\[2mm]
 [\bm{\widetilde H}_{{\rm DL},l}^{(t)}[k]]_{n_{\rm AC},n_{\rm BS}} &=
 \left[ \bm{\bar a}^*_{\rm AC}({\widetilde \mu}_l^{\rm AC},{\widetilde \nu}_l^{\rm AC},k)
 \right]_{n_{\rm AC}}  [\bm{H}_{{\rm DL},l}^{(t)}[k]]_{n_{\rm AC},n_{\rm BS}}
 \left[ \bm{\bar a}_{\rm BS}({\widetilde \mu}_l^{\rm BS},{\widetilde \nu}_l^{\rm BS},k) \right]_{n_{\rm BS}} . \tag{58} \label{eq_H_ttdu_k_dl_mn}\\[2mm]
 \bm{\widetilde A}_{{\rm DL},l}[k] &= \left( \bm{a}_{\rm AC}(\mu_l^{\rm AC},\nu_l^{\rm AC}) \circ
 \bm{\bar a}_{\rm AC}(\mu_l^{\rm AC},\nu_l^{\rm AC},k) \circ \bm{\bar a}^*_{\rm AC}({\widetilde \mu}_l^{\rm AC},
 {\widetilde \nu}_l^{\rm AC},k) \right) \nonumber \\
 &\quad \times \left( \bm{a}_{\rm BS}(\mu_l^{\rm BS},\nu_l^{\rm BS}) \circ \bm{\bar a}_{\rm BS}(\mu_l^{\rm BS},\nu_l^{\rm BS},k)
 \circ \bm{\bar a}^*_{\rm BS}({\widetilde \mu}_l^{\rm BS},{\widetilde \nu}_l^{\rm BS},k) \right)^{\rm H} \nonumber \\
 &= \bm{A}_{{\rm DL},l}[k] \circ \left( \bm{\bar a}^*_{\rm AC}({\widetilde \mu}_l^{\rm AC},{\widetilde \nu}_l^{\rm AC},k)
 \bm{\bar a}^{\rm T}_{\rm BS}({\widetilde \mu}_l^{\rm BS},{\widetilde \nu}_l^{\rm BS},k) \right) . \tag{60} \label{eq_A_ttdu_dl_k_2}
\end{align}
\hrulefill
\end{figure*}

 Taking all $N_{\rm AC}$ and $N_{\rm BS}$ antennas of THz UM-MIMO arrays at aircraft
 and the $l$th BS into consideration, the complete DL spatial-frequency channel matrix
 at the $k$th subcarrier of the $n$th OFDM symbol, i.e., $\bm{H}_{{\rm DL},l}^{[n]}[k]$
 in (\ref{eq_y_dl}), can be then formulated as
\setcounter{equation}{51}
\begin{align}\label{eq_H_k_dl_2} % eq 52
 \bm{H}_{{\rm DL},l}^{[n]}[k] =& \sqrt{G_l} \alpha_l e^{\textsf{j} {2\pi \psi_{l,k} (n-1)T_{\rm sym}}} e^{-\textsf{j} {2\pi \left({\textstyle{k-1 \over K}}-{\textstyle{1 \over 2}}\right)f_s \tau_l}} \nonumber\\
 &\times \bm{A}_{{\rm DL},l}[k] ,
\end{align}
 where the DL array response matrix $\bm{A}_{{\rm DL},l}[k]\! \in\! \mathbb{C}^{N_{\rm AC}\!\times\! N_{\rm BS}}$ associated with the array response vectors at aircraft and the $l$th BS is given by (\ref{eq_A_dl_k_2}) on the top of this page.
%\begin{align}\label{eq_A_dl_k_2} % eq 53
% \bm{A}_{{\rm DL},l}[k] =&\ \bm{a}_{\rm AC}(\mu_l^{\rm AC},\nu_l^{\rm AC},k)
% \bm{a}_{\rm BS}^{\rm H}(\mu_l^{\rm BS},\nu_l^{\rm BS},k) \nonumber \\
% =& \left( \bm{a}_{\rm AC}(\mu_l^{\rm AC},\nu_l^{\rm AC}) \circ
% \bm{\bar a}_{\rm AC}(\mu_l^{\rm AC},\nu_l^{\rm AC},k) \right)\!
% \left( \bm{a}_{\rm BS}(\mu_l^{\rm BS},\nu_l^{\rm BS}) \circ
% \bm{\bar a}_{\rm BS}(\mu_l^{\rm BS},\nu_l^{\rm BS},k) \right)^{\rm H} \nonumber \\
% {\mathop=\limits^{(a)}}& \left( \bm{a}_{\rm AC}(\mu_l^{\rm AC},\nu_l^{\rm AC})
% \bm{a}^{\rm H}_{\rm BS}(\mu_l^{\rm BS},\nu_l^{\rm BS}) \right)
% \circ \left( \bm{\bar a}_{\rm AC}(\mu_l^{\rm AC},\nu_l^{\rm AC},k)
% \bm{\bar a}^{\rm H}_{\rm BS}(\mu_l^{\rm BS},\nu_l^{\rm BS},k) \right),
%\end{align}
 In (\ref{eq_A_dl_k_2}), we have used the identity
 $(\bm{a}\! \circ\!\bm{b})(\bm{c}\! \circ\! \bm{d})^{\rm H}\! =\!
 (\bm{a}\bm{c}^{\rm H})\! \circ\! (\bm{b}\bm{d}^{\rm H})$ \cite{Liu_JISS08} in equation $(a)$.

\section{Proof of Lemma 1}

 After compensating the antenna transmission delay via the ideal TTDU module, the compensated
 ($n_{\rm AC},n_{\rm BS}$)th element of DL spatial-delay domain passband channel matrix
 $\bm{\bar H}_{{\rm DL},l}^{(t)}(\tau)$ in (\ref{eq_H_tau_dl_mn}), denoted by
 $[\bm{\widetilde{\bar H}}_{{\rm DL},l}^{(t)}(\tau)]_{n_{\rm AC},n_{\rm BS}}$, can be expressed as
\setcounter{equation}{53}
\begin{align}\label{eq_H_ttdu_tau_dl_mn} % eq 54
 [\bm{\widetilde{\bar H}}_{{\rm DL},l}^{(t)}(\tau)]_{n_{\rm AC},n_{\rm BS}} =&\!\
 \delta(\tau - {\widetilde \tau}_l^{[n_{\rm AC}]}) \circledast [\bm{\bar H}_{{\rm DL},l}^{(t)}(\tau)]_{n_{\rm AC},n_{\rm BS}} \nonumber\\
 &\circledast \delta(\tau - {\widetilde \tau}_l^{[n_{\rm BS}]}) ,
\end{align}
 where $\circledast$ represent the linear convolution operation,
 and ${\widetilde \tau}_l^{[n_{\rm AC}]}$ and ${\widetilde \tau}_l^{[n_{\rm BS}]}$
 are the compensated transmission delays yielded by TTDUs at aircraft and BSs,
 respectively, denoted by (\ref{eq_tau_tidle_AC}) and (\ref{eq_tau_tidle_BS}), respectively, on the top of this page.
%\begin{align} % eqs 55,56
% {\widetilde \tau}_l^{[n_{\rm AC}]} =& \left((n_{\rm AC}^{\rm h}\!-\!1)d
% \sin({\widetilde \theta}_l^{\rm AC})\cos({\widetilde \varphi}_l^{\rm AC})
% \!+\!(n_{\rm AC}^{\rm v}\!-\!1)d\sin({\widetilde \varphi}_l^{\rm AC})\right)/c , \label{eq_tau_tidle_AC}\\[-1mm]
% {\widetilde \tau}_l^{[n_{\rm BS}]} =& -\left((n_{\rm BS}^{\rm h}\!-\!1)d
% \sin({\widetilde \theta}_l^{\rm BS})\cos({\widetilde \varphi}_l^{\rm BS})
% \!+\!(n_{\rm BS}^{\rm v}\!-\!1)d\sin({\widetilde \varphi}_l^{\rm BS})\right)/c . \label{eq_tau_tidle_BS}
%\end{align}
 Similar to (\ref{eq_H_f_dl_mn}), by taking the Fourier transform of (\ref{eq_H_ttdu_tau_dl_mn})
 and the down-conversion, the baseband frequency domain response of
 $[\bm{\widetilde{\bar H}}_{{\rm DL},l}^{(t)}(\tau)]_{n_{\rm AC},n_{\rm BS}}$ is given by
\setcounter{equation}{56}
\begin{align}\label{eq_H_ttdu_f_dl_mn} % eq 57
 [\bm{\widetilde H}_{{\rm DL},l}^{(t)}(f)]_{n_{\rm AC},n_{\rm BS}} =&\!\
 e^{-\textsf{j} {2\pi f {\widetilde \tau}_l^{[n_{\rm AC}]}}} [\bm{H}_{{\rm DL},l}^{(t)}(f)]_{n_{\rm AC},n_{\rm BS}} \nonumber\\
 &\times e^{-\textsf{j} {2\pi f {\widetilde \tau}_l^{[n_{\rm BS}]}}} .
\end{align}
 The spatial-frequency channel coefficient at the $k$th subcarrier
 $[\bm{\widetilde H}_{{\rm DL},l}^{(t)}[k]]_{n_{\rm AC},n_{\rm BS}}$
 can be then written as (\ref{eq_H_ttdu_k_dl_mn}) on the top of this page.
%\begin{align}\label{eq_H_ttdu_k_dl_mn} % eq 58
% [\bm{\widetilde H}_{{\rm DL},l}^{(t)}[k]]_{n_{\rm AC},n_{\rm BS}} =&
% \left[ \bm{\bar a}^*_{\rm AC}({\widetilde \mu}_l^{\rm AC},{\widetilde \nu}_l^{\rm AC},k)
% \right]_{n_{\rm AC}}  [\bm{H}_{{\rm DL},l}^{(t)}[k]]_{n_{\rm AC},n_{\rm BS}}
% \left[ \bm{\bar a}_{\rm BS}({\widetilde \mu}_l^{\rm BS},{\widetilde \nu}_l^{\rm BS},k) \right]_{n_{\rm BS}} .
%\end{align}
 Finally, by collecting all $N_{\rm AC}$ and $N_{\rm BS}$ antennas of THz UM-MIMO arrays
 at aircraft and the $l$th BS, the compensated DL spatial-frequency channel matrix
 at the $k$th subcarrier of the $n$th OFDM symbol, i.e., $\bm{\widetilde H}_{{\rm DL},l}^{[n]}[k]$
 in (\ref{eq_H_ttdu_k_dl}), can be formulated as
\setcounter{equation}{58}
\begin{align}\label{eq_H_ttdu_k_dl_2} % eq 59
 \bm{\widetilde H}_{{\rm DL},l}^{[n]}[k] =& \sqrt{G_l} \alpha_l
 e^{\textsf{j} {2\pi \psi_{l,k} (n-1)T_{\rm sym}}}
 e^{-\textsf{j} {2\pi \left({\textstyle{k-1 \over K}}-{\textstyle{1 \over 2}}\right)f_s \tau_l}} \nonumber\\
 &\times \bm{\widetilde A}_{{\rm DL},l}[k] ,
\end{align}
 where the compensated DL array response matrix $\bm{\widetilde A}_{{\rm DL},l}[k]$ is given by (\ref{eq_A_ttdu_dl_k_2}) on the top of this page.
%\begin{align}\label{eq_A_ttdu_dl_k_2} % eq 60
% \bm{\widetilde A}_{{\rm DL},l}[k] =& \left( \bm{a}_{\rm AC}(\mu_l^{\rm AC},\nu_l^{\rm AC}) \circ
% \bm{\bar a}_{\rm AC}(\mu_l^{\rm AC},\nu_l^{\rm AC},k) \circ \bm{\bar a}^*_{\rm AC}({\widetilde \mu}_l^{\rm AC},
% {\widetilde \nu}_l^{\rm AC},k) \right) \nonumber \\
% \ & \times \left( \bm{a}_{\rm BS}(\mu_l^{\rm BS},\nu_l^{\rm BS}) \circ \bm{\bar a}_{\rm BS}(\mu_l^{\rm BS},\nu_l^{\rm BS},k)
% \circ \bm{\bar a}^*_{\rm BS}({\widetilde \mu}_l^{\rm BS},{\widetilde \nu}_l^{\rm BS},k) \right)^{\rm H} \nonumber \\
% =&\ \bm{A}_{{\rm DL},l}[k] \circ \left( \bm{\bar a}^*_{\rm AC}({\widetilde \mu}_l^{\rm AC},{\widetilde \nu}_l^{\rm AC},k)
% \bm{\bar a}^{\rm T}_{\rm BS}({\widetilde \mu}_l^{\rm BS},{\widetilde \nu}_l^{\rm BS},k) \right) .
%\end{align}

 The proof of Lemma 1 is completed.

\end{appendices}

\ifCLASSOPTIONcaptionsoff
  \newpage
\fi

\vspace{-1.0cm}
%%%%%%%%%%%%%%%%%%% IEEEbiography %%%%%%%%%%%%%%%%%%%

% Anwen Liao
%\begin{IEEEbiography}[{\includegraphics[width=0.9in,height=1.1in,clip,keepaspectratio]{PhotoLiao/PhotoLiao.eps}}]
% {Anwen Liao} (S'19) received the B.S. and M.S. degrees from the Beijing Institute of
% Technology, Beijing, China, in 2015 and 2018, respectively. He is currently pursuing
% the Ph.D. degree in the School of Information and Electronics of Beijing Institute of
% Technology, Beijing, China. His research interests include massive MIMO systems,
% millimeter-wave communications, hybrid beamforming, unmanned aerial vehicle
% communications, etc.
%\end{IEEEbiography}

\vspace{-1.0cm}

\end{document}